\documentclass[seceq]{ptptex}

\usepackage{graphicx}
\usepackage{wrapft}

\title{
Core-Collapse Supernovae%
}

\subtitle{Reflections and Directions}    

\author{
Hans-Thomas \textsc{Janka}$^{1}$,\\ 
Florian \textsc{Hanke}$^{1}$, 
Lorenz \textsc{H\"udepohl}$^{1}$,
Andreas \textsc{Marek}$^{1}$,\\ 
Bernhard \textsc{M\"uller}$^{1}$, 
and
Martin \textsc{Obergaulinger}$^{2}$
}

\inst{
$^1$Max Planck Institute for Astrophysics, Karl-Schwarzschild-Str.~1,\\
85748 Garching, Germany\\
$^2$Departament d{\'{}}Astronomia i Astrof{\'i}sica, Universitat de
Val{\`e}ncia,\\ Edifici d{\'{}}Investigaci{\'o} Jeroni Mu{\~n}oz, C/
Dr.~Moliner, 50,\\ E-46100 Burjassot (Val{\`e}ncia), Spain
}

\abst{
Core-collapse supernovae are among the most fascinating phenomena 
in astrophysics and provide a formidable challenge for theoretical
investigation.
They mark the spectacular end of the lives of massive stars and, 
in an explosive eruption, release as much energy as the sun produces
during its whole life. A better understanding of the astrophysical 
role of supernovae as birth sites of neutron stars, black holes, and 
heavy chemical elements, and more reliable predictions of the observable
signals from stellar death events are tightly linked to the solution of
the long-standing puzzle how collapsing stars achieve to explode. In 
this article our current knowledge of the processes that contribute
to the success of the explosion mechanism are concisely reviewed. 
After a short overview of the sequence of stages of stellar core-collapse
events, the general properties of the progenitor-dependent neutrino 
emission will be briefly described. Applying sophisticated neutrino transport
in axisymmetric (2D) simulations with general relativity as well as
in simulations with an approximate treatment of relativistic effects, 
we could find successful
neutrino-driven explosions for a growing set of progenitor 
stars. First results of three-dimensional (3D) models have been
obtained, and magnetohydrodynamic simulations demonstrate that
strong initial magnetic fields in the pre-collapse core can foster
the onset of neutrino-powered supernova explosions even in 
nonrotating stars. These results are discussed in the
context of the present controversy about the value of 2D simulations
for exploring the supernova mechanism in realistic 3D environments,
and they are interpreted against the background of the current 
disagreement on the question whether the standing accretion
shock instability (SASI) or neutrino-driven convection is the 
crucial agency that supports the onset of the explosion.
}

\PTPindex{483, 421, 423, 415, 451, 452, 242} 

\begin{document}

\maketitle

\section{Supernova theory in a nutshell}
\label{sec:theory}

\begin{wrapfigure}{r}{\halftext}
 \centerline{\includegraphics[width=\halftext,angle=270]{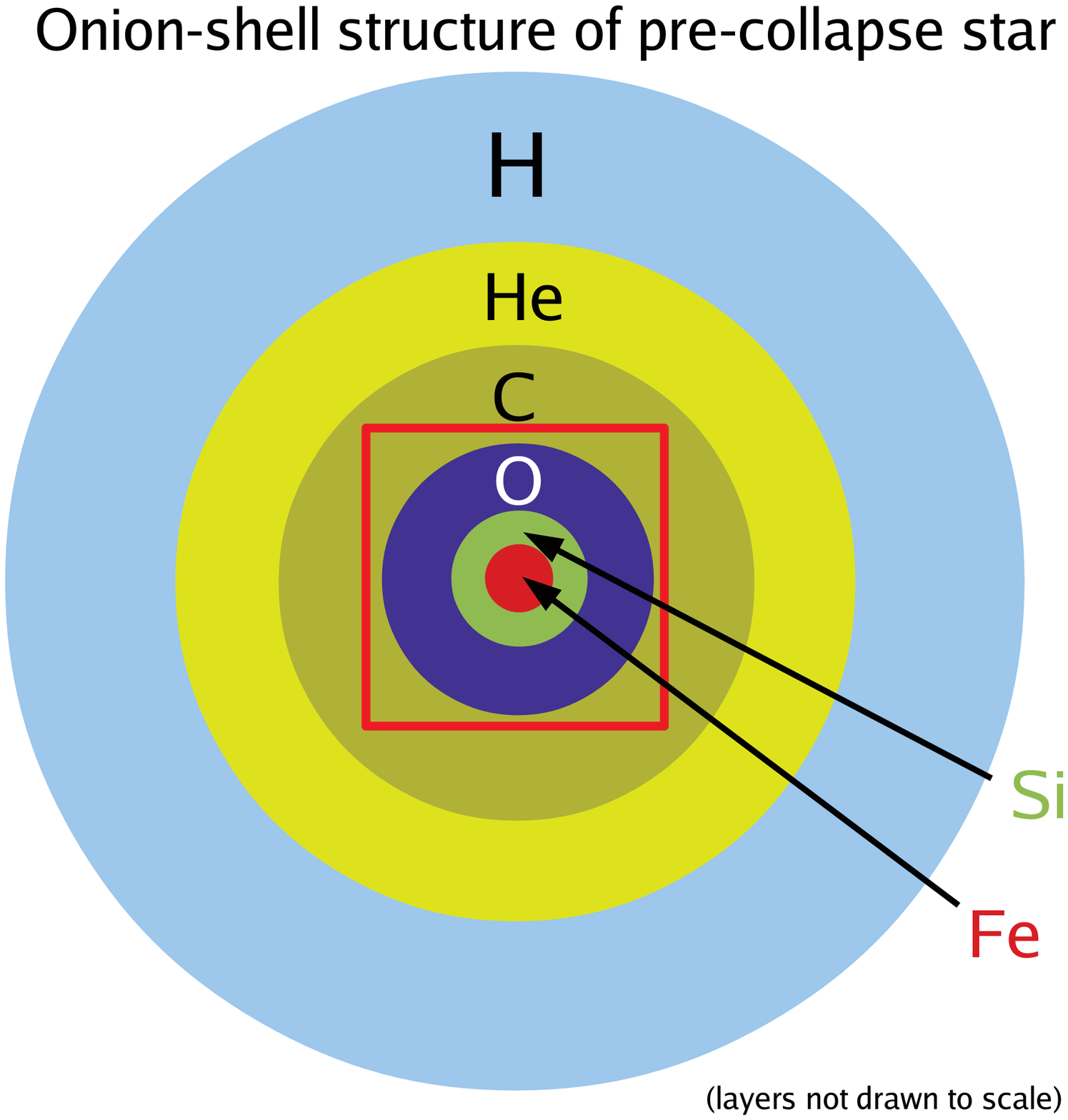}}
\caption{Schematic onion-shell structure of a supernova progenitor
  star before core collapse. Only the main elemental constituents of
  the different composition shells, which contain the products and ashes
  of the sequence of nuclear burning stages, are indicated. Note that the
  radial thickness of the layers is not drawn to scale. The red square
  indicates the inner region zoomed into in Fig.~\ref{fig:snphases}.}
  \label{fig:prog-onion}
\end{wrapfigure}

Massive stars in the range between $\sim$8\,$M_\odot$ and 
several 10\,$M_\odot$ develop low-entropy cores, in which
relativistic electrons dominate the pressure. Heavy nuclei
yield only a small, though important, contribution to providing 
stabilization against the inward pull of gravity.
The core consists of the final products of the star's nuclear
burning history. It is surrounded by concentric shells 
that, from outside inward, contain the successively heavier 
ashes of all previous burning stages (Fig.~\ref{fig:prog-onion}).

\begin{figure}
  \centerline{\includegraphics[width=0.45\textwidth,angle=270]{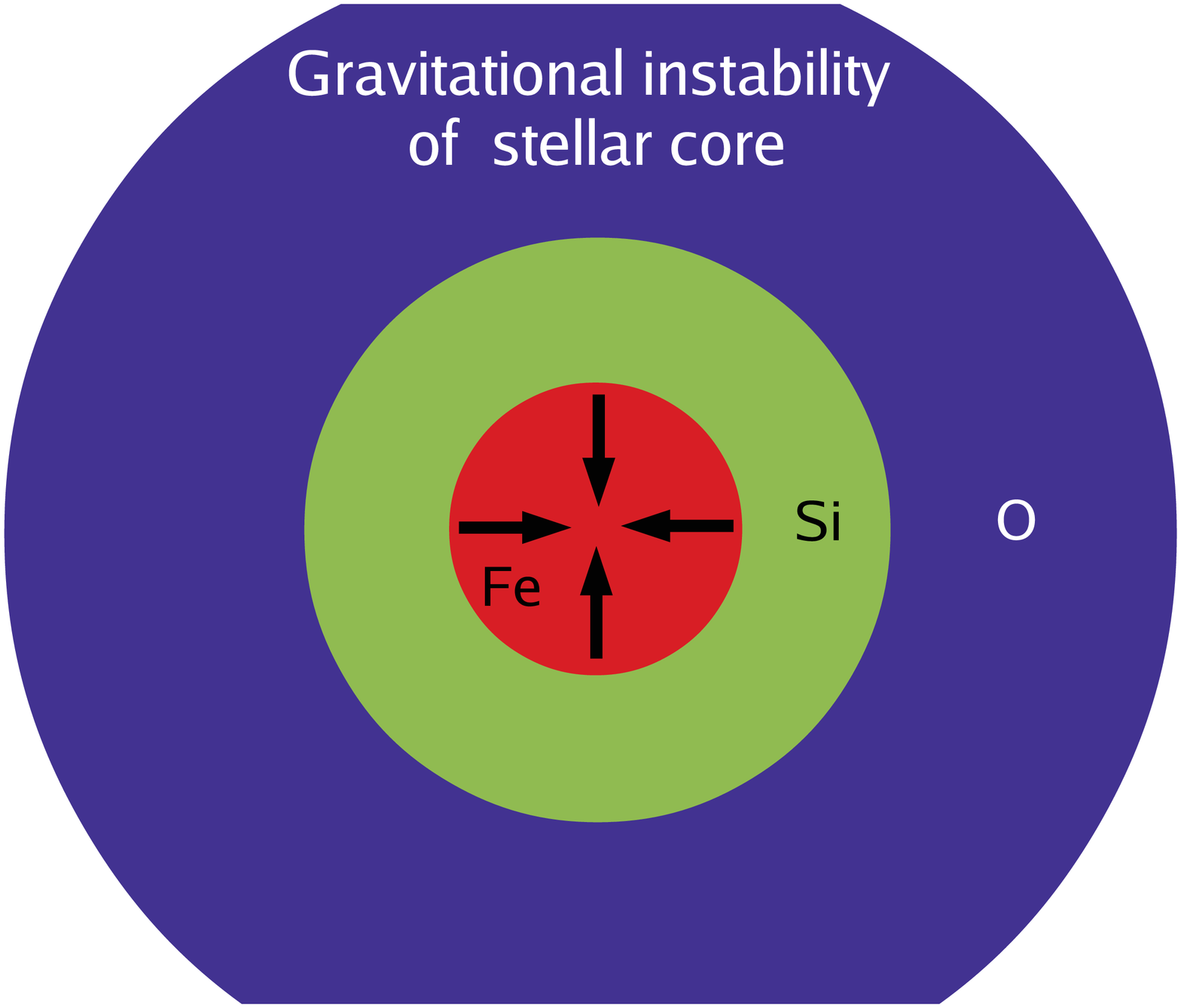}
              \includegraphics[width=0.45\textwidth,angle=270]{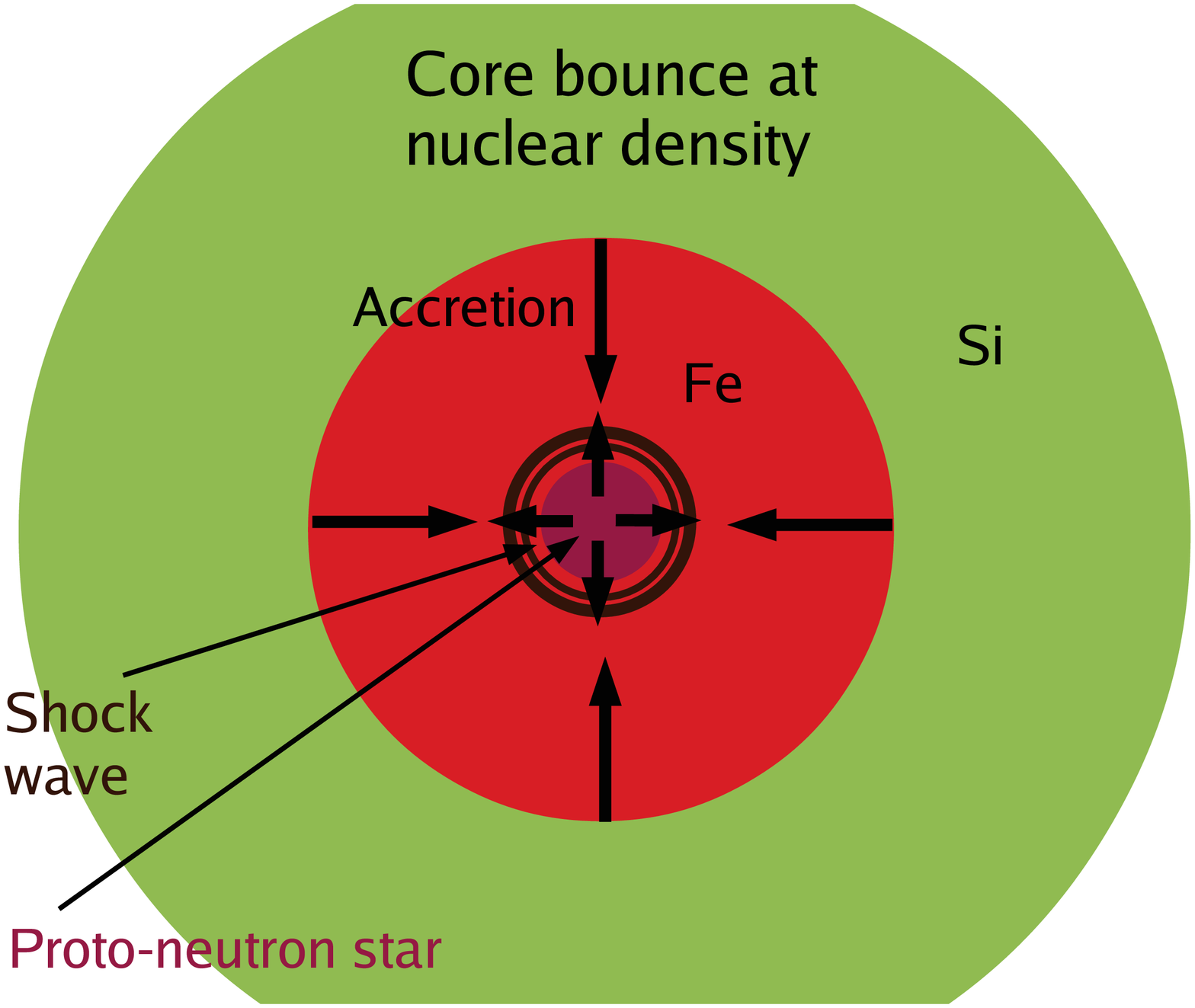}}\vspace{5pt}
  \centerline{\includegraphics[width=0.45\textwidth,angle=270]{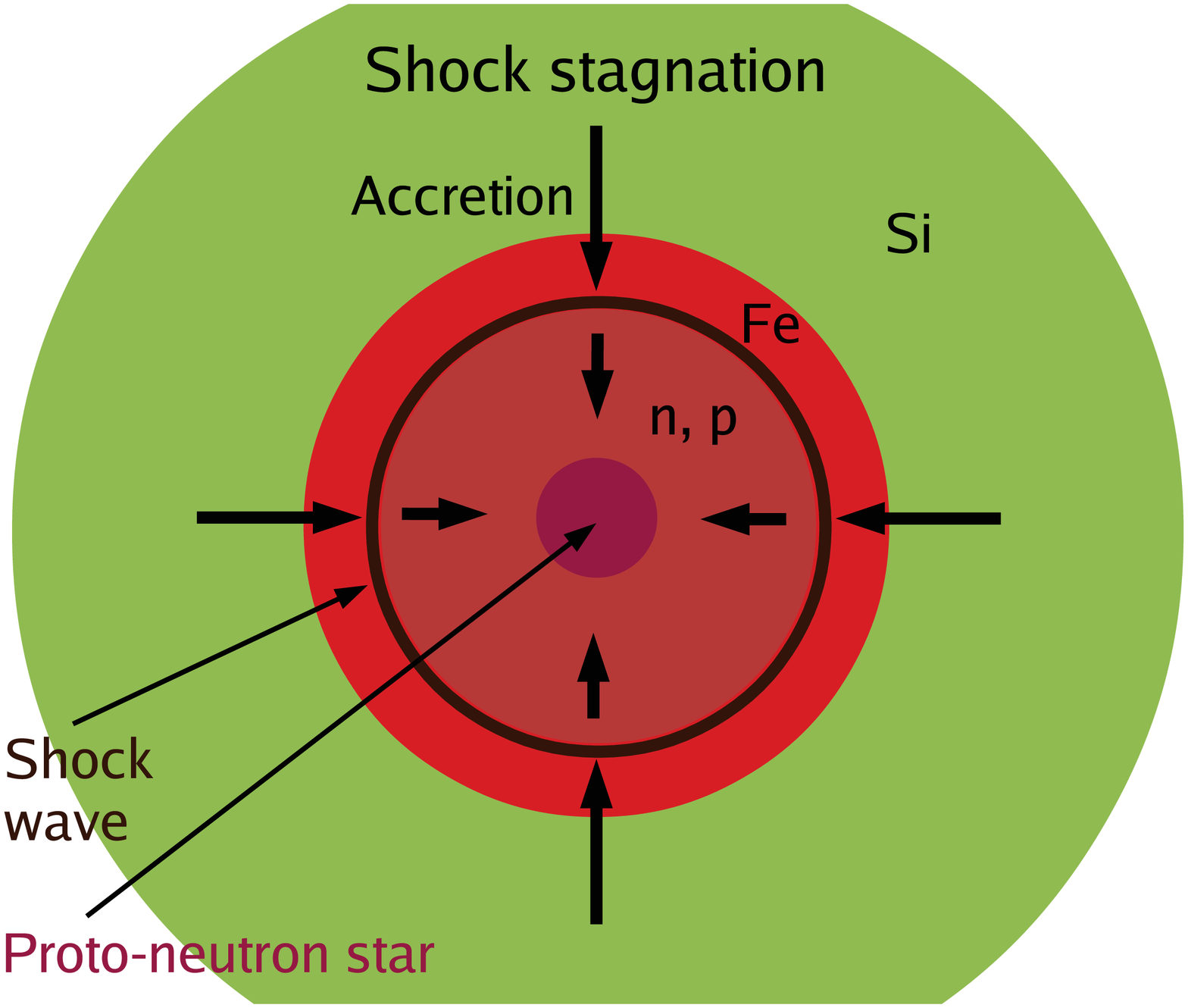}
              \includegraphics[width=0.45\textwidth,angle=270]{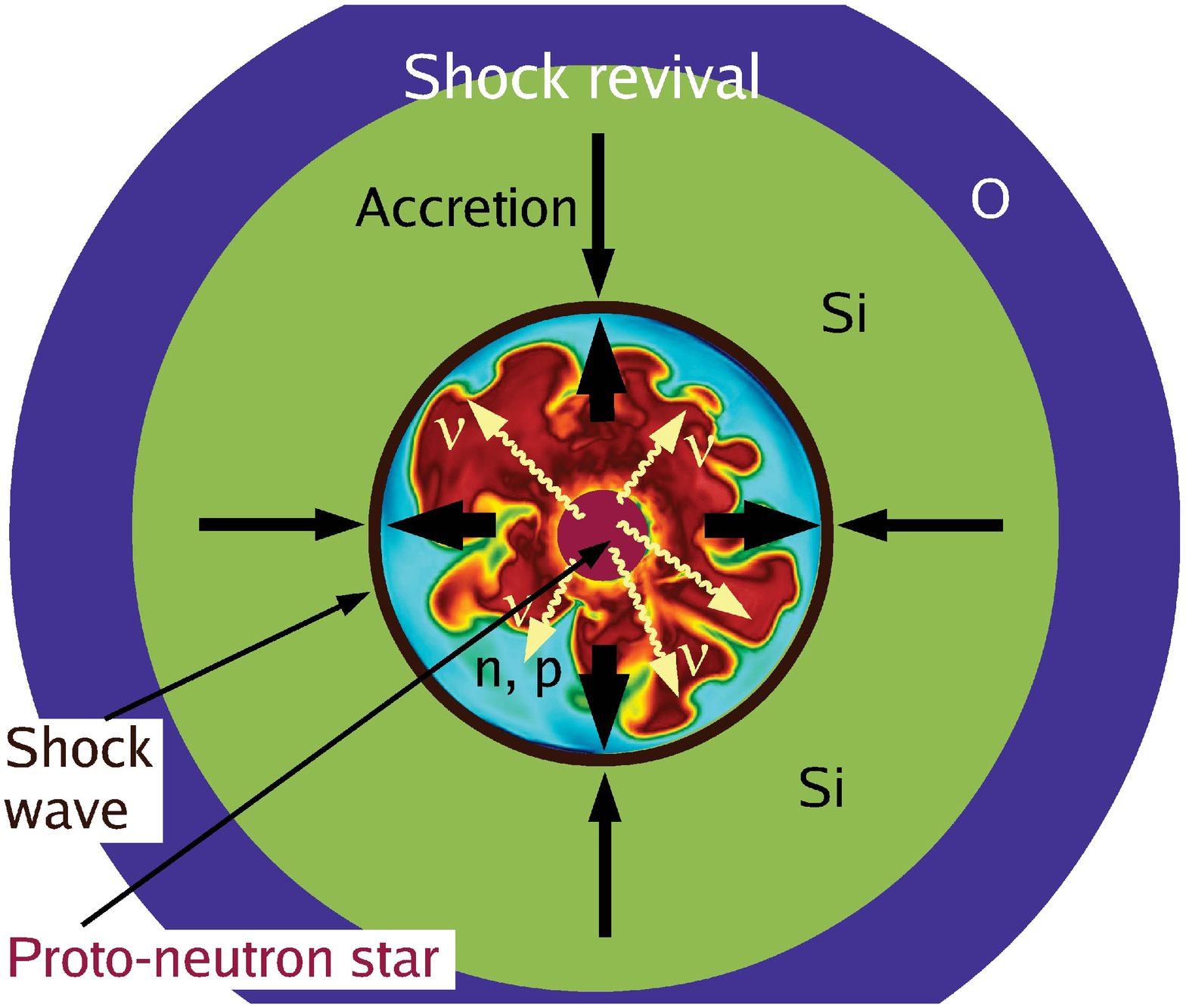}}\vspace{5pt}
  \centerline{\includegraphics[width=0.45\textwidth,angle=270]{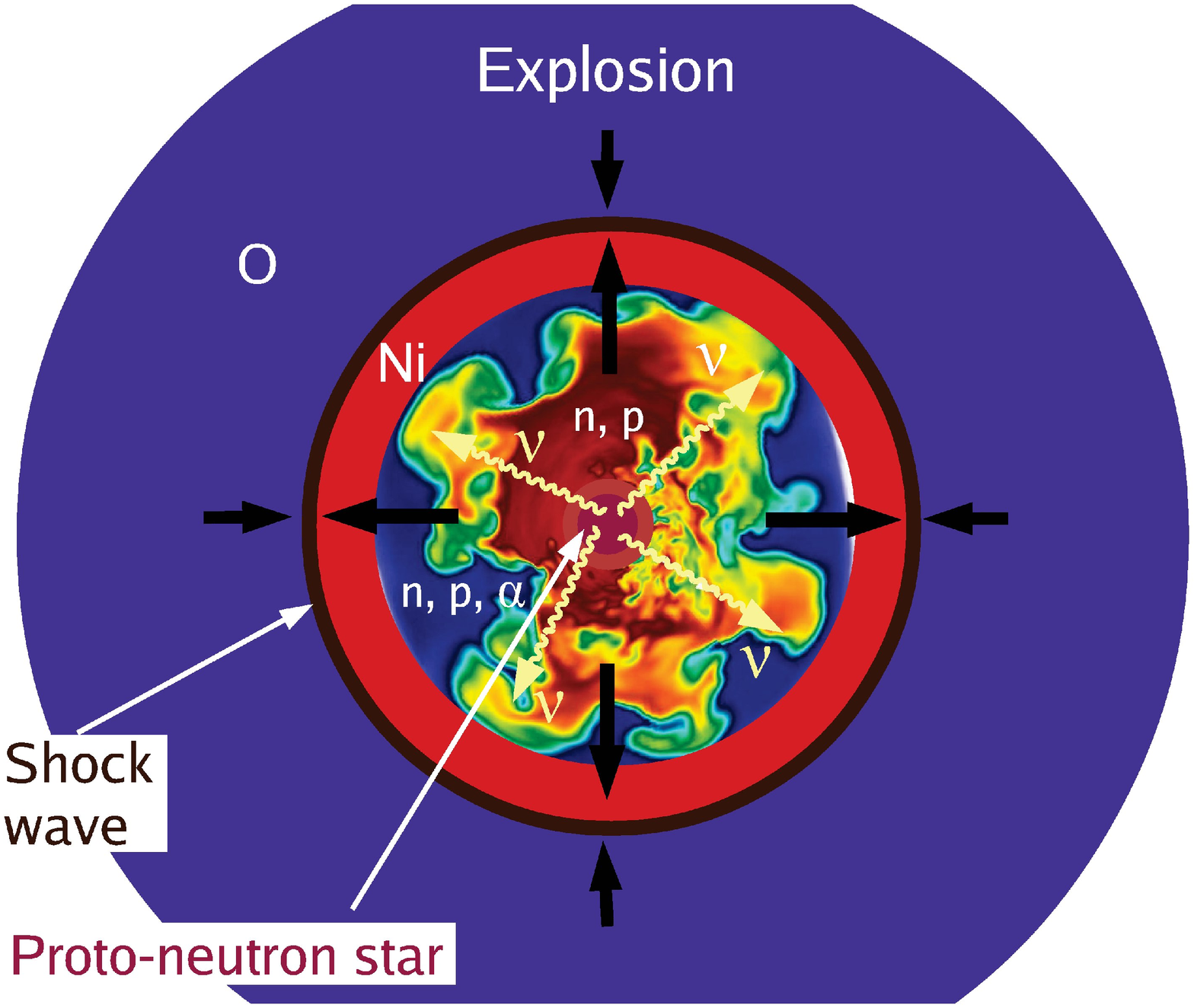}
              \includegraphics[width=0.45\textwidth,angle=270]{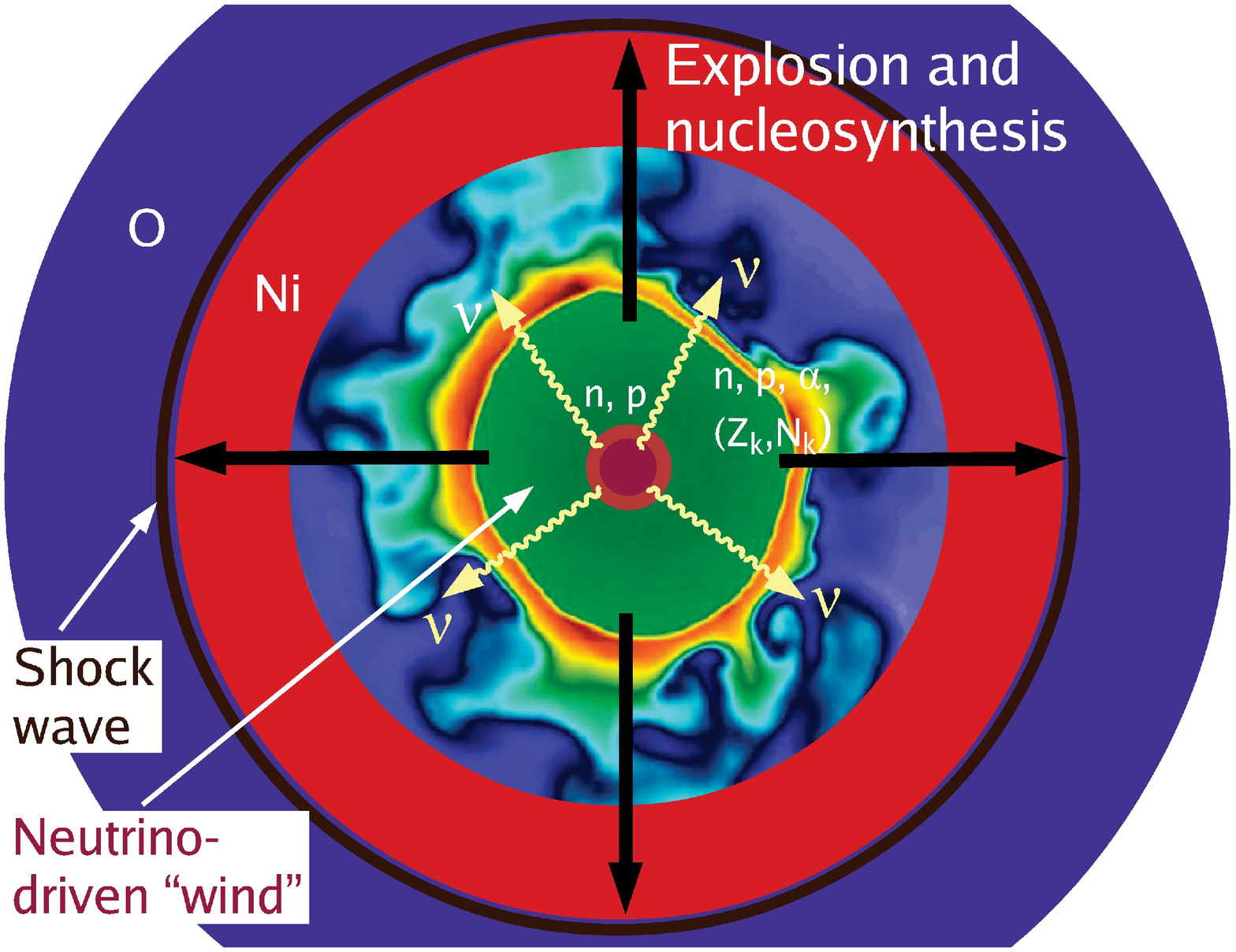}}
  \caption{Schematic representation of the evolution stages from the onset
           of stellar core collapse ({\em top left}) to the development of
           a supernova explosion on a scale of several 1000 kilometers.
           The displayed intermediate stages show the moment of core bounce
           and shock formation ({\em top right}), shock stagnation and onset
           of quasi-stationary accretion ({\em middle left}), beginning of
           the reexpansion of the shock wave (``shock revival'',
           {\em middle right}), and acceleration of the explosion
           ({\em bottom left}). Nickel formation is indicated in the
           matter heated by the outgoing shock, but also the rising
           bubbles of neutrino-heated ejecta and the essentially
           spherically symmetric neutrino-driven wind ({\em bottom right})
           are interesting sites for nucleosynthesis.}
  \label{fig:snphases}
\end{figure}

Shell burning leads to a continuous growth of the mass of the 
central core until gravitational instability finally sets in.
At this time the core resembles a hot white dwarf close to its
maximum mass of the order of the Chandrasekhar mass. 
It has a typical diameter of about 3000\,km,
a central temperature around $10^{10}\,$K (or approximately 1\,MeV), 
a central density of several $10^9\,$g\,cm$^{-3}$, an entropy of 
$\sim$0.7 to $\sim$1\,$k_{\mathrm B}$ per nucleon ($k_{\mathrm B}$ is 
Boltzmann's constant), and a proton-to-baryon ratio of around 0.45.
The accelerating contraction and ultimately the collapse of the 
degenerate core is initiated by the shift of nuclear statistical
equilibrium (NSE). As the temperature in the contracting core 
increases, high-energy photons produce a growing number of 
$\alpha$ particles and free nucleons. Thermal energy is thus
consumed to overcome the binding energy of heavy nuclei. This
endothermic process, which partially disintegrates the
iron-group material that had been assembled during the final
burning stage of the progenitor, lowers the effective adiabatic 
index\footnote{The effective (dynamically relevant) adiabatic 
index is defined as the logarithmic density derivative 
of the pressure, $\left (\partial\ln P/\partial\ln \rho\right )_m$,
along a fluid element's trajectory,
averaged over the volume of the collapsing core. It
governs the transition to gravitational instability and the
collapse dynamics.\cite{rf:ShapiroTeukolsky.1983}} 
below the critical value 
($\approx$4/3) for gravitational instability. 
With growing density, electron captures on heavy nuclei
and on free protons
become more and more important. The corresponding loss of 
lepton number by the production and escape of electron 
neutrinos softens the pressure increase with density even
further and accelerates the collapse (Fig.~\ref{fig:snphases},
top left panel). 

Neutrinos become trapped in the collapsing matter at a density 
of roughly $10^{12}$\,g\,cm$^{-3}$, at which conditions their
diffusion timescale out of the core begins to exceed the
freefall timescale of the gas. Despite
neutrino trapping, the collapse cannot be stopped before
nuclear matter density ($\sim$$2.7\times 10^{14}$\,g\,cm$^{-3}$)
is reached and the equation of state stiffens because of 
repulsive contributions to the nucleon interaction potential. 
When the homologously and subsonically collapsing inner core 
decelerates and rebounds into the surrounding, supersonically 
infalling layers, sound waves steepen into a shock front
(Fig.~\ref{fig:snphases}, top right panel). This shock front
expands into the overlying Fe-core material, but is 
quickly damped by energy losses due to the dissociation of 
Fe-group nuclei into free nucleons, which extracts a thermal
energy of about 8.8 MeV per nucleon from the postshock matter.
Only 1--2\,ms after shock formation, the velocities downstream
of the shock have thus become negative. Nevertheless, the shock
continues to propagate outward in mass and radius because initially
the high mass accretion rate of up to several solar masses per 
second leads to the accumulation of a thick layer of dense matter
behind the shock. Only when the mass accretion rate has 
decayed sufficiently and the hot, bloated mantle of the 
proton-neutron star begins to shrink in response to the lepton
number and energy loss through neutrinos, 
the radial shock expansion comes to a halt
and the shock forms a stagnant accretion shock at a radius 
between 100 and 200\,km
(Fig.~\ref{fig:snphases}, middle left panel). Quasi-stationary
conditions apply later on with only slow changes of the 
mass accretion rate, $\dot M$, neutron star mass $M_\mathrm{ns}$
and radius $R_\mathrm{ns}$, and neutrino emission parameters
(luminosity $L_\nu$ and mean spectral energy $\left\langle
\epsilon_\nu\right\rangle$). In nonexploding spherically
symmetric (i.e., one-dimensional, 1D) simulations the
shock retreats and its radius follows the contraction of the
nascent neutron star roughly according to the
relation.\cite{rf:Janka.2012}
\begin{equation}
R_\mathrm{s} \propto
\frac{(L_\nu\left\langle\epsilon_{\nu}^2\right\rangle)^{4/9}
R_\mathrm{ns}^{16/9}}{\dot M^{2/3}M_\mathrm{ns}^{1/3}}
\, .
\label{eq:shockradius}
\end{equation}
High mass accretion rates therefore tend to damp the shock
expansion while neutrino-energy deposition behind the shock, 
which depends on the product
$L_\nu\left\langle\epsilon_{\nu}^2\right\rangle$\footnote{The
energy transfer by neutrinos scales linearly with the neutrino 
luminosity and the average interaction cross section. The
latter increases roughly with the luminosity-averaged
square of the neutrino energy.}, can drive shock 
expansion. This issue will be elaborated on further below.

In order to successfully launch a supernova explosion, some
mechanism is necessary by which the stalled shock can be revived. 
Such a mechanism needs to tap the huge reservoir of gravitational
binding energy that is released during the formation of a neutron
star. During the infall of the stellar core the energy is first
converted to internal energy by hydrodynamic forces (i.e., 
compression and the viscous 
dissipation of kinetic energy in matter decelerated in the
accretion shock). The degeneracy and thermal energy of electrons 
and nucleons thus stored in the proto-neutron
is subsequently radiated away by neutrinos over a timescale of 
many seconds. 

\begin{wrapfigure}{r}{\halftext}
 \centerline{\includegraphics[width=0.45\textwidth,angle=270]{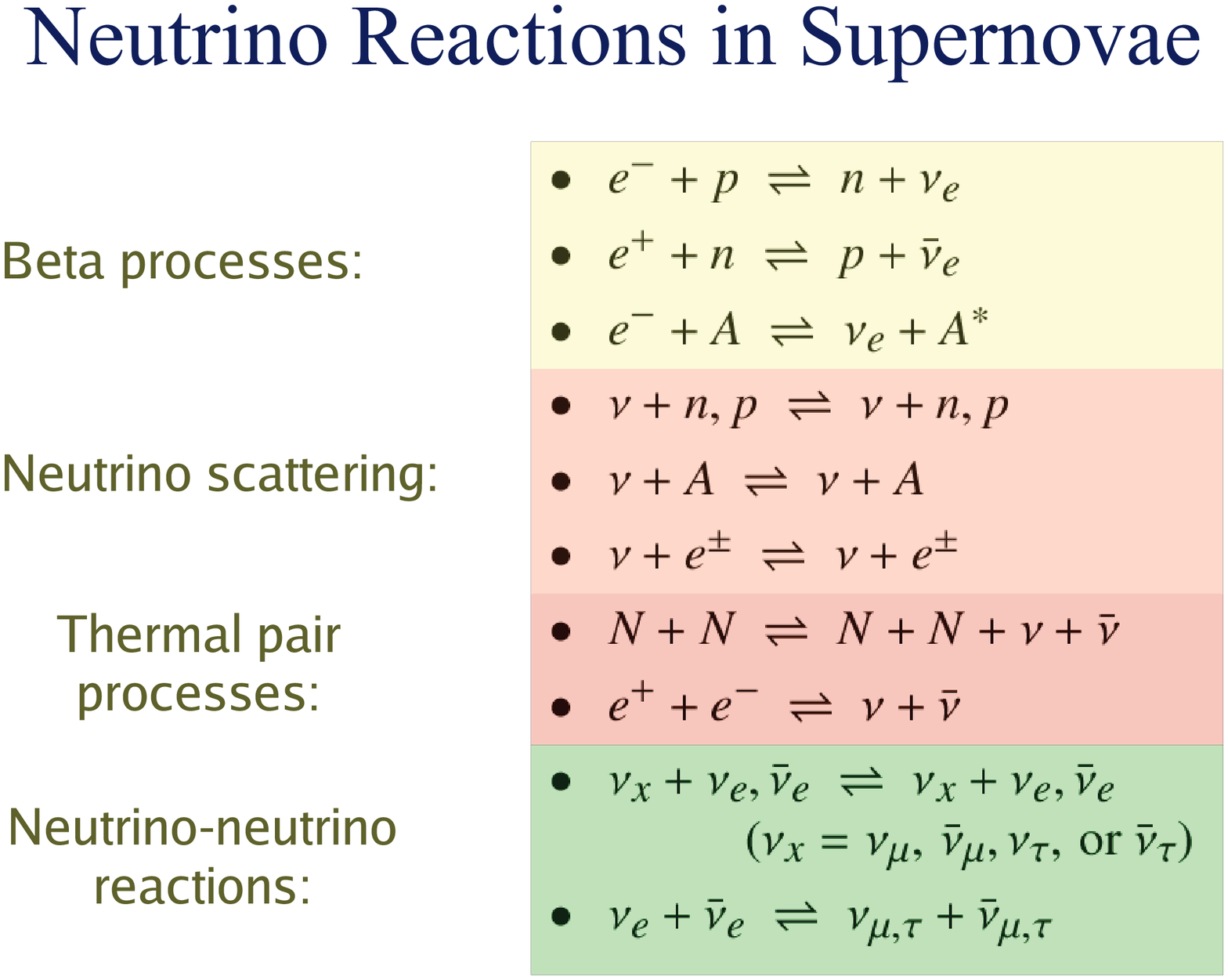}}
\caption{Summary of important neutrino reactions in the supernova core
  and/or nascent neutron
  star\cite{rf:RamppJanka.2002,rf:Buras.etal.2006a,rf:Mueller.etal.2010}.
  The symbol $\nu$ can mean any type of
  neutrino, $A$ represents an atomic nucleus, and $N$ means neutron
  ($n$) or proton ($p$).}
  \label{fig:neutrinos}
\end{wrapfigure}

Deep in the highly degenerate neutron-star interior electron
neutrinos, $\nu_e$, are first produced by electron captures on
protons. On their diffusive propagation towards the 
neutrinosphere, these electron 
neutrinos lose some of their energy in absorption-reemission
processes as well as in scattering reactions with electrons and
free neutrons and protons (Fig.~\ref{fig:neutrinos}). This effect
together with the gravitational settling and compression of the
outer layers of the proto-neutron star initially leads to 
rising temperatures before after some seconds cooling sets in.
Since the degeneracy is partially lifted in the hot proto-neutron
star mantle, the secondary production
of electron antineutrinos, $\bar\nu_e$, by positron captures on 
neutrons becomes possible. Neutrino-antineutrino pairs of all 
three flavors are created by thermal processes, i.e.,
nucleon-nucleon bremsstrahlung and electron-positron annihilation. 
Pure neutrino reactions (Fig.~\ref{fig:neutrinos})
also contribute to the shaping of the emitted spectra
of muon and tau neutrinos and antineutrinos ($\nu_\mu$,
$\bar\nu_\mu$, $\nu_\tau$, $\bar\nu_\tau$)\cite{rf:Keil.etal.2003}, 
which are not produced by fast beta reactions and thus are less
tightly coupled to the stellar medium.

Even a small fraction of the huge energy reservoir 
of several $10^{53}$\,ergs carried away by neutrinos 
is already sufficient to account for the canonical explosion
energy of a core-collapse supernova, which ranges between
some $10^{50}$\,erg to around $10^{51}$\,erg.
It may appear astonishing that the explosion selects an
energy scale that is 2--3 orders of magnitude lower than 
the reservoir of available energy. The energy scale of the
explosion, however, is not set by the neutron star binding
energy but by the structure of the progenitor configuration.
The degenerate iron core closely resembles a white dwarf, embedded
in the more or less dense stratification of concentric stellar
shells containing lighter nuclear burning products. The 
energy scale of the 
explosion is determined by the binding energy of the 
progenitor layers in immediate vicinity of the initial mass 
cut between proto-neutron star and supernova ejecta\footnote{Note
that the collapse roughly conserves the total energy of the 
infalling shells so that the energetic considerations are
possible for the progenitor conditions.}. For typical
progenitor stars the binding energy of the silicon shell and 
overlying layers is of the order of $10^{50}$--$10^{51}$\,ergs,
similar to the binding energy of the pre-collapse Fe core.
Any self-regulated mechanism will tend to deposit an energy of 
this magnitude: Once matter has received this amount of energy, 
it will tend to become unbound and will expand away from what is
going to become the bifurcation region between compact remnant
and supernova ejecta.

The {\em delayed neutrino-driven explosion 
mechanism}\cite{rf:BetheWilson.1985} is such a 
self-regulated process, because after absorbing energy in
interactions with neutrinos, the heated matter expands away from
the neutrino-heating region. This naturally
limits the energy input. Neutrinos are produced in huge numbers
in the dense and very hot interior of the nascent neutron star
and leave the neutrinospheric region with roughly 
black-body spectra and temperatures of typically 3--6 MeV.
This is considerably hotter than the gas in the layer between
neutrinosphere and stalled shock. Neutrino heating and cooling
in this region is dominated by the absorption of electron neutrinos
and antineutrinos on free neutrons and protons and the inverse of
these processes,
\begin{eqnarray}
\nu_e + n &\longleftrightarrow& p + e^- \, ,\label{eq:betanue} \\
\bar\nu_e + p &\longleftrightarrow& n + e^+ \, .
\label{eq:betaantinue}
\end{eqnarray}
The corresponding rates per nucleon $q_{\nu}^+$ and 
$q_{\nu}^-$ for the energy 
input and loss of the stellar medium are approximately given
by:\cite{rf:Janka.2001}
\begin{eqnarray}
q^+_{\nu_e+\bar\nu_e} &\approx& 160\,\,
\frac{L_{\nu}}{10^{52}\,\mathrm{erg/s}}
\left (\frac{r}{100\,\mathrm{km}}\right )^{-2} 
\left (\frac{k_\mathrm{B}T_{\nu}}{4\,\mathrm{MeV}}\right )^2\
\mathrm{MeV\,s^{-1}\ per\ nucleon,} \label{eq:rate1}\\
q^-_{\nu_e+\bar\nu_e} &\approx& 145\,
\left (\frac{k_\mathrm{B}T_\nu}{2\,\mathrm{MeV}}\right )^6\
\mathrm{MeV\,s^{-1}\ per\ nucleon.} 
\label{eq:rate2}
\end{eqnarray}
Here $L_\nu$ and $T_\nu$ are the luminosity and spectral temperature
of either $\nu_e$ or $\bar\nu_e$ (whose emission characteristics
during the shock revival phase are similar), the squared 
radius $r$ measures the geometric dilution of the flux leaving 
the neutrinosphere, and $T(r)$ is the local gas temperature. 
The heating and cooling rates in Eqs.~(\ref{eq:rate1}) and
(\ref{eq:rate2}) have roughly the same magnitude. This means that
both processes are in tight competition. Indeed, the neutrinosphere
is surrounded by a cooling layer, in which neutrino 
losses dominate. Since, however, the temperature 
declines roughly like $r^{-1}$, the cooling rate falls off
with $r^{-6}$ and therefore much steeper than the heating rate,
which follows basically an $r^{-2}$ dependence. 
Consequently, there must be a ``gain radius'' $R_\mathrm{g}$, 
outside of which neutrino heating becomes stronger than
neutrino cooling.\cite{rf:BetheWilson.1985}

Neutrinos can deposit considerable amounts of
energy in the layer between $R_\mathrm{g}$ and 
the shock position at radius $R_\mathrm{s}$, where most of the 
nuclei are dissociated into free nucleons (the preheating in
the rapidly infalling, undissociated material ahead of the shock 
is small). The optical depth for $\nu_e$ and $\bar\nu_e$ 
absorption in the gain layer between $R_\mathrm{g}$ and $R_\mathrm{s}$
can be estimated to be\cite{rf:Janka.2012} 
\begin{equation}
\tau_\mathrm{gain} \approx 0.026\,\left ( 
\frac{k_\mathrm{B}T_\nu}{4\,\mathrm{MeV}}\right )^{\! 2}
\left (\frac{\dot M}{0.1\,M_\odot/\mathrm{s}}\right )
\left ( \frac{R_\mathrm{s}}{200\,\mathrm{km}} \right )^{\! 3/2}
\left ( \frac{R_\mathrm{g}}{100\,\mathrm{km}} \right )^{\! -2}
\left (\frac{M_\mathrm{ns}}{1.5\,M_\odot} \right )^{\! -1/2} \,,
\label{eq:optdepth}
\end{equation}
where $\dot M$ and $M_\mathrm{ns}$ are again the rate of mass 
accretion by the shock and the neutron star mass, respectively.
For typical mass accretion rates of 
$\sim$0.1--0.3\,$M_\odot$\,s$^{-1}$ and neutrino spectral
temperatures $k_\mathrm{B}T_\nu \approx 4$--6\,MeV the optical
depth is 0.05--0.1, which means that up to about 10\% of the 
through-going neutrinos can be captured by a neutron or proton.

If the neutrino energy deposition is sufficiently strong, the
stalled shock can be revived to potentially initiate a 
successful supernova explosion (Fig.~\ref{fig:snphases}, 
middle right panel). The threshold for runaway conditions can be 
coined in terms of a critical luminosity\cite{rf:BurrowsGoshy.1993}, 
$L_{\nu,\mathrm{c}}(\dot M)$ (for $\nu = \nu_e$ or $\bar\nu_e$),
which depends on the mass accretion rate of the shock.
A steady accretion shock cannot be maintained when the neutrino 
luminosity exceeds this critical value (for a very detailed analysis,
see Ref.~\citen{rf:Fernandez.2012}). The threshold value increases
with higher $\dot M$, because neutrino heating has to overcome the
higher ram pressure of the infalling material. Due to its stronger
gravity also a larger neutron star mass $M_\mathrm{ns}$ 
requires a higher neutrino luminosity for shock revival.
The dependence of $L_{\nu,\mathrm{c}}$ on $\dot M$ and $M_\mathrm{ns}$
can be analytically estimated on grounds of simple considerations. 
Shock expansion can become strong when neutrinos heat the postshock 
matter faster than it can be advected with the accretion 
flow from the shock downward 
through the gain radius into the cooling layer, where the gas 
loses its energy by neutrino 
emission.\cite{rf:JankaKeil.1998,rf:Thompson.etal.2005,rf:Buras.etal.2006b} \
This requirement is confirmed by all existing numerical simulations
(e.g., \citen{rf:Buras.etal.2006b,rf:MurphyBurrows.2008,rf:MarekJanka.2009,rf:Mueller.etal.2012a,rf:Mueller.etal.2012b,rf:Dolence.etal.2012}).
Thus requiring the neutrino-heating timescale to be shorter than the 
advection timescale through the gain layer one obtains for the
critical luminosity:\cite{rf:Janka.2012}
\begin{equation}
L_{\nu,\mathrm{c}}(\dot M) \propto 
\beta^{-2/5}\,\dot M^{2/5}\,M_\mathrm{ns}^{4/5} \,,
\label{eq:critlum}
\end{equation}
where $\beta$ parametrizes the ratio of postshock density to preshock
density\footnote{The numerical factor in the scaling relation is found
to be $\sim$(5--6)$\times 10^{52}$\,erg/s for $\beta \sim 10$,
$\dot M = 1\,M_\odot$/s, and $M_\mathrm{ns} = 1.5\,M_\odot$.}.
This functional dependence nicely fits the critical curves obtained in
Ref.~\citen{rf:BurrowsGoshy.1993},
but it predicts considerably steeper power-law dependences on 
$M_\mathrm{ns}$ and $\dot M$ than the critical explosion condition
of Eq.~(11) in Ref.~\citen{rf:Dolence.etal.2012}. It should be noted,
however, that in the latter reference the optical depth was used as 
a constant free fit parameter but should actually be expressed as
function of the fundamental quantities ($M_\mathrm{ns}$, $\dot M$)
that govern the overall dynamics and structure of the postshock layer.
 
Convective overturn in the neutrino-heating
region has been recognized to support the onset of the 
explosion\cite{rf:Herant.etal.1994,rf:Burrows.etal.1995,rf:JankaMueller.1996}
and to lower the critical 
luminosity.\cite{rf:JankaMueller.1996,rf:YamasakiYamada.2006,rf:MurphyBurrows.2008,rf:Nordhaus.etal.2010,rf:Hanke.etal.2012} \
Nonradial and partially turbulent 
mass motions do not only stretch the dwell time of 
matter in the gain layer. Convective downdrafts carry postshock
material to the immediate vicinity of the 
gain radius, where neutrino heating is strongest.
Moreover, the outward rise and expansion cooling of neutrino-heated
gas in buoyant high-entropy bubbles reduce the energy
loss by reemission of neutrinos and push the shock
farther out. The residency time of matter in the
gain layer is thus prolonged even more. 
This combination of favorable circumstances is
crucial for the development of runaway conditions.
Nonradial oscillations and sloshing motions of the shock, which
are associated with the growth of the {\bf s}tanding {\bf a}ccretion 
{\bf s}hock
{\bf i}nstability (SASI\cite{rf:Blondin.etal.2003}), seem to have
a similarly favorable influence on the conditions for neutrino-driven
explosions.\cite{rf:Scheck.etal.2008,rf:MarekJanka.2009,rf:Mueller.etal.2012a,rf:Mueller.etal.2012b} \ 
All these effects reduce the critical value of the neutrino 
luminosity in multi-dimensional models compared to the 
1D case.\cite{rf:JankaMueller.1996,rf:MurphyBurrows.2008,rf:Nordhaus.etal.2010,rf:Hanke.etal.2012,rf:Dolence.etal.2012} 

When the blast wave takes off, explosive burning leads to the
production of $^{56}$Ni and other radioactive species in the
shock-heated silicon and/or oxygen layers (Fig.~\ref{fig:snphases},
bottom left panel). Recombination of nucleons to alpha particles
and Fe-group nuclei in the neutrino-heated high-entropy plumes
contributes to the nucleosynthetic yields, and the neutrino-driven
baryonic outflow (``neutrino wind''), which is shed off the surface
of the nascent neutron star by neutrino heating to fill the
surroundings of the compact remnant after accretion has ended
(Fig.~\ref{fig:snphases}, bottom right panel),
is considered as an interesting site for the formation of 
trans-iron nuclei up to $A \sim 110$ in a weak r-process and
of p-rich isotopes in the neutrino-proton process, depending on
whether the wind develops a neutron or proton excess (for a 
review, see Ref.~\citen{rf:ArconesThielemann.2012}). 
Both the electron
fraction, $Y_e$, and the entropy in the early neutrino-heated ejecta 
and in the neutrino-driven wind have a crucial
influence on the nucleosynthesis, and both are set by the neutrino 
interactions of Eqs.~(\ref{eq:betanue}) and (\ref{eq:betaantinue}). 
Therefore they
depend sensitively on the emission properties (luminosities
and spectra) of the $\nu_e$ and $\bar\nu_e$ radiated by the 
forming neutron star and on the expansion dynamics of the 
ejecta (which determines the time interval of intense 
neutrino interactions). Also neutrino oscillations, in particular
collective neutrino flavor transformations, can have an impact on
the neutron-to-proton ratio that develops in the ejecta on their
way away from the neutrino source 
(e.g., Ref.~\citen{rf:Tamborra.etal.2012}).

It should be noted that the explosion energy of the supernova
is {\em not} determined at the instant when the explosion sets
in; in particular it is not given by the energy neutrinos have
transferred to the postshock matter up to this time. Instead,
the total energy (i.e., internal plus gravitational plus the 
initially much smaller kinetic energy) 
of the gain layer is close to zero
when the shock begins to propagate outward, which means that this 
layer is only marginally unbound.\cite{rf:Janka.2001,rf:Fernandez.2012} \
This is observed in all numerical models of neutrino-driven 
explosions (e.g., 
\citen{rf:Scheck.etal.2006,rf:MarekJanka.2009,rf:Mueller.etal.2012a}).
Instead, the explosion energy builds up only on a longer timescale
of several 100\,ms to more than a second. There are several 
sources that contribute to the final explosion energy
(see results and discussions in Refs.~\citen{rf:Scheck.etal.2006,rf:Ugliano.etal.2012,rf:MarekJanka.2009}).
First, neutrinos continue to heat ``cool'' gas that is freshly accreted
through the shock and channelled towards the gain radius in 
convective downdrafts to replace there the hot matter that expands 
in high-entropy bubbles driving the shock expansion.\cite{rf:MarekJanka.2009} \
In addition, the recombination of free nucleons to $\alpha$ particles
and heavy nuclei in the ejecta releases up to $\sim$9\,MeV per
nucleon and is a very efficient source of energy for the developing 
explosion (0.1\,$M_\odot$ of nucleonic matter can provide up to
about $1.7\times 10^{51}$\,erg of recombination energy)
as pointed out in Ref.~\citen{rf:Scheck.etal.2006}.
It should be noted that the production of free nucleons by 
nuclear photodisintegration in the infalling matter ---either
when the gas passes the accretion shock or when it is compressed and 
neutrino heated in the accretion downflows--- mostly
taps the gravitational binding energy and only to a smaller
extent is fuelled by neutrino absorption. The conversion of 
gravitational binding energy to nuclear photodisintegration energy 
during
the infall is therefore an important energy storage whose contents
are released when the matter is subsequently reejected
and cools in the explosive outflow. Similarly, the
neutrino-driven wind blown off the hot neutron star's surface
gains power from neutrino heating as well as nuclear recombination.
Its energy can yield a significant contribution to the energy budget 
of the supernova explosion. Scheck et al.\cite{rf:Scheck.etal.2006} 
found that $\sim$30--70\% of the blast-wave energy can be provided by 
this long-term outflow from the proto-neutron star (with higher
relative importance for stronger explosions).
In contrast, nuclear burning of silicon and oxygen to nickel
in the shock-heated outer layers 
yields only a smaller amount of extra energy for the supernova;
the production of 0.1\,$M_\odot$ of iron-group material (a typical
number for normal core-collapse supernovae) releases
only about $10^{50}$\,erg. During the first $\sim$1--3\,s seconds of
the explosion all these energy souces have to provide enough energy
to unbind the overlying stellar layers (whose
binding energy can range between $\lesssim$$10^{50}$\,erg
and about $10^{51}$\,erg, depending on the core mass and 
compactness of the progenitor star) and, beyond that, to account for the 
measurable kinetic energy of the supernova. If the blast-wave energy
is too low, some of the matter swept up by the outgoing shock
will not be able to escape to infinity and will fall back onto the
compact remnant. Because much or even most of the nucleosynthesized
$^{56}$Ni may thus be accreted instead of being expelled, such 
fallback supernovae are expected to be faint and hard to observe.
It is therefore empirically unclear which fraction of stellar core 
collapses might belong to such types of events.

\begin{figure}
 \centerline{\includegraphics[width=0.7\textwidth,angle=270]{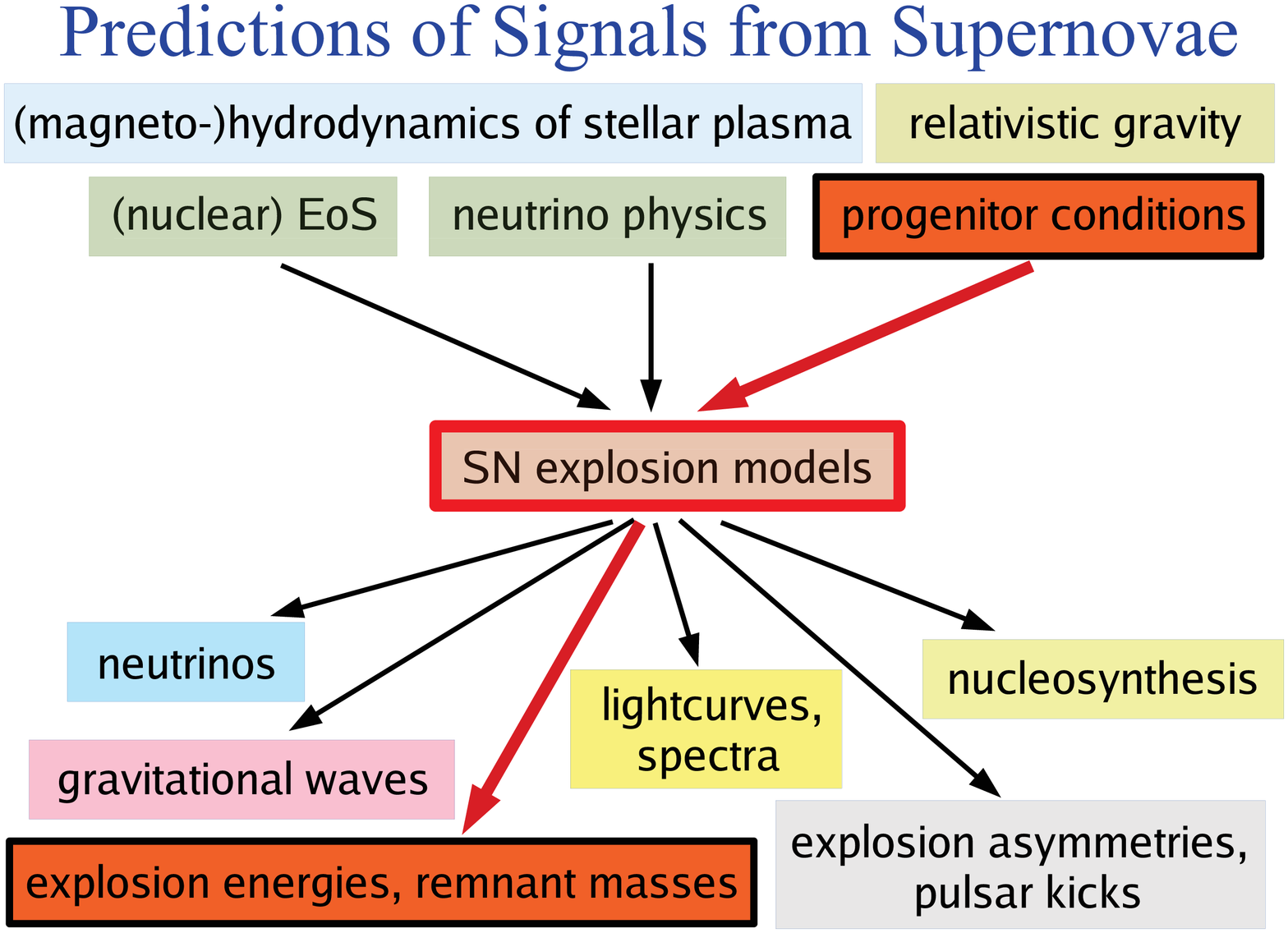}}
\caption{Input and output from supernova explosion models. The predicted
  signals of stellar explosions depend on the complex interplay of a
  wide variety of input physics and require the knowledge of the initial
  conditions in the progenitor stars.}
  \label{fig:signals}
\end{figure}

Sophisticated explosion models with all necessary theoretical
ingredients are essential for a deeper
understanding of observable phenomena and measurable signals 
associated with the death of massive stars and therefore also for
developing a better definition of the role of supernovae in the 
astrophysical context (Fig.~\ref{fig:signals}). One important 
aspect of the latter is the progenitor-explosion-remnant connection,
which includes theoretical calculations of supernova energies, 
nucleosynthetic yields, and compact object masses in dependence of the
properties of the progenitor stars. This problem is linked to the 
unsolved question which progenitors leave black holes behind instead
of neutron stars. Self-consistent numerical simulations 
are especially needed when one aims for
reliable predictions of neutrino signals and gravitational waves,
which could be detected in the case of a future galactic event
and would serve as valuable direct probes of the dynamical processes 
and thermodynamic conditions in the supernova core. As described
above, also the conditions for the explosive creation of heavy
elements are set by the interaction of the ejecta with the
intense neutrino fluxes from the nascent neutron star. 
Moreover, the production of 
radioactive nuclei like $^{56}$Ni as well as the nucleosynthetic 
reprocessing of the star's composition layers by the outgoing shock
wave depend sensitively on the explosion energy and the location 
of the mass cut that separates the compact remnant from the
supernova ejecta. Hydrodynamic instabilities during the first
second of the explosion are responsible for the observed 
large-scale explosion asymmetries that can lead to high pulsar recoil
velocities.\cite{rf:Scheck.etal.2006,rf:Wongwathanarat.etal.2010,rf:Nordhaus.etal.2010b,rf:Nordhaus.etal.2012,rf:Wongwathanarat.etal.2012} \
By deforming the supernova shock 
they also seed the growth of secondary mixing instabilities at the 
composition shell interfaces after the passage of the explosion
shock. These large-scale radial mixing processes destroy the 
onion-skin structure of the progenitor and carry heavy elements
with high velocities from the region of their formation deep
in the stellar core into the helium and hydrogen layers and, in 
turn, sweep hydrogen and helium inward in velocity as well as
radial space. This compositional mixing and the large-scale 
explosion asymmetries have important consequences for the shape
of supernova lightcurves, the characteristics of the supernova spectra,
and the time evolution of the electromagnetic emission in different 
wave bands.

It is clear that final answers for the wealth of questions linked
to all these issues will ultimately require a consistent as well as
consolidated solution of the explosion mechanism of core-collapse
supernovae. Admittedly, supernova theory is not yet there, but it 
has made good progress in directions that are promising for bringing
us closer to the goal of our efforts. Despite the incompleteness of 
the present understanding of the problem at the heart of exploding 
stars, it still seems illuminative to explore
possible implications and thus to move forward in assembling
the pieces of a great puzzle, in which both theoretical and 
observational bricks need to be interweaved.
While some adopt a pessimistic point of view
and concentrate on the empty half of the glass, lamenting about
a still imperfect match of modeling results 
with measured supernova properties and
proclaiming solicitousness about what they, ostensibly, perceive
as a defocussing and deception of the field by
``myths that have crept into modern 
discourse''\cite{rf:Burrows.2012}, we prefer to look at current
developments from a more optimistic perspective and to extract
motivation from the fact that the glass seems to be half full
at least. In this spirit a recent
review article\cite{rf:Janka.2012} has highlighted advances
that have happened over the past ten years. But also new challenges 
have emerged from the ambitious work of researchers around the
globe, whose important contributions to various aspects of the
field have refined our picture of stellar
core collapse and the associated physical processes.

In the following brief, focussed overview, 
we shall summarize some of the latest results 
obtained by the Garching group and its collaborators. Although 
core-collapse theory is in rapid flow and the field is
enthusiastically rushing towards more realism by advancing
the models from two dimensions to three-dimensional space,
basic physics that plays a role in the explosion mechanism does 
not depend on the considered dimension. Progress that has been
achieved in the past, although necessarily obtained with  
constrained setups like axisymmetric (2D) simulations, 
will nevertheless provide the foundations 
for a better understanding of the crucial ingredients in a 
working supernova mechanism. In this sense we hope that the
contents of this
article will not grow stale too fast, despite the temporary
nature of many modeling aspects and of conclusions that can 
be drawn still only from a narrowed perspective.

\section{Generic properties of the neutrino emission}
\label{sec:neutrinoemission}

It is now commonly accepted that the delayed neutrino heating
mechanism is unable to yield explosions in spherical symmetry
except for progenitors stars with O-Ne-Mg cores instead of 
Fe cores. These stars at the lower end of the mass range for 
supernova progenitors reach highly degenerate conditions already
after central carbon burning so that electron captures on Mg, Na,
and Ne begin to reduce the effective adiabatic index and enforce the
collapse.\cite{rf:Nomoto.1984} \ Since these progenitors
possess an extremely steep density gradient around the core,
the shock accelerates outward in response to the rapidly 
decaying mass accretion rate [cf.\ Eq.~(\ref{eq:shockradius})] 
and neutrino energy deposition powers an outflow sufficiently
strong to unbind the dilute He- and H-layers.\cite{rf:Kitaura.etal.2006} \
The energies of such explosions are, however, low (around
$10^{50}$\,erg, see Fig.~\ref{fig:expenergies} and 
Ref.~\citen{rf:Janka.etal.2008}), and only little
nickel is ejected (some $10^{-3}\,M_\odot$; Ref.~\citen{rf:Wanajo.etal.2011}),
for which reason the supernovae must be expected to be faint. 

\begin{wrapfigure}{r}{\halftext}
 \centerline{\includegraphics[width=\halftext]{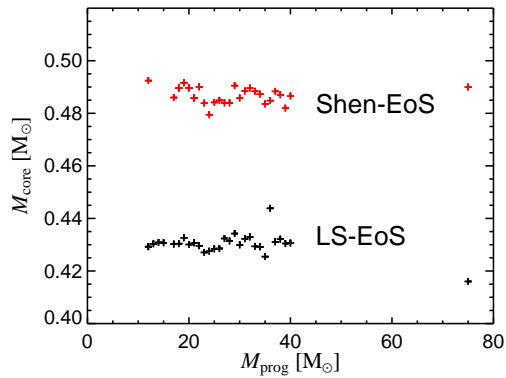}}
\caption{Inner core masses enclosed by the shock-formation position
  in iron-core-collapse simulations of a large set of (solar metallicity)
  progenitor stars.\cite{rf:Woosley.etal.2002} \
  The point of shock formation (defined by the location
  where the postshock entropy first reaches 3\,$k_\mathrm{B}$ per nucleon)
  depends on the deleptonization during collapse, which hardly varies
  with the mass of the progenitor. Also the EoS has an influence, 
  connected to different collapse timescales and composition differences.
  We used the EoS of Ref.~\citen{rf:Shen.etal.1998} in comparison to the EoS
  of Ref.~\citen{rf:LattimerSwesty.1991} with compressibility modulus 
  $K = 180$\,MeV.}
  \label{fig:coremasses}
\end{wrapfigure}

In all other progenitors multi-dimensional flows have turned
out to be crucial for getting explosions. If large-scale 
nonsphericity, e.g.\ rotational deformation, is absent in the 
progenitor core, asymmetries must grow from small initial
perturbations by hydrodynamic instabilities. During core infall,
however, the conditions are not favorable for such a growth. 
The contracting flow remains essentially spherical (on large
scales) up to core
bounce. Moreover, during infall electron captures on heavy nuclei 
and free protons (the former dominate during most of the evolution)
maintain a very similar structure of the homologously collapsing
inner core, and the electron fraction converges to nearly the 
same central value ($Y_{e,\mathrm{c}} = 0.25$--0.27 for a 
central lepton fraction $Y_\mathrm{lep,c} = 0.28$--0.30;
Ref.~\citen{rf:Marek.PhD.2007}) in different progenitors
despite a small initial spread of values in the core center
before collapse ($Y_{e,\mathrm{i}}\approx 0.425$--0.445) and
considerable differences of the initial central entropy
($s \approx 0.6$--$1.2\,k_\mathrm{B}$ per nucleon). 
Since the mass of the homologous core scales with the 
instantaneous Chandrasekhar mass, 
$M_\mathrm{ic} \propto M_\mathrm{CH}\propto Y_e^2$,
the baryonic mass enclosed by the shock formation radius 
is also very similar over a wide mass range of progenitor
stars (from the same modeling set; Fig.~\ref{fig:coremasses}). 
The shock formation position is defined here by the location
where the postshock entropy first reaches 3\,$k_\mathrm{B}$ per
nucleon. The degree of deleptonization during collapse, 
however, depends on the duration of the infall until neutrino
trapping and on composition differences (especially the free 
proton fraction), both of which are connected to 
equation of state (EoS) properties (a detailed study can 
be found in Ref.~\citen{rf:Marek.PhD.2007}). 
Therefore the shock formation point is 
somewhat different for different EoSs ($\sim$0.43\,$M_\odot$ 
vs.\ $\sim$0.49\,$M_\odot$ in Fig.~\ref{fig:coremasses}).
In any case, however, the bounce shock is launched deep inside
the stellar iron core, which makes it impossible for the shock
to overcome the energy losses by nuclear photodisintegration 
processes in the massive, overlying shells of iron. 
With the small masses of the homologous inner core in current
models, successful supernova explosions by the 
hydrodynamical bounce-shock mechanism are ruled out.

\begin{figure}[!]
  \centerline{\includegraphics[width=\textwidth]{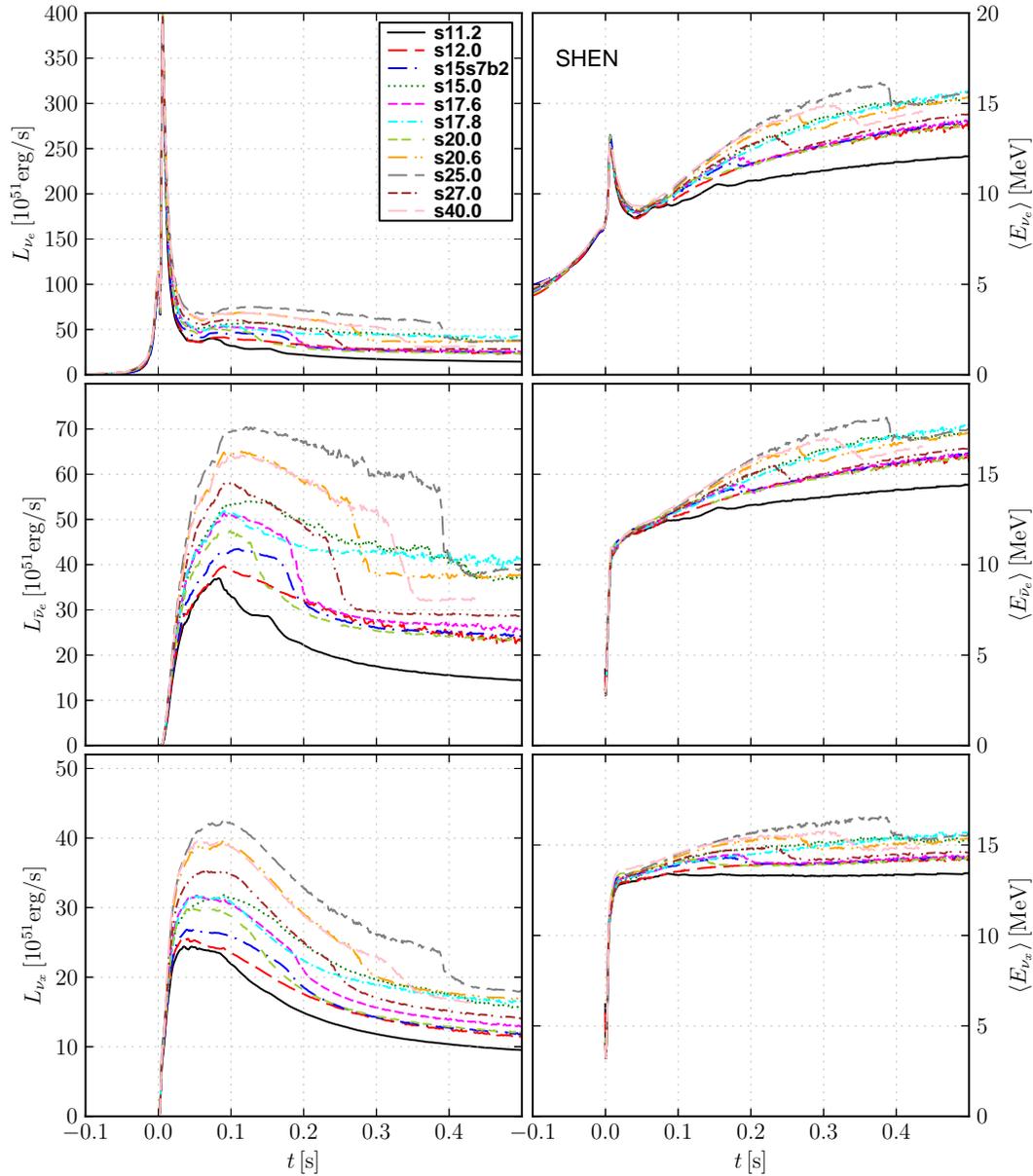}}
  \caption{Time evolution of luminosities ({\em left}) and mean
   energies (defined as ratio of energy
   flux to number flux; {\em right}) for $\nu_e$ ({\em top}),
   $\bar\nu_e$ ({\em middle}), and one kind of heavy-lepton neutrino
   (all treated equally; {\em bottom}) as measured in the lab frame
   at infinity. Time is normalized to the moment of core bounce.
   The results are plotted for
   (nonexploding) 1D simulations of a set of solar-metallicity
   progenitors\cite{rf:Woosley.etal.2002} and an older
   15\,$M_\odot$ progenitor (s15s7b2; Ref.~\citen{rf:WoosleyWeaver1995}),
   applying the EoS of Ref.~\citen{rf:Shen.etal.1998}.}
  \label{fig:neutrinotimeshen}
\end{figure}

\begin{figure}[!]
  \centerline{\includegraphics[width=\textwidth]{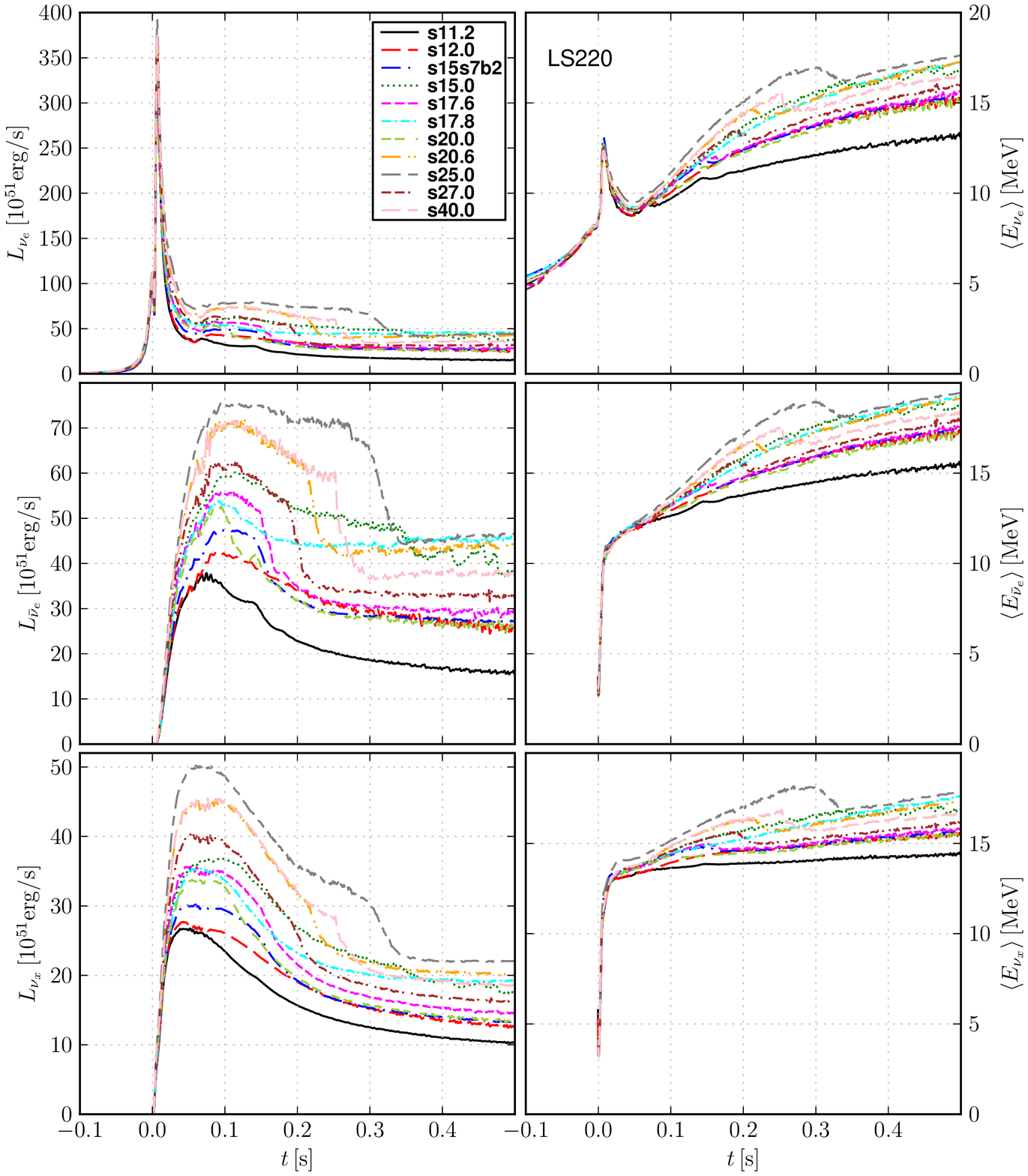}}
  \caption{Same as Fig.~\ref{fig:neutrinotimeshen}, but for
   (nonexploding) 1D simulations with the
   EoS of Ref.~\citen{rf:LattimerSwesty.1991} using a compressibility 
   modulus of $K = 220$\,MeV. This EoS is softer than the one used in
   Fig.~\ref{fig:neutrinotimeshen}.}
  \label{fig:neutrinotimels}
\end{figure}

After a short phase of maximum strength, the bounce shock
is weakened again by the drain of nuclear energy. A negative
entropy gradient emerges in the shock deceleration region,
which quickly decays in a short (typically 30--50\,ms)
phase of ``prompt postshock convection''. This phase makes
a gravitational wave signal\cite{rf:Mueller.etal.2012c} but
has only a weak influence on the neutrino emission. 
Somewhat later, neutrino heating behind the shock grows in
strength because the postshock temperature drops and the
mean energies of the radiated neutrinos continuously rise
with time (Figs.~\ref{fig:neutrinotimeshen} and 
\ref{fig:neutrinotimels}). A gain layer forms and 
a local entropy maximum builds up just outside the gain radius
and creates the necessary (and in most cases sufficient) 
condition for the onset of convective overturn in the accretion
layer behind the stalled shock. First Rayleigh-Taylor fingers 
show up typically 80--100\,ms after bounce, but it can take 
roughly another $\sim$100\,ms before the accretion 
flow in the postshock layer becomes largely perturbed by 
convective downdrafts and buoyant plumes. On a similar 
timescale also the SASI grows and can manifest itself
(at least in two dimensional (2D, i.e., axisymmetric) models)
by pronounced sloshing
motions of the shock surface (for a detailed numerical study
of neutrino-driven convection and the SASI and their 
interaction in 2D core-collapse simulations, see
Refs.~\citen{rf:Scheck.etal.2008,rf:Mueller.etal.2012b}).
Besides the convective overturn in the neutrino-heating
layer, convection also develops in the deleptonization
region inside the 
neutrinospheres.\cite{rf:Buras.etal.2006b,rf:Dessart.etal.2006}. \
Associated effects on the neutrino emission, for example
an increase of the heavy-lepton neutrino fluxes and a 
reduction of the $\bar\nu_e$ flux compared to the
nonconvective proto-neutron star, however, are also small in
the first $\sim$100\,ms after core bounce.\cite{rf:Buras.etal.2006b} \

During this early postbounce period of up to about 100\,ms,
spherically symmetric models therefore yield reliable information
about the neutrino emission properties. We display corresponding
information on the radiated luminosities and mean spectral
energies (for an observer at rest at a large distance from the
source) for all neutrino species in 
Figs.~\ref{fig:neutrinotimeshen} and \ref{fig:neutrinotimels}.
(Muon and tau neutrinos and antineutrinos have very similar 
opacities and are lumped together to one kind of heavy-lepton
neutrino, $\nu_x$.)
In the first case the relatively stiff nuclear 
(EoS) of Ref.~\citen{rf:Shen.etal.1998} was used, in the second case the 
considerably softer one\footnote{Somewhat unconventionally,
we use the terms ``stiff'' and ``soft'' for the EoS to classify
the {\em radius evolution} of the forming
neutron star with the corresponding EoS, ``soft'' meaning that 
the nascent remnant {\em contracts faster} to its final radius.
Note that a ``soft'' EoS in our sense {\em neither} means that
the maximum neutron star mass is low {\em nor} 
that the final neutron star radius is particularly small.
Physically, the maximum mass of the neutron star is
determined by the supernuclear properties of cold neutron
star matter, whereas the radius evolution of the nascent 
compact object strongly depends on how the conditions in
matter at subnuclear densities depend on changes of the
temperature and of the neutron-to-proton ratio (both are evolving 
in response to neutrino emission).}
of Ref.~\citen{rf:LattimerSwesty.1991}. These images should 
be compared with
Fig.~3 in Ref.~\citen{rf:OConnorOtt.2012}. Although our set
of core-collapse calculations and the one of 
Ref.~\citen{rf:OConnorOtt.2012} employ
different series of progenitor models, basic features of the 
neutrino signal are generic and, even more, general trends 
should be comparable even beyond 100\,ms after bounce, despite
the disregard of multi-dimensional effects.

It should be noted that our calculations were performed with
the \textsc{Prometheus-Vertex} code with general relativistic
corrections\cite{rf:RamppJanka.2002}, which performed very well
in tests against fully relativistic
treatments.\cite{rf:Liebendoerfer.etal.2005,rf:Marek.etal.2006,rf:Mueller.etal.2010} \
The energy-dependent three-flavor neutrino transport module
is based on a two-moment closure scheme with a variable Eddington
factor derived from a model-Boltzmann equation. The transport
includes velocity-dependent terms, energy-bin coupling,
and the full set of neutrino opacities\cite{rf:Buras.etal.2006a} 
(see Fig.~\ref{fig:neutrinos}). 

Although in Ref.~\citen{rf:OConnorOtt.2012} a variety
of simplifications were applied (e.g., a subset of neutrino
processes was used and velocity-dependent effects were
ignored in the transport), the overall evolution of the 
neutrino emission properties of $\nu_e$ and $\bar\nu_e$
is reproduced relatively well. This outcome is connected
to the facts that on the one hand electron-flavor neutrinos mainly
interact by the charged-current absorption and emission 
processes and on the other hand 
the neutrinos escape from regions near the
proto-neutron star surface, where the matter is essentially
at rest. Features that depend on the motion of the stellar
medium, however, are absent in the results of 
Ref.~\citen{rf:OConnorOtt.2012}. One example is the 
characteristic local minimum of the 
$\nu_e$ luminosity shortly after core bounce and before the
luminosity rises steeply to its prominent maximum.
In this feature the luminosity drops from
$\sim$$1.15\times 10^{53}$\,erg\,s$^{-3}$ to 
$\sim$$0.7\times 10^{53}$\,erg\,s$^{-3}$ (which can easily be 
seen only on a zoom to its location). This dip is formed 
about 1\,ms after bounce and approximately another millisecond
before shock breakout from the neutrinosphere (where the 
optical depth decreases to about unity). At this time 
the shock is still deep in the optically thick region and
the neutrinos produced behind the shock are still trapped. 
Most of the $\nu_e$ release therefore comes from the 
compressed, unshocked and semitransparent layer around
100\,km. When the matter in this layer falls inward with
increasingly higher speed, the emission gets more and more
redshifted due to the Doppler effect and, moreover, an 
increasing fraction of the neutrinos is advected inward
with the gas flow. These effects lead to the transient
luminosity reduction 
before the rise to the shock-breakout burst sets in.

The $\nu_e$ signal through core collapse, bounce, and
shock breakout is extremely similar for all progenitors
and also for different EoSs.\cite{rf:Kachelriess.etal.2005} \
Even during the first $\sim$100\,ms after bounce the
progenitor stars have little influence on the mean
energies of the radiated neutrinos of all kinds, although
the luminosities already begin to exhibit a spread that 
reflects the different mass infall rates in the cores of
different stars. More compact progenitor 
cores\footnote{Compactness is measured by the ratio of mass
to corresponding radius enclosing this mass. Higher compactness
means that more mass falls inward in a certain time. For
a definition and discussion of the compactness in the context
of black hole formation, see Ref.~\citen{rf:OConnorOtt.2011}.}
lead to higher mass accretion rates of the nascent neutron
star, which in turn produces higher neutrino luminosities 
supplied by the emission of neutrinos from the hot, 
lepton-rich, freshly accreted material. Higher stellar
core compactness systematically correlates with higher
luminosities of all neutrino species.\cite{rf:OConnorOtt.2012} 

At times $t \gtrsim 100$\,ms post bounce, consequences of 
multi-dimensional
effects on the neutrino emission cannot be ignored any
longer.\cite{rf:Buras.etal.2006b,rf:MarekJanka.2009,rf:Marek.etal.2009} \
The radiated neutrino signal is then affected by convection
inside the neutron star, which raises the 
heavy-lepton neutrino fluxes, reduces the $\bar\nu_e$ emission,
and decreases the mean energies
of the escaping neutrinos\cite{rf:Buras.etal.2006b,rf:Marek.etal.2009}.
The neutrino emission is also modified and modulated by
nonspherical accretion downflows, which are associated
with anisotropic, transient, short ``bursts'' of accretion
luminosity.\cite{rf:Marek.etal.2009,rf:Ott.etal.2008,rf:Brandt.etal.2011,rf:MuellerE.etal.2012} \
Nevertheless, some general properties can be discussed that
are shared by 1D as well as angle-averaged 2D and 3D results.

At $t \gtrsim 100$\,ms post bounce also the mean energies
of the radiated neutrinos begin to exhibit a spread that
reflects the core compactness. Higher core compactness and
higher mass accretion rates do not only produce higher 
neutrino luminosities but also larger mean energies of the
emitted neutrinos. While before $t \sim 100$\,ms p.b.\ the
nascent neutron stars in all progenitors are still relatively
similar, the different mass accretion rates in stars with
different core compactnesses cause the proto-neutron star 
masses to become increasingly different. For this reason the 
mean energies develop a growing spread with time: More massive 
proto-neutron stars possess hotter neutrinospheres and the
mean neutrino energies increase more steeply than in the case
of less massive neutron stars in less compact progenitors.
The collapsing cores of such less compact stars feed the growth 
of the neutron star mass with rapidly decreasing rates at
late postbounce times.
Interestingly, in all cases (and for both stiff and soft
neutron star EoS) there is a point during the 
late accretion phase when the mean energy of $\bar\nu_e$
crosses and subsequently exceeds the mean energy of the 
heavy-lepton neutrinos. The physical reason for this 
effect was discussed in Ref.~\citen{rf:Marek.etal.2009}.
After the explosion has taken off and the accretion has
ended, the mean energies adopt again the well known hierarchy,
where $\left\langle \epsilon_{\nu_e}\right\rangle <
\left\langle \epsilon_{\bar\nu_e}\right\rangle <
\left\langle \epsilon_{\nu_x}\right\rangle$.

It should be noticed that in Ref.~\citen{rf:OConnorOtt.2012}
---presumably because of the approximations in the 
neutrino transport used there--- considerably 
higher mean neutrino energies are predicted for the 
later postbounce evolution (compare
Fig.~3 in Ref.~\citen{rf:OConnorOtt.2012} with
Figs.~\ref{fig:neutrinotimeshen} and \ref{fig:neutrinotimels}
in our paper\footnote{Note that the comparison between the
different transport solvers in Fig.~2 of 
Ref.~\citen{rf:OConnorOtt.2012} was done with a reduced
set of neutrino opacities instead of the more sophisticated
and larger set of neutrino reactions usually
applied in the simulations of the Garching group.
Moreover, the approximate description of general relativistic
gravity used in the \textsc{Prometheus-Vertex} 
calculations of Ref.~\citen{rf:Liebendoerfer.etal.2005} 
(referred to in Ref.~\citen{rf:OConnorOtt.2012}) was
subsequently improved by Marek et al.\cite{rf:Marek.etal.2006},
whose preferred description was shown to lead to much 
better agreement with fully relativistic
results.\cite{rf:Mueller.etal.2010}}).
In particular, also the differences between the core-collapse
models with soft and stiffer EoS tended to be larger in 
Ref.~\citen{rf:OConnorOtt.2012}. 
The authors of the latter paper proposed to use
the cumulative neutrino energy from the postbounce accretion
phase, which correlates with the mass accretion rate of the
progenitor, to obtain information of the core structure 
of the collapsing star in the case of a neutrino measurement
from a future galactic supernova. In order to break the degeneracy 
between nuclear EoS and progenitor-core compactness (a higher 
neutrino luminosity and number flux, for example, can be 
caused by a larger mass accretion rate or by a softer EoS),
they suggested to combine the measurement of the cumulative 
energy (or cumulative event number) with the information about
the mean energy of the neutrinos. In view of the weak 
sensitivity of the mean energy to the EoS (cf.\ 
Figs.~\ref{fig:neutrinotimeshen} and \ref{fig:neutrinotimels})
such a discrimination will be quite an ambitious undertaking.

\begin{figure}[htb]
\vspace{10pt}
  \parbox{\halftext}{
          \centerline{\includegraphics[width=\halftext]{profile_comparison.eps}}
          \caption{Pre-collapse density profiles of progenitor stars used for the
          2D explosion models of Figs.~\ref{fig:entropytime}--\ref{fig:expenergies}.
          The circles mark the boundaries of the Fe-core and the Si/Si+O interface
          in the solar-metallicity
          11.2\,$M_\odot$ (s11.2), 25\,$M_\odot$ (s25.0), and 27\,$M_\odot$
          (s27.0) models\cite{rf:Woosley.etal.2002}, in the 15\,$M_\odot$
          (s15s7b2) model\cite{rf:WoosleyWeaver1995}, and in the ultra-metal poor
          ($10^{-4}$ solar) 8.1\,$M_\odot$ progenitor\cite{rf:Heger.2011},
          whereas they
          indicate the Fe/Si and Si/O+N transitions in the zero-metallicity
          9.6\,$M_\odot$ (z9.6) star.\cite{rf:Heger.2011} \ In the 8.8\,$M_\odot$
          progenitor with O-Ne-Mg core\cite{rf:Nomoto.1984} the circles denote
          the composition changes from NSE to Ne+O, then to Ne+Mg+O, to C+O,
          and finally to He.}
          \label{fig:densityprofiles}}
          \hfill
  \parbox{\halftext}{
          \centerline{\includegraphics[width=\halftext]{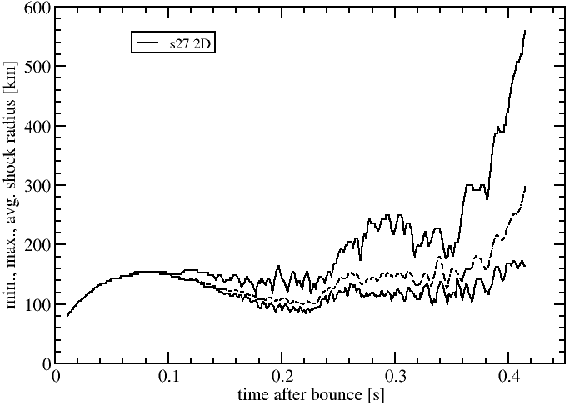}}
          \caption{Postbounce evolution of the maximum, minimum and average
          shock radius in a 2D simulation of the 27\,$M_\odot$ progenitor
          of Fig.~\ref{fig:densityprofiles} with
          the \textsc{Prometheus-Vertex} code, which employs an approximate
          treatment of general relativistic effects in the gravitational
          potential and neutrino transport. As in the fully relativistic
          2D simulation shown in Fig.~\ref{fig:entropytime},
          bottom right panel, a neutrino-driven explosion takes
          place, although the runaway shock expansion sets in somewhat
          later and after less vigorous bipolar shock oscillations than in
          the relativistic calculation.}
          \label{fig:27msunexplosion}}
\end{figure}

\section{Relativistic explosion models in two dimensions}
\label{sec:2Dexplosions}

Successful neutrino-driven explosions cannot be obtained in 1D 
with the neutrino heating associated with the neutrino emission
properties of present models as discussed in 
Sect.~\ref{sec:neutrinoemission}. The Garching group, however,
has recently found explosions for a growing set of 
progenitor stars with different masses (8.1, 8.8, 9.6,
11.2, 15, and 27\,$M_\odot$) and metallicities
(Fig.~\ref{fig:densityprofiles}) in 2D simulations with the
general relativistic \textsc{CoCoNuT-Vertex} hydrodynamics 
and neutrino transport
code.\cite{rf:Mueller.etal.2010,rf:Mueller.etal.2012a,rf:Mueller.etal.2012b,rf:Mueller.etal.2012c} \
These results can be considered as a basic confirmation of the 
self-consistent 2D explosion models of 11.2\,$M_\odot$ and
15\,$M_\odot$ stars computed with the 
\textsc{Prometheus-Vertex} code by Buras et al.\cite{rf:Buras.etal.2006b}
and Marek \& Janka.\cite{rf:MarekJanka.2009} \ Also the
27\,$M_\odot$ explosion has recently also been reproduced
(Fig.~\ref{fig:27msunexplosion})
with the \textsc{Prometheus-Vertex} tool, which employs a
relativistic approximation of the gravity potential\cite{rf:Marek.etal.2006}
and relativistic corrections in the transport 
module.\cite{rf:RamppJanka.2002} \ Despite differences
in details, e.g.\ in the postbounce dynamics and
explosion times of 2D
models with general relativistic and approximate relativistic
treatment, there seems to be overall agreement and compatibility
of the results obtained with these two different numerical 
schemes\footnote{The \textsc{Prometheus-Vertex} and 
\textsc{CoCoNuT-Vertex} codes do not only differ in the
treatment of general relativity, they also employ largely
different hydrodynamics solvers, cf.\ Refs.~\citen{rf:RamppJanka.2002}
and \citen{rf:Mueller.etal.2010}.}.

\begin{figure}
  \centerline{\includegraphics[width=\halftext]{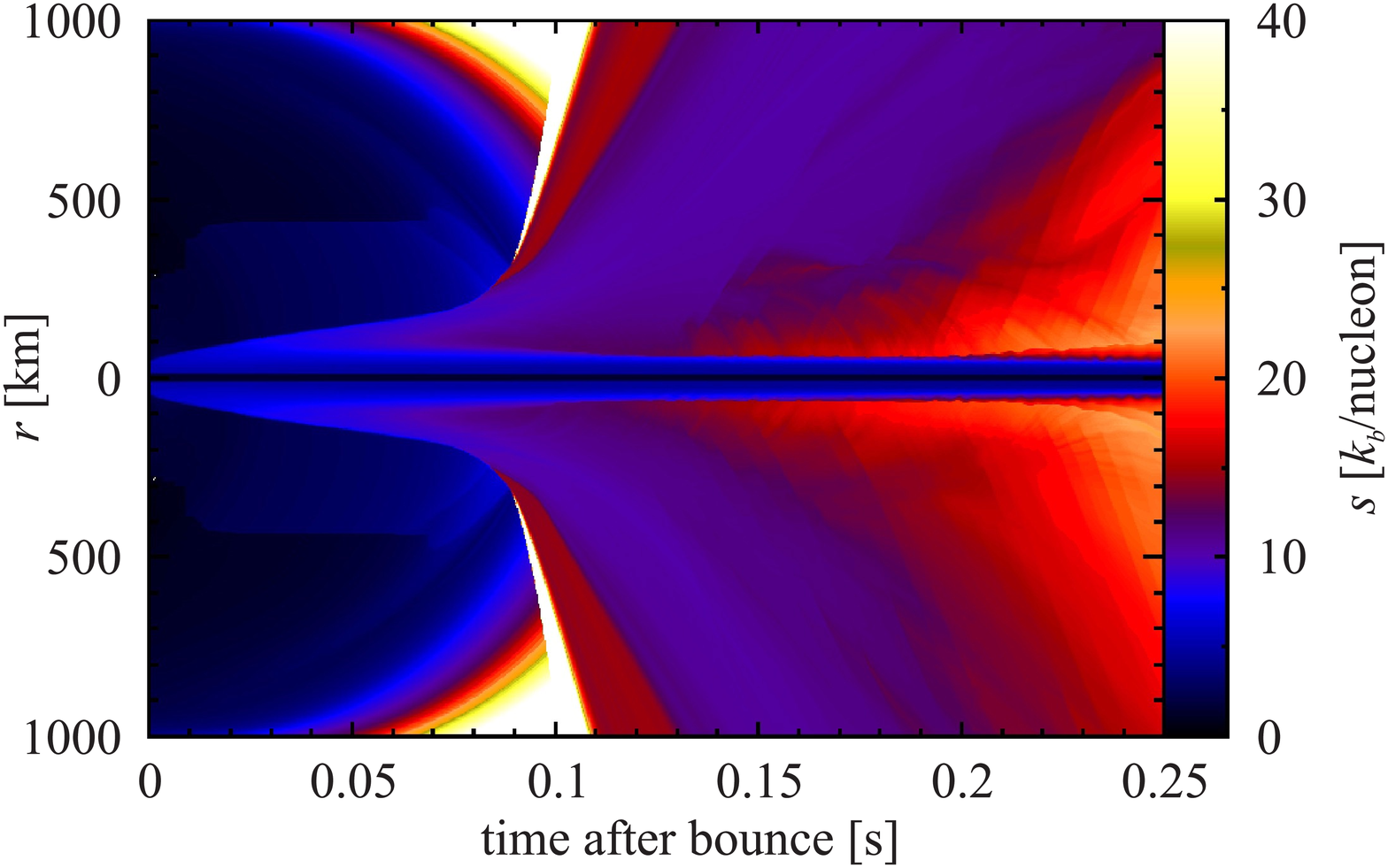}\ \
              \includegraphics[width=\halftext]{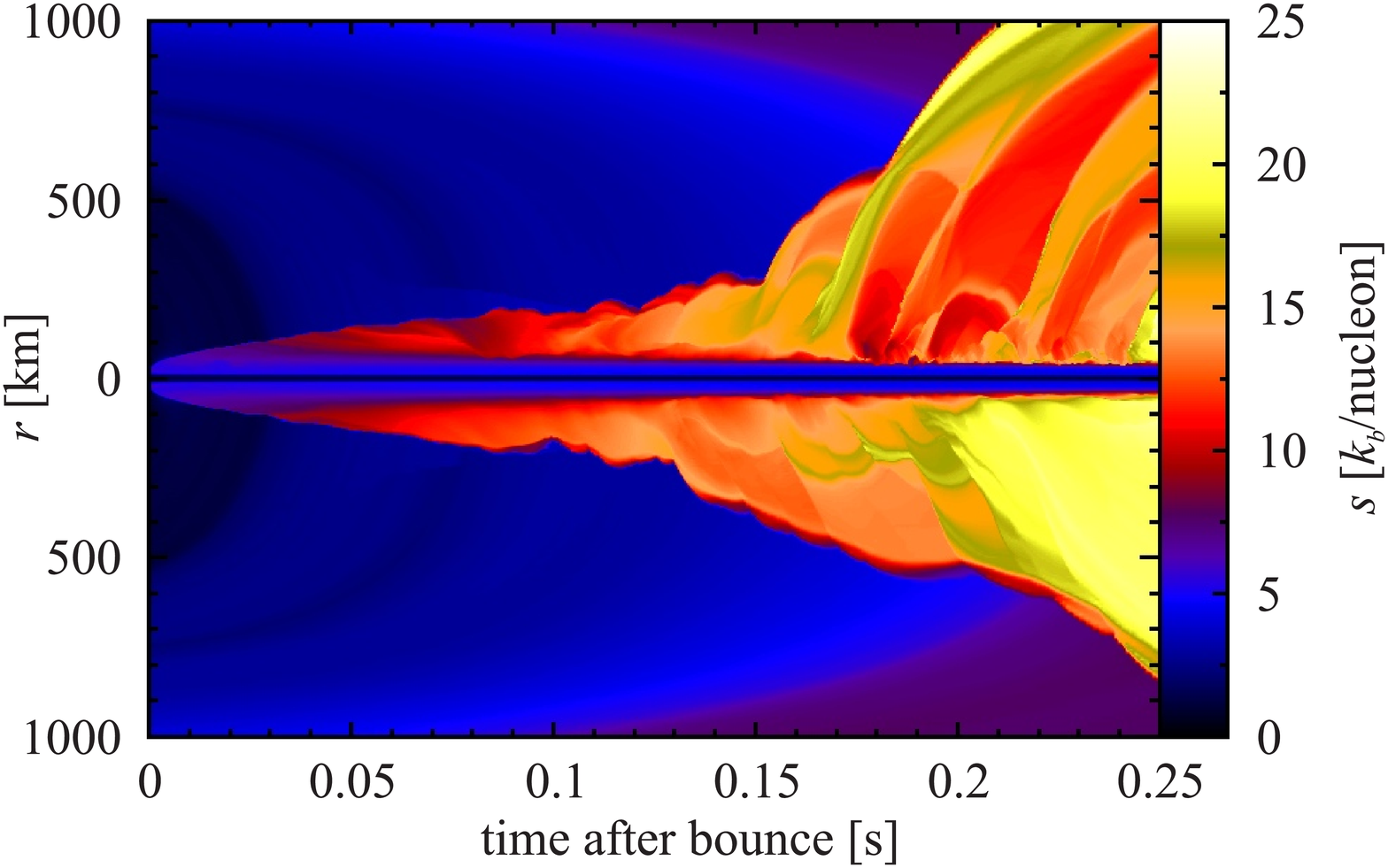}}\vspace{5pt}
  \centerline{\includegraphics[width=\halftext]{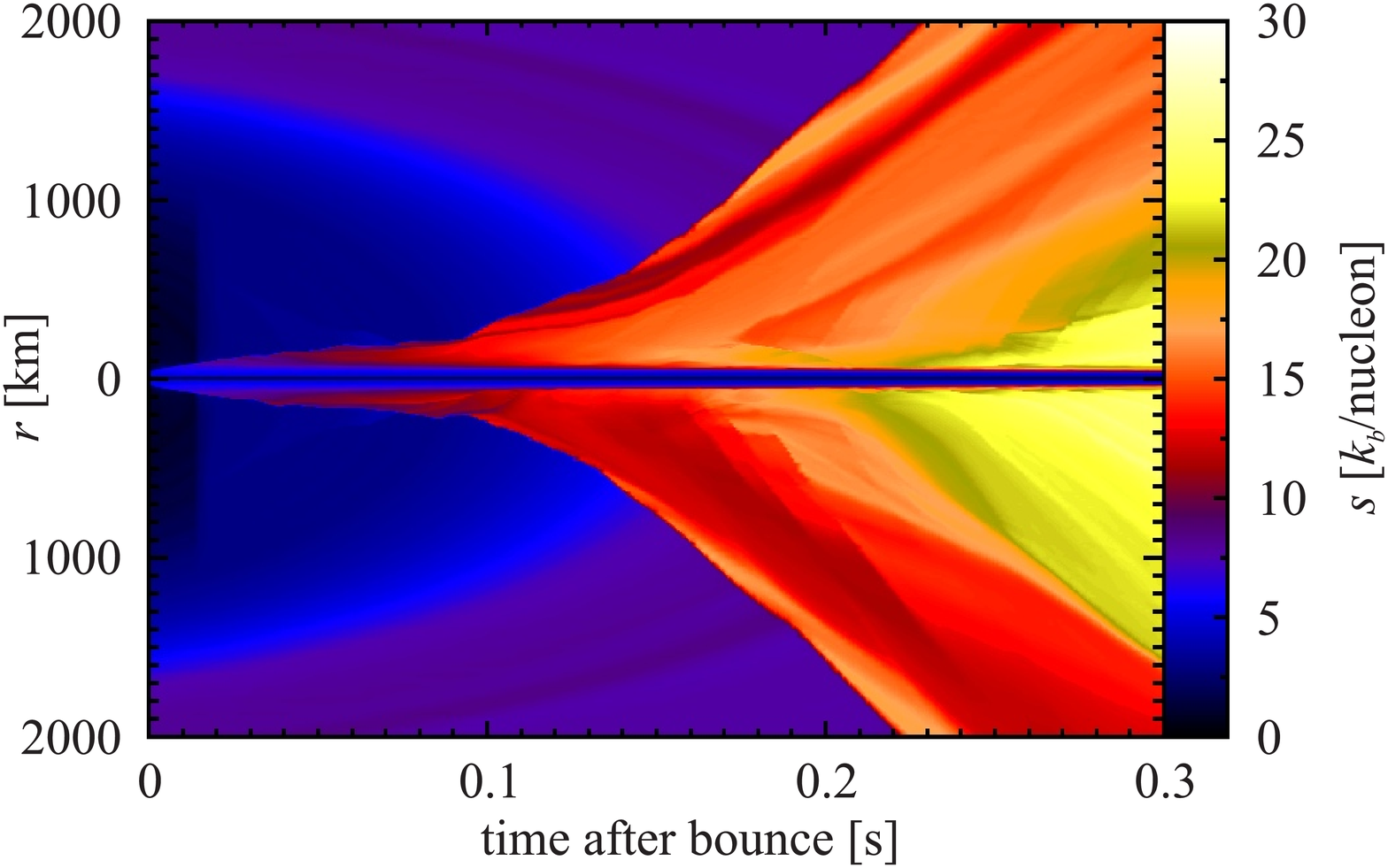}\ \
              \includegraphics[width=\halftext]{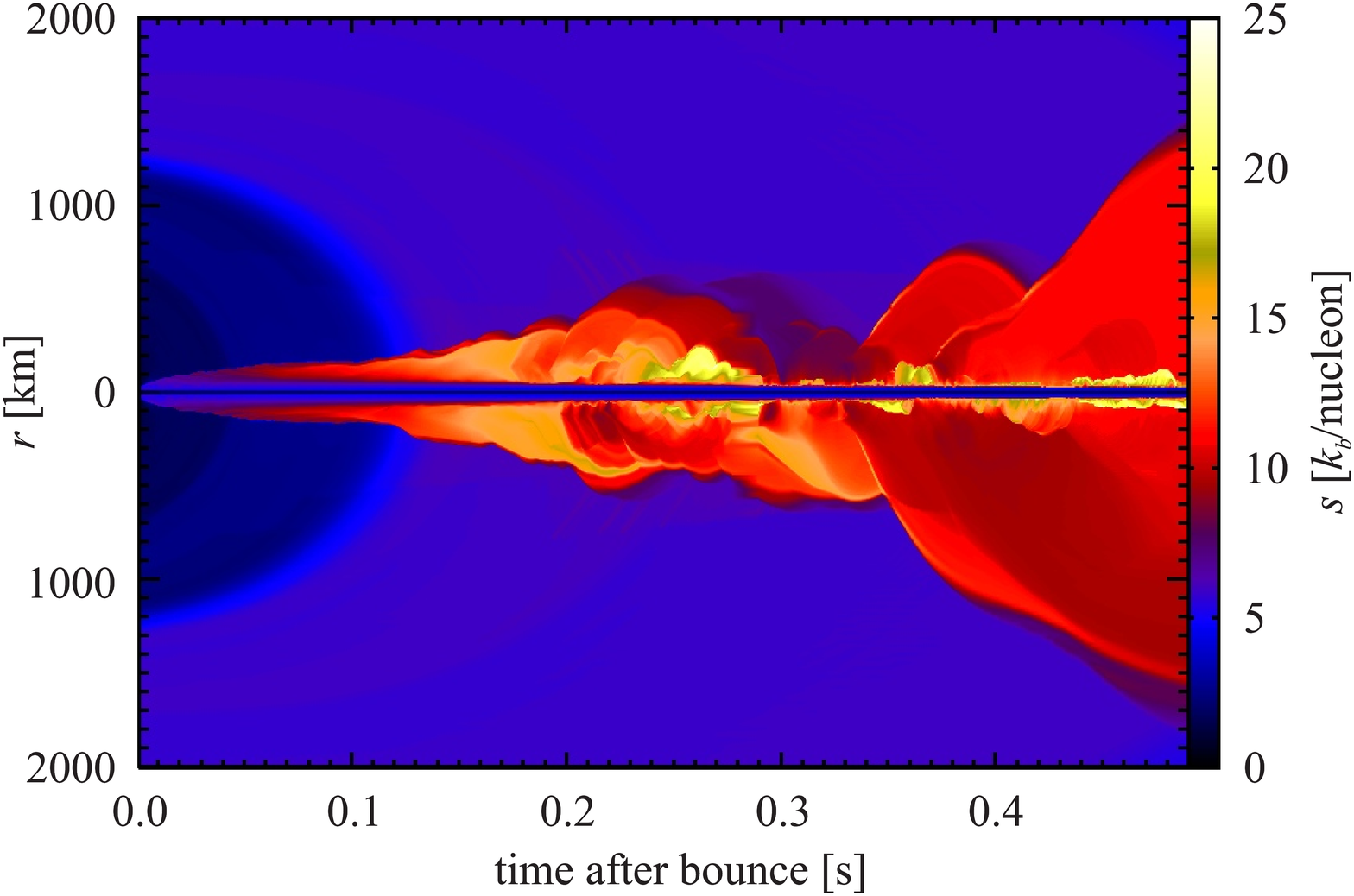}}\vspace{5pt}
  \centerline{\includegraphics[width=\halftext]{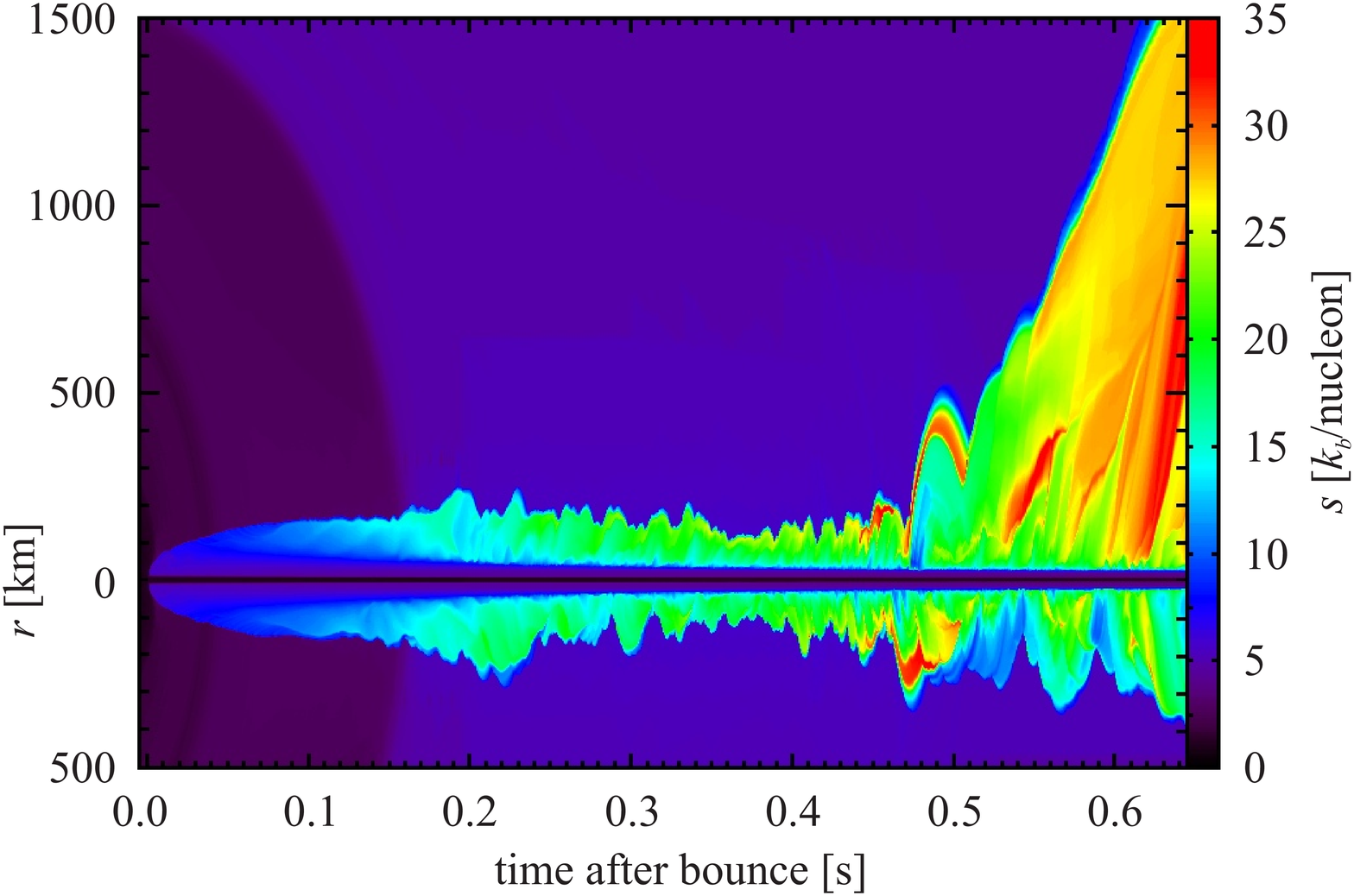}\ \
              \includegraphics[width=\halftext]{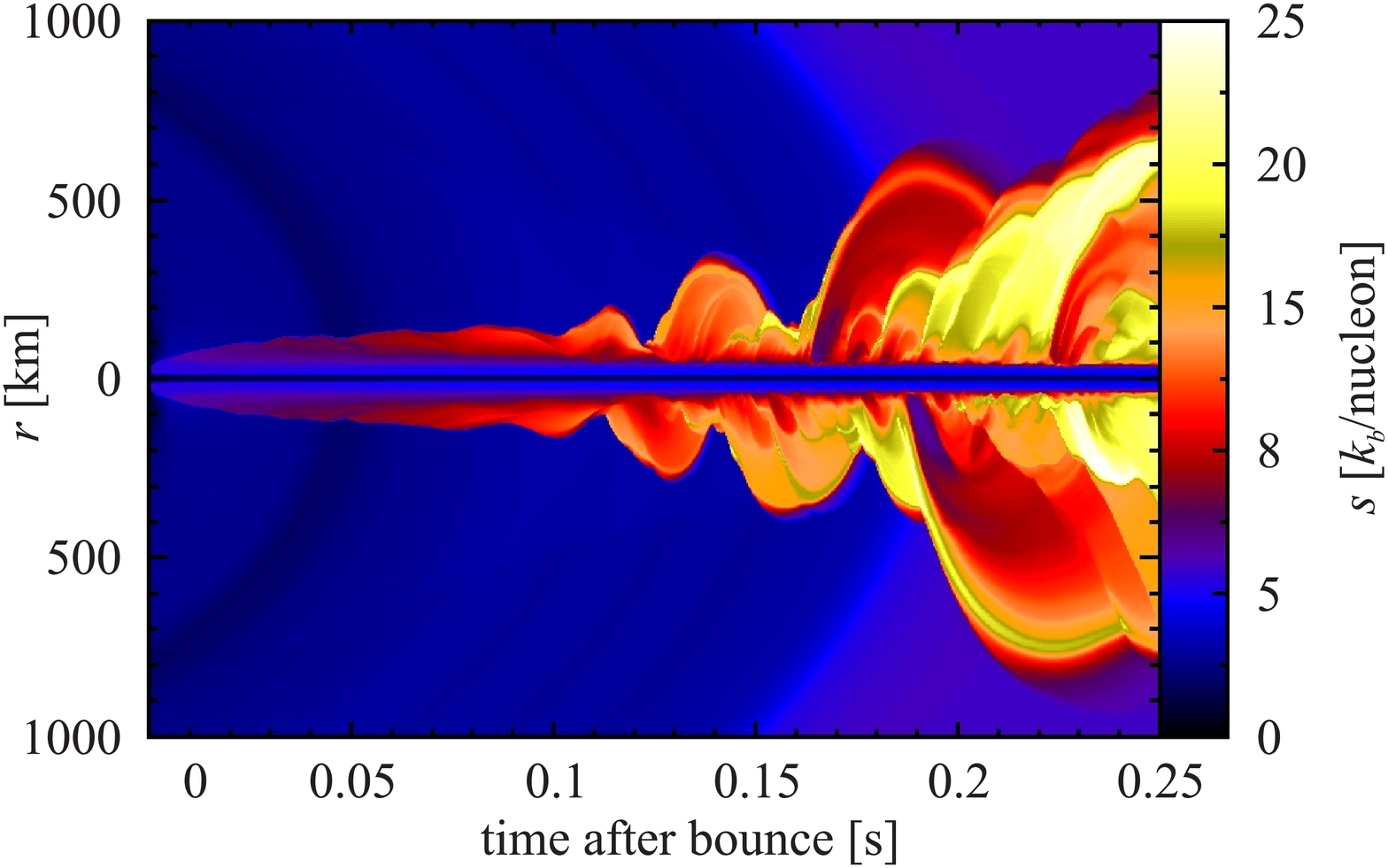}}
  \caption{Time evolution after core bounce of the profiles of entropy per nucleon
           in north-polar and south-polar directions for 2D (general relativistic)
           explosion 
           simulations\cite{rf:Mueller.etal.2012a,rf:Mueller.etal.2012b,rf:Mueller.etal.2012c} 
           of the 8.8\,$M_\odot$ ONeMg-core progenitor
           and the 8.1\,$M_\odot$, 9.6\,$M_\odot$, 11.2\,$M_\odot$,
           15\,$M_\odot$ and 27\,$M_\odot$ Fe-core stars with different
           metallicities ({\em from top left to bottom right})
           shown in Fig.~\ref{fig:densityprofiles}. Dark blue and black
           represent low-entropy unshocked matter while green, yellow,
           and red indicate hot, high-entropy matter. Bipolar shock 
           oscillations suggest SASI activity.}
  \label{fig:entropytime}
\end{figure}

\begin{figure}
  \centerline{\includegraphics[width=\halftext]{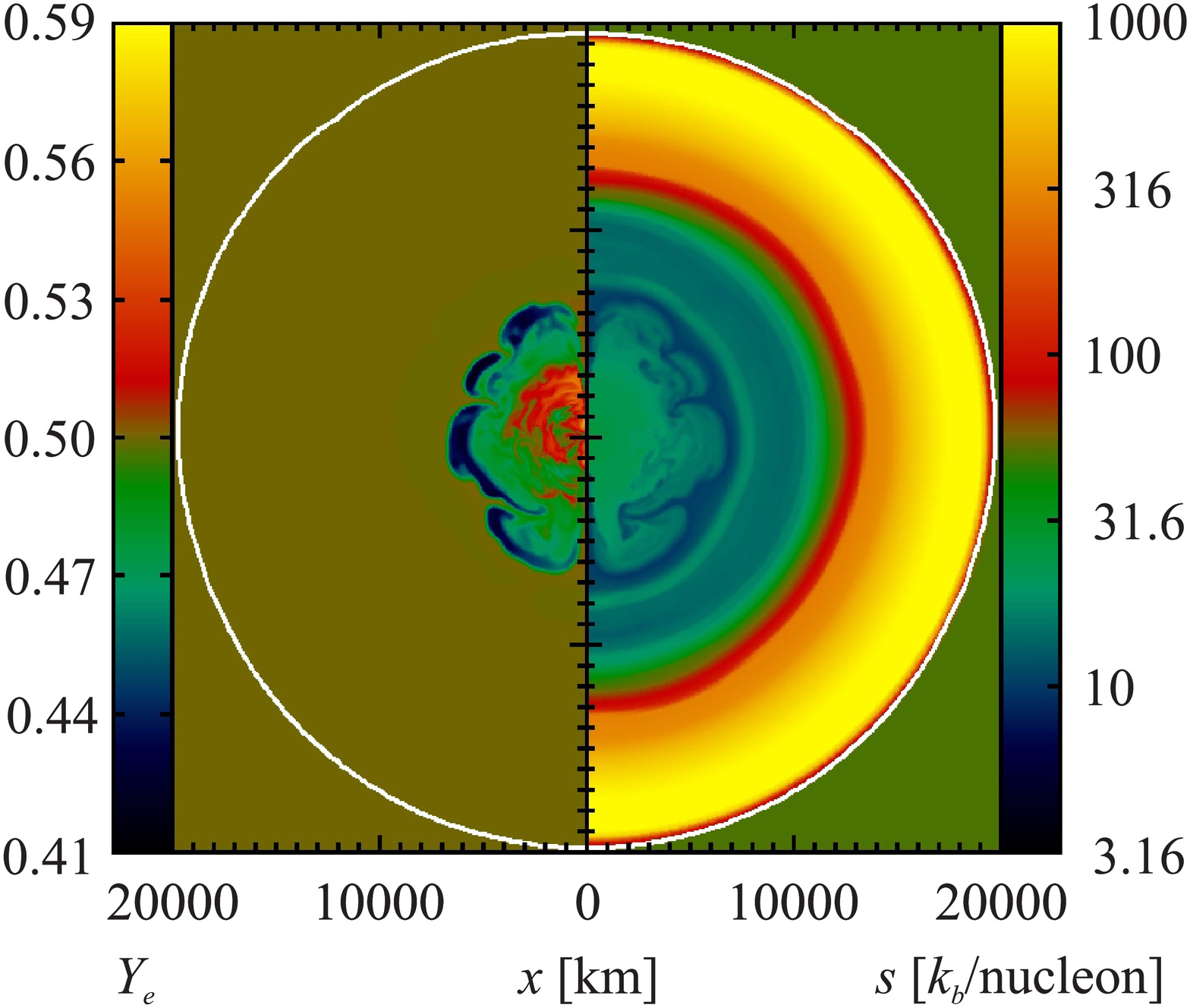}\ \
              \includegraphics[width=\halftext]{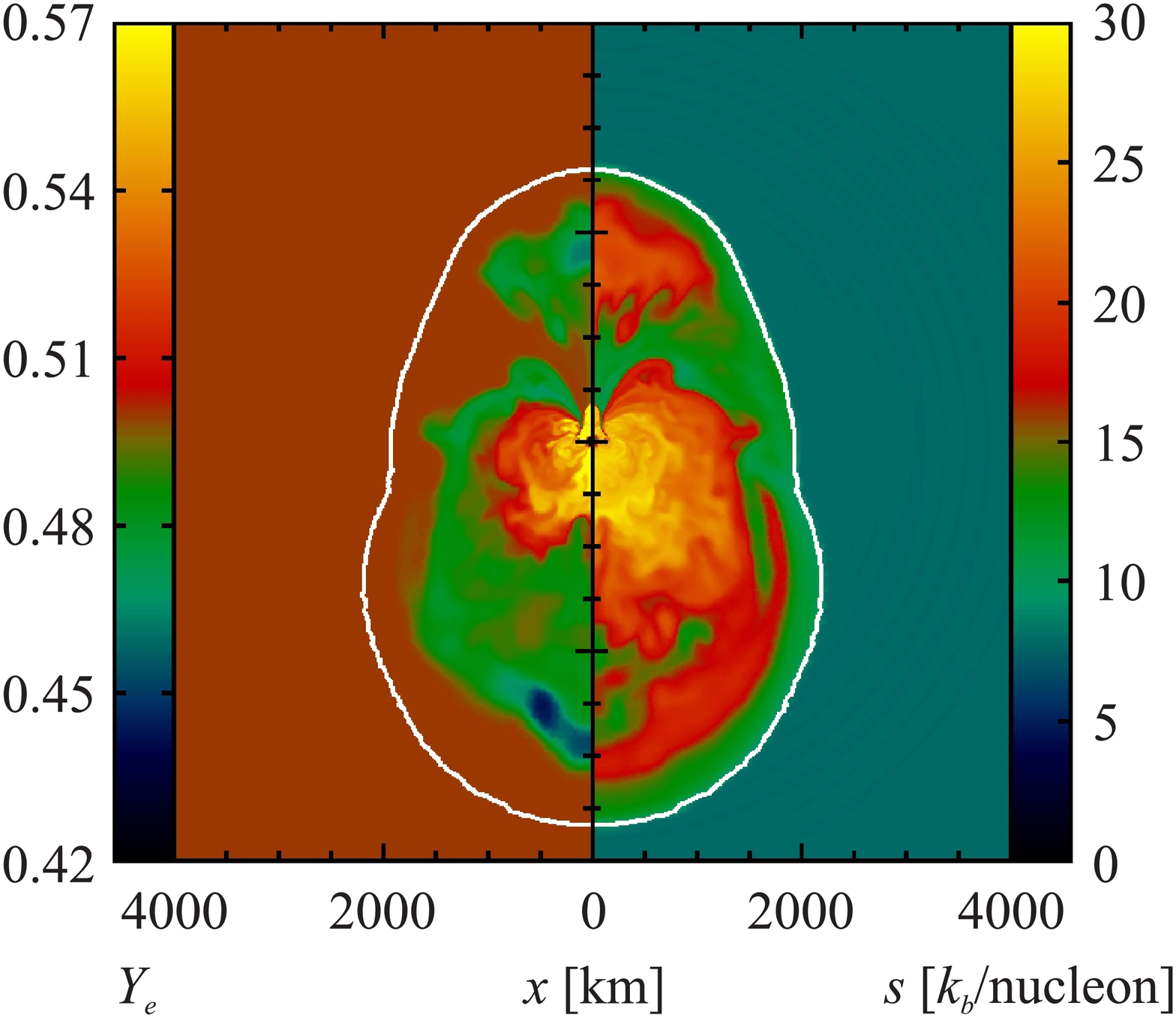}}\vspace{5pt}
  \centerline{\includegraphics[width=\halftext]{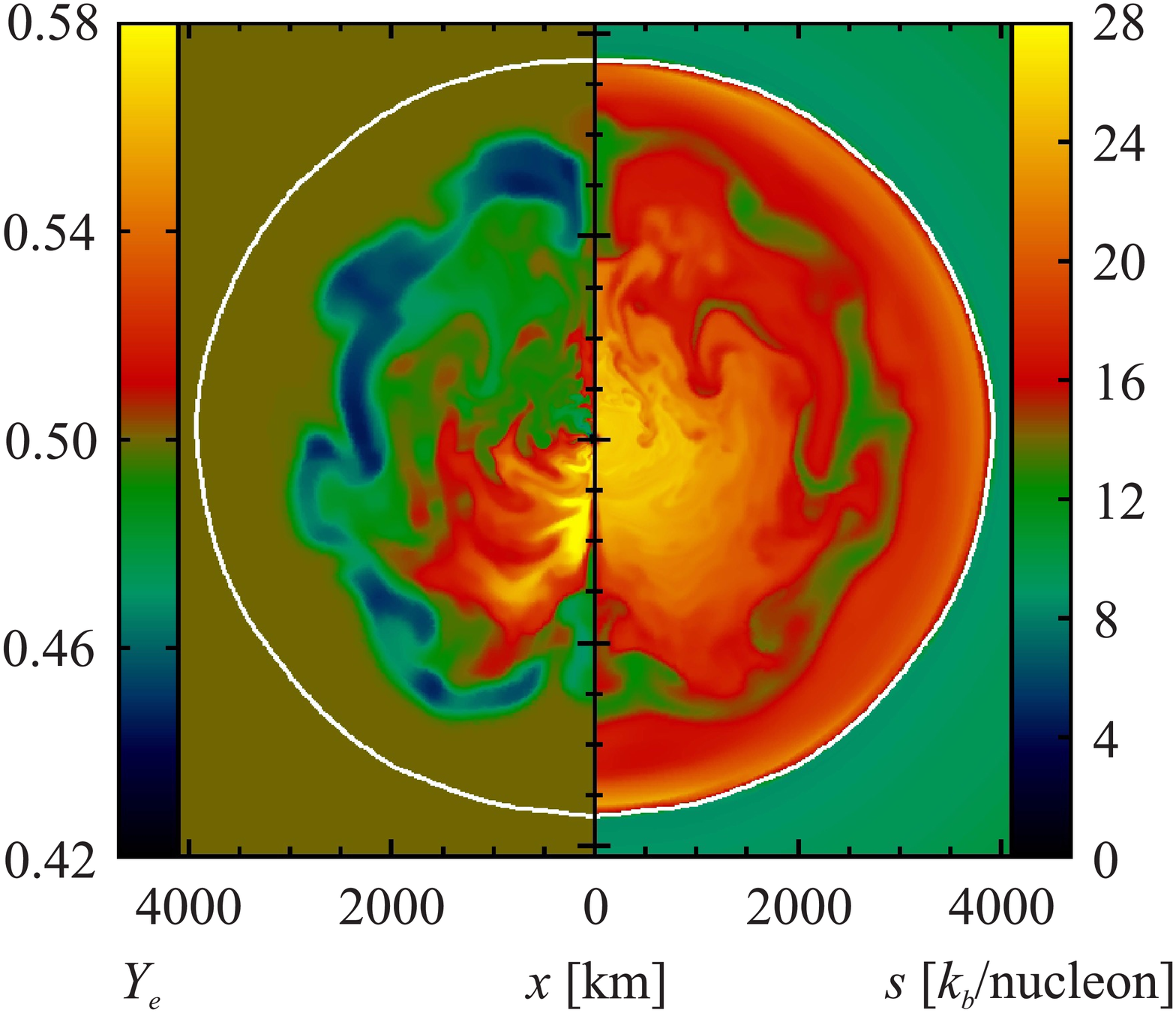}\ \
              \includegraphics[width=\halftext]{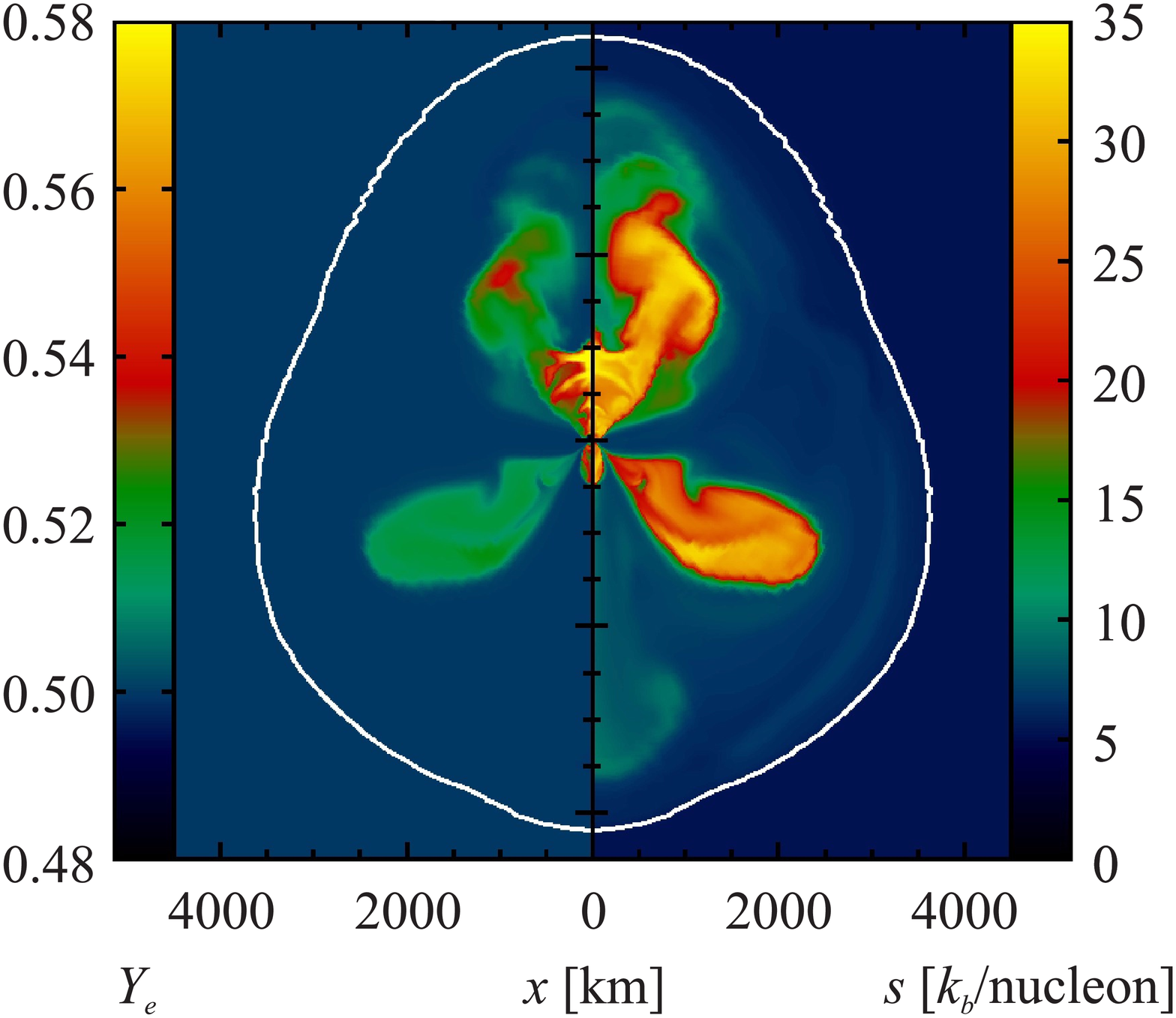}}\vspace{5pt}
  \centerline{\includegraphics[width=\halftext]{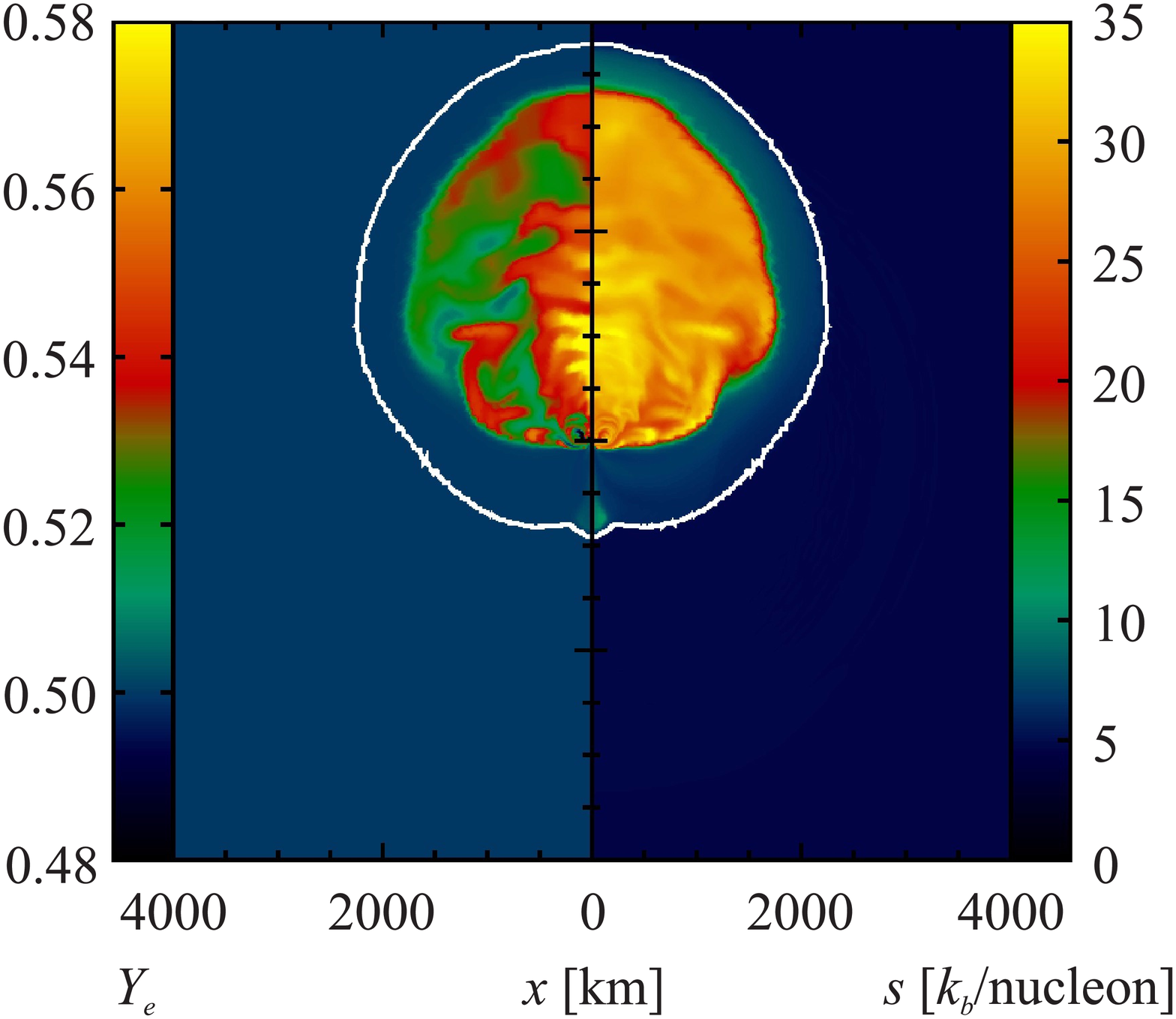}\ \
              \includegraphics[width=\halftext]{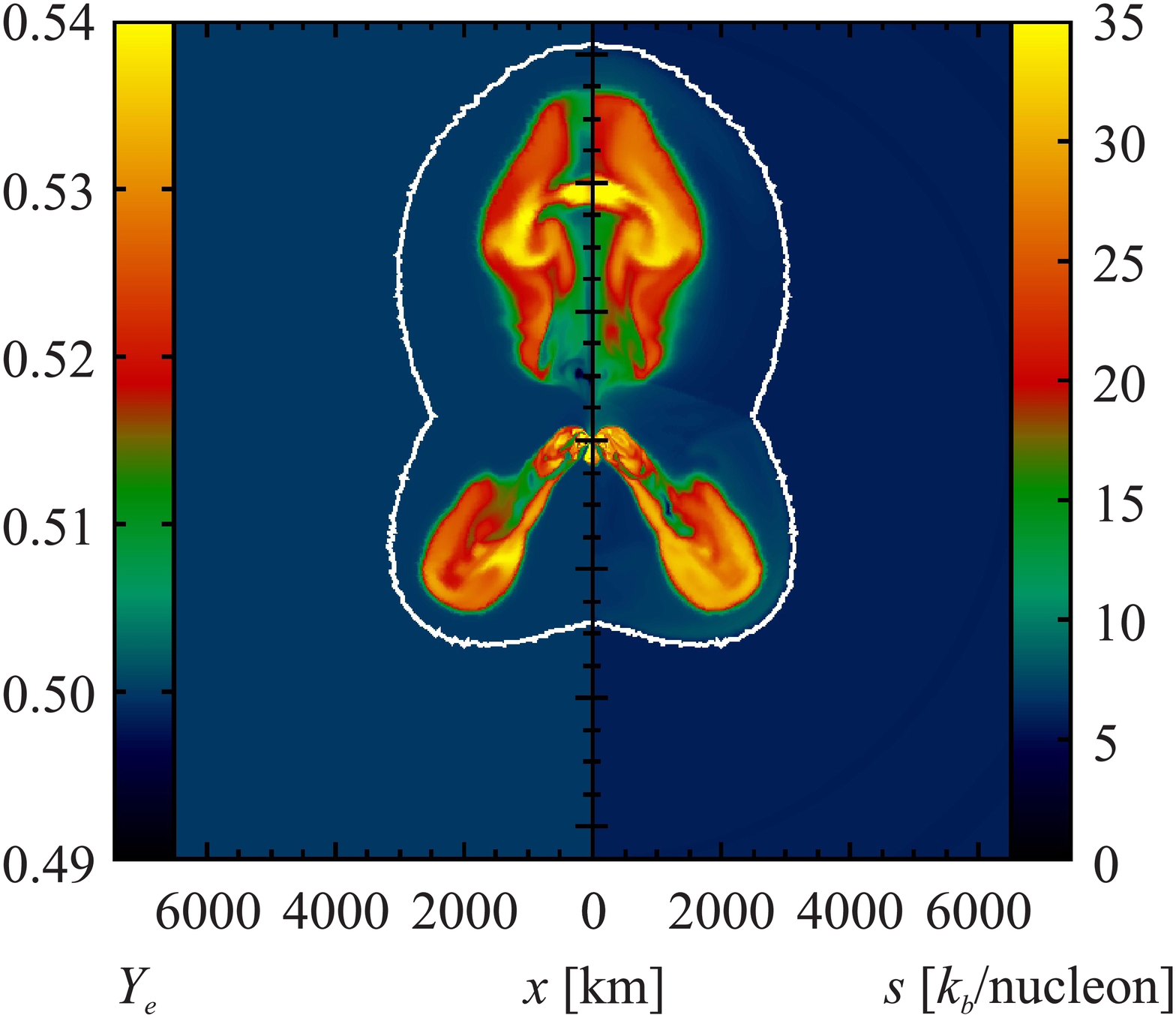}}
  \caption{Cross-sectional distributions of electron fraction
           (left half panels) and
           entropy per nucleon (right half panels) for the simulations
           shown in Fig.~\ref{fig:entropytime} ({\em from top left to 
           bottom right} for 8.8, 8.1, 9.6, 11.2, 15, and 27\,$M_\odot$
           progenitors) at postbounce times near
           the end of the simulations: $t_\mathrm{pb} = 365$\,ms,
           330\,ms, 318\,ms, 920\,ms, 775\,ms, and 790\,ms, respectively.
           The supernova shock is marked by a thin white line.
           High-entropy bubbles of neutrino-heated, expanding matter
           drive aspherical shock expansion in most cases.}
  \label{fig:yeentropysnaps}
\end{figure}

\begin{figure}[htb]
  \parbox{\halftext}{
          \centerline{\includegraphics[width=\halftext]{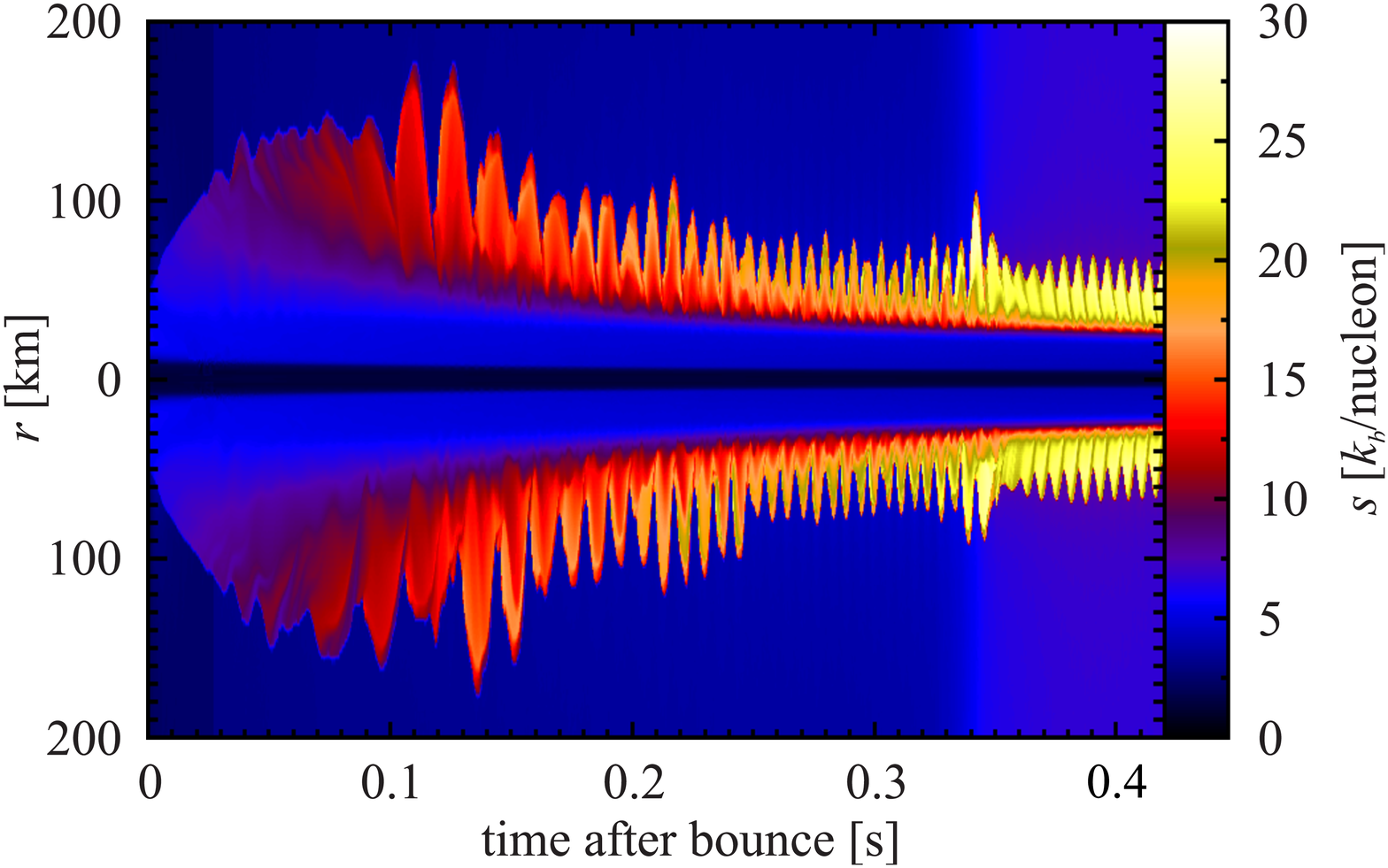}}
          \caption{Same as Fig.~\ref{fig:entropytime}, but for the
          nonexploding solar-metallicity 25\,$M_\odot$ star of
          Fig.~\ref{fig:densityprofiles}. Alternating north-polar
          and south-polar shock excursions indicate violent SASI sloshing
          motions.}
          \label{fig:entropytime25}}
          \hfill
  \parbox{\halftext}{
          \centerline{\includegraphics[width=\halftext]{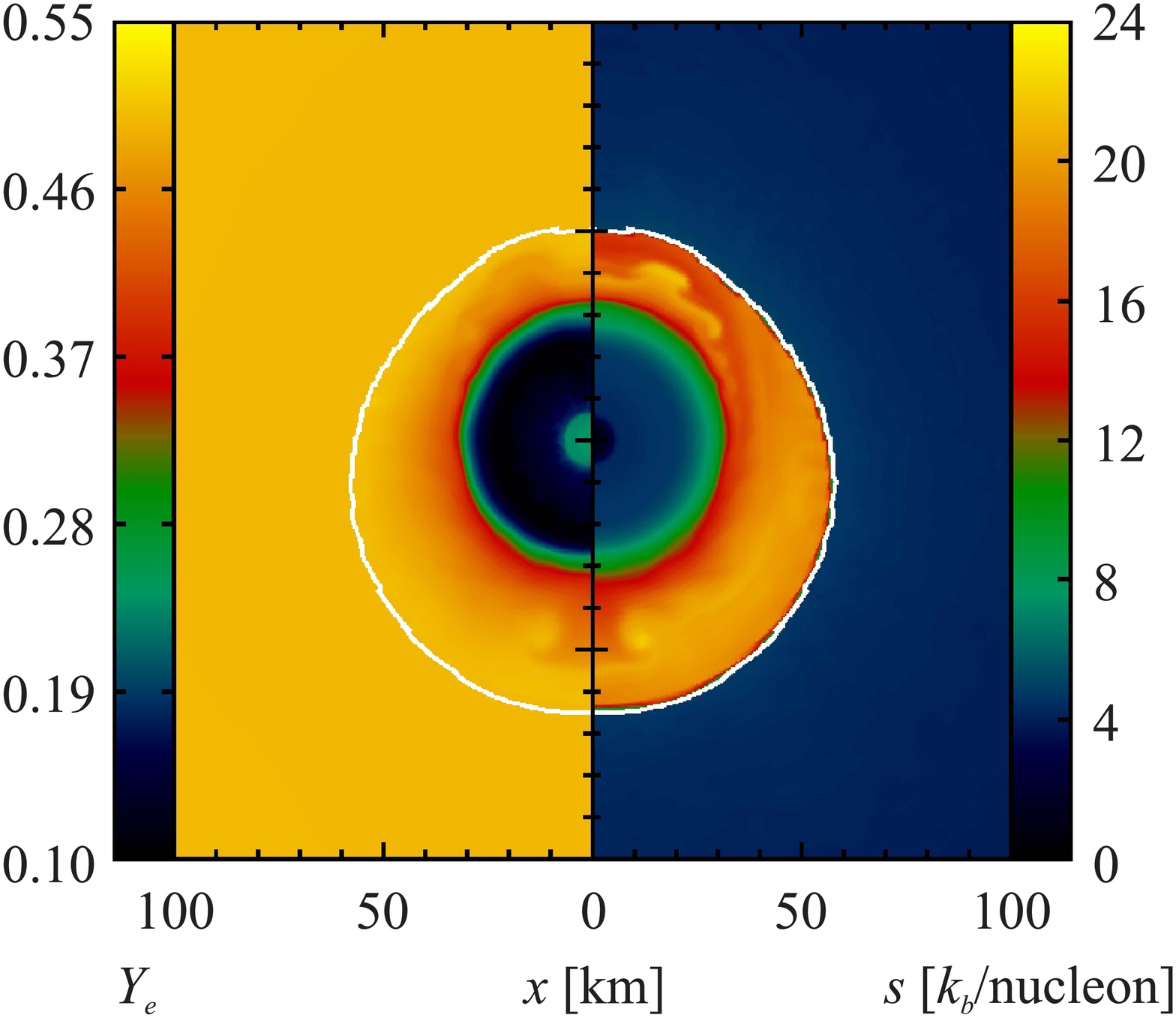}}
          \caption{Same as Fig.~\ref{fig:yeentropysnaps}, but for the
          nonexploding solar-metallicity 25\,$M_\odot$ model of
          Fig.~\ref{fig:entropytime25} at 309\,ms post bounce.
          The postshock layer is very narrow and the rapid infall
          velocities of the fluid between shock and gain radius
          suppress the growth of neutrino-driven convection. The
          entropy structures in the postshock layer are mainly 
          linked to Kelvin-Helmholtz instability
          associated with strong shear effects at the interface of 
          differentially moving regions in the sloshing flow.}
          \label{fig:yeentropysnaps25}}
\end{figure}

\begin{figure}[htb]
\vspace{10pt}
  \parbox{\halftext}{
          \centerline{\includegraphics[width=\halftext]{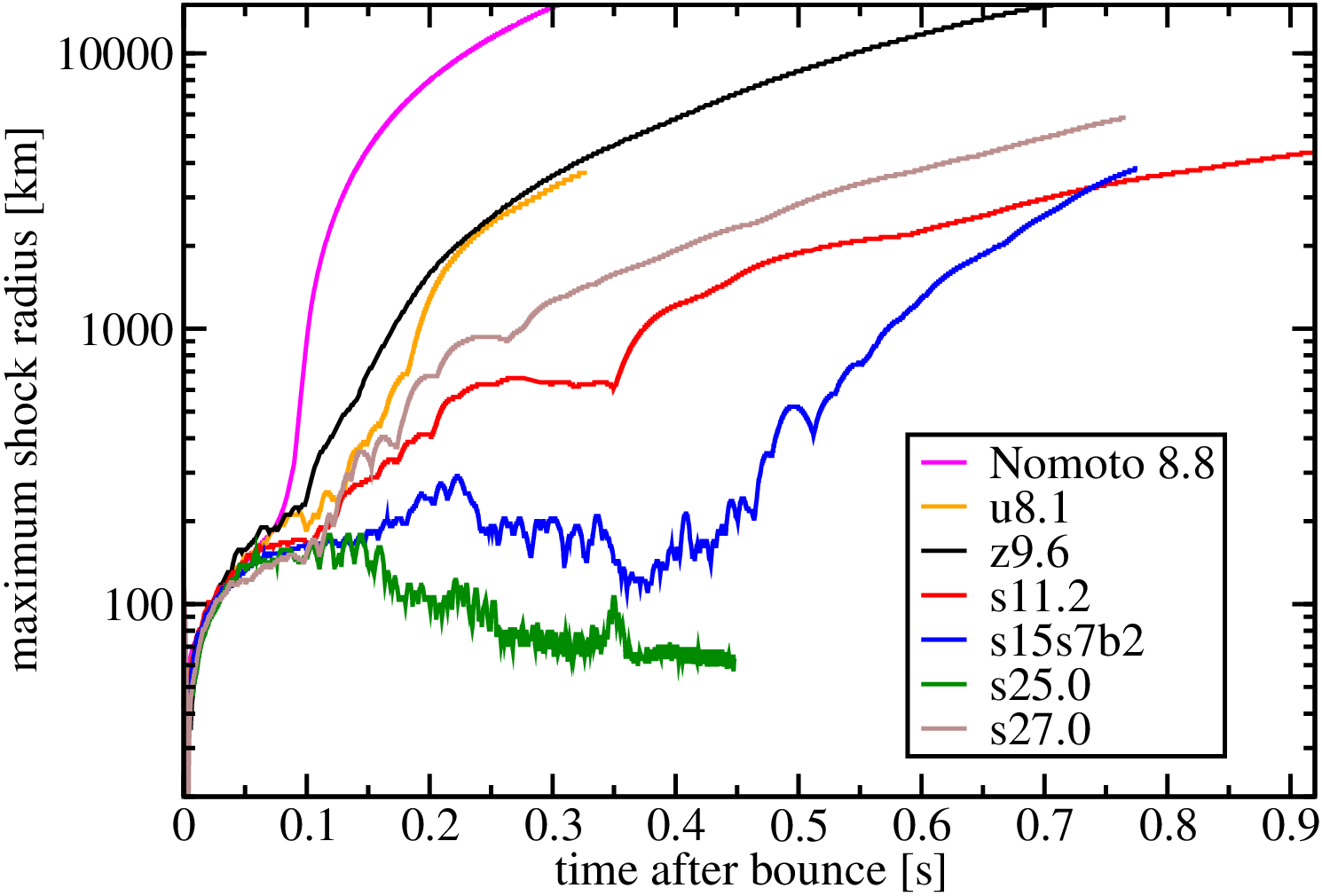}}
          \caption{Maximum shock radii versus time for the set of
          2D explosion models of Figs.~\ref{fig:entropytime} and
          \ref{fig:yeentropysnaps} and for the nonexploding 2D model
          of Figs.~\ref{fig:entropytime25} and
          \ref{fig:yeentropysnaps25}.}
          \label{fig:shockradii}}
          \hfill
  \parbox{\halftext}{
          \centerline{\includegraphics[width=\halftext]{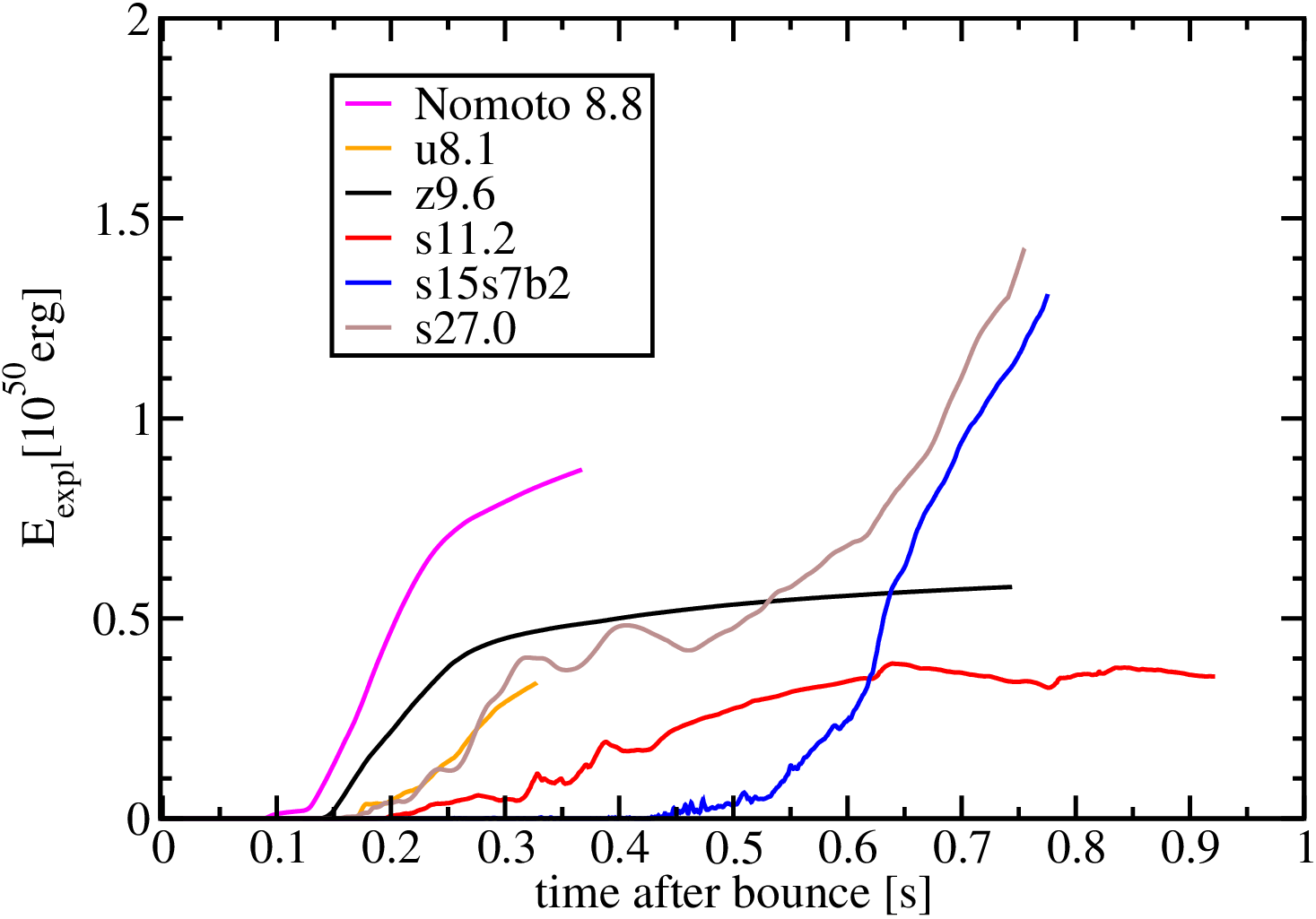}}
          \caption{Evolution of the ``diagnostic'' explosion energy
          (i.e., the energy of outward expanding postshock matter with
          positive specific energy) versus time for the set of models of
          Fig.~\ref{fig:shockradii}.}
          \label{fig:expenergies}}
\end{figure}

Entropy profiles in the north and south polar directions 
as functions of postbounce time are displayed for the relativistic
explosion 
models in Fig.~\ref{fig:entropytime}. Corresponding $Y_e$ and 
entropy distributions for representative instants 
near the end of the simulated postbounce periods are given in
Fig.~\ref{fig:yeentropysnaps}\footnote{Note that the explosion
simulation for the 8.8\,$M_\odot$ progenitor with O-Ne-Mg
core was performed with the relativistic approximations of
the \textsc{Prometheus-Vertex} scheme.}. Of the set of 
progenitors investigated so far, only the 25\,$M_\odot$ did not
show any tendency of an explosion and looks very disfavorable for
a final success when the computational run was stopped at 450\,ms 
after bounce (Fig.~\ref{fig:entropytime25}). 
This can be seen in Fig.~\ref{fig:shockradii},
which shows a very small maximum radius of the shock in the
25\,$M_\odot$ case, while in all other relativistic 2D models
the shock expands with high velocities of at least
6000--7000\,km\,s$^{-1}$, in most cases much faster. The
explosions develop on different timescales, depending on the
different density profiles and locations of the composition 
shell interfaces, which determine the 
decay of the expansion-damping mass accretion rate. 
(Remember that according to Eq.~(\ref{eq:critlum}) the critical 
luminosity is larger for higher $\dot M$ and bigger
neutron star mass, which also grows faster for high
$\dot M$.) The decrease of the
mass accretion rate with time competes with the
progenitor-dependent evolution of the neutrino emission 
properties (cf.\ Figs.~\ref{fig:neutrinotimeshen} and 
\ref{fig:neutrinotimels}), which decide about the strength
of the neutrino energy deposition. If the critical condition
for an explosion in 2D is met at some point, the model makes
the transition to a runaway expansion of the supernova shock.
In the 25\,$M_\odot$ star with its highest compactness 
(Fig.~\ref{fig:densityprofiles}) 
the large mass infall rate (and quickly
growing neutron star mass) prevent an explosion until 
the end of our computation.

In Fig.~\ref{fig:expenergies} the ``diagnostic energies'' are
given for the successful cases. The diagnostic energy at
each time is defined as the total energy (internal plus
gravitational plus kinetic) of all postshock material with
positive radial velocity and and positive total specific energy.
This is not yet the explosion energy of the supernova, because the
positive energies may still increase by ongoing neutrino heating
of accreted and then reejected gas, by additional energy 
associated with the neutrino-driven wind
material, by the recombination of free nucleons and
$\alpha$ particles to heavy nuclei, and by nuclear burning in the
shock-heated layers (see Sect.~\ref{sec:theory}). Moreover
the binding energy of the outer stellar layers (still ahead
of the shock) will contribute on the negative side and has to
be overcome to make the star unbound for an explosion with
excess kinetic energy at infinity. While this binding energy
is negligible for the 8.8\,$M_\odot$ progenitor (whose 
O-Ne-Mg core is surrounded by an extremely dilute and loosely 
bound H-shell) and small for the lower-mass iron-core stars
(i.e., less than $\sim$$3\times 10^{48}$\,erg in the 8.1\,$M_\odot$
case and even less for the 9.6\,$M_\odot$ star), the
binding energy of the stellar mantle and envelope can be
appreciable for more massive progenitors.
In the 11.2\,$M_\odot$ and 15\,$M_\odot$ models, for example,
the binding energies of the preshock shells are
$7.5\times 10^{49}$\,erg and $2.6\times 10^{50}$\,erg,
respectively, whereas the still available recombination
energies of the postshock matter are $2\times 10^{49}$\,erg
and 1--$2\times 10^{50}$\,erg.\cite{rf:Mueller.etal.2012a} \
In the case of the 15\,$M_\odot$ and 27\,$M_\odot$ models
the diagnostic energies are still steeply increasing at the
end of the simulations and the terminal values cannot be
guessed. In contrast, the diagnostic energies of the other
models begin to saturate. The explosions will therefore become
relatively weak, and a lot of fallback has to be expected.

\section{Magnetically supported explosions in nonrotating progenitors}
\label{sec:MHDexplosions}

Magnetic fields are known to be strongly amplified 
during the secular postbounce evolution of rapidly 
differentially rotating, collapsing stellar cores. Besides
compression upon infall, the amplification is mainly achieved
by field winding in shear layers and the magnetorotational
instability (MRI), see e.g., 
Refs.~\citen{rf:Meier.etal.1976,rf:Akiyama.etal.2003}.
One may ask, however, how strong the fields have to be, and
how strong the corresponding initial fields,
if magnetic effects are to have an impact on the development
of the explosion in nonrotating (or very slowly rotating)
stellar cores. Slow core rotation is predicted by evolution 
models of magnetized, massive stars\cite{rf:Heger.etal.2005}
because of angular momentum loss mainly during the red-giant
phase. With predicted pre-collapse spin periods of more than
hundred seconds in the iron core, the
postshock layer attains rotation periods of hundreds of 
milliseconds to many seconds and possesses only tiny amounts
of free energy of rotation (which is the reservoir to be tapped
for field amplification). In this case neither wrapping nor the
MRI are efficient mechanisms to enhance the field strength.

\begin{wrapfigure}{r}{\halftext}
 \centerline{\includegraphics[width=\halftext]{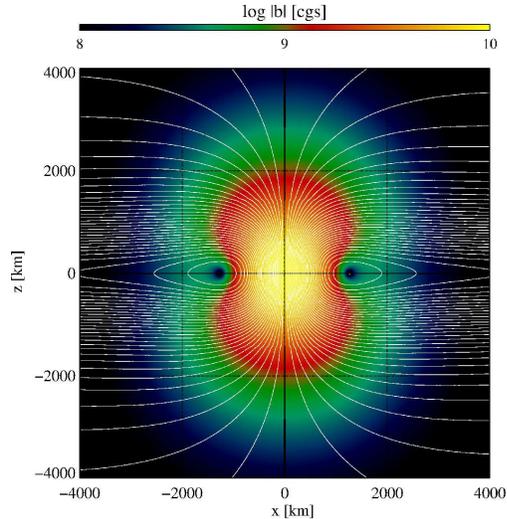}}
\caption{Field lines and field strength (color-coded) with central
 value of $B_{\mathrm{c}} = 10^{10}\,$G assumed for the initial
 (pre-collapse) field configuration of a 15\,$M_\odot$
 progenitor.\cite{rf:Woosley.etal.2002} \ Initial configurations
 with other central field values used in our simulations are
 obtained by simple scaling. In our nonrotating
 stellar cores we have adopted a purely poloidal field. This is
 clearly artificial and besides the experimental purpose
 the assumed absence of stellar rotation to create toroidal
 field components might serve as justification. A high-resolution
 version of this figure is available upon request.
 }
\label{fig:inifield}
\end{wrapfigure}

Obergaulinger \& Janka\cite{rf:ObergaulingerJanka.2011} have recently
investigated the effects of magnetic fields on the shock evolution
in nonrotating stellar cores\footnote{The results discussed here
are based on simulations with higher resolution and a 
two-dimensional, multi-energy-group, two-moment closure scheme for
neutrino transport, which was upgraded relative to the models in
Ref.~\citen{rf:ObergaulingerJanka.2011}.
The improvements will be entered in a revised version of the 
latter paper.}. In this case the amplification of 
the fields in the gas flow relies on two processes:
\begin{itemize}
\item 
  Compression of the fields against the magnetic
  pressure increases the magnetic energy density $e_{\mathrm{mag}}$; 
  magnetic energy is created by this process at a rate
  $s_{\mathrm{cmp}} = - e_{\mathrm{mag}} \vec \nabla \cdot \vec v$.
\item 
  Stretching and folding of the field lines, in which case 
  the energy density of the field can be amplified at a rate of
  $s_{\mathrm{str}} = B^i B^j \nabla_j v_i$.
  This process is the main ingredient of a turbulent dynamo.
\end{itemize}
The amplification by compression takes place in the 
converging flow of the unshocked, infalling layers, in the
flow passing the shock, and in convective downdrafts in the
postshock region. Turbulent mass motions due to neutrino-driven 
convection and the SASI\cite{rf:Endeve.etal.2010,rf:Endeve.etal.2012}
lead to further amplification by dynamo action in the region between
stalled shock and protoneutron star.

\begin{figure}
 \centerline{\includegraphics[width=0.5\textwidth]{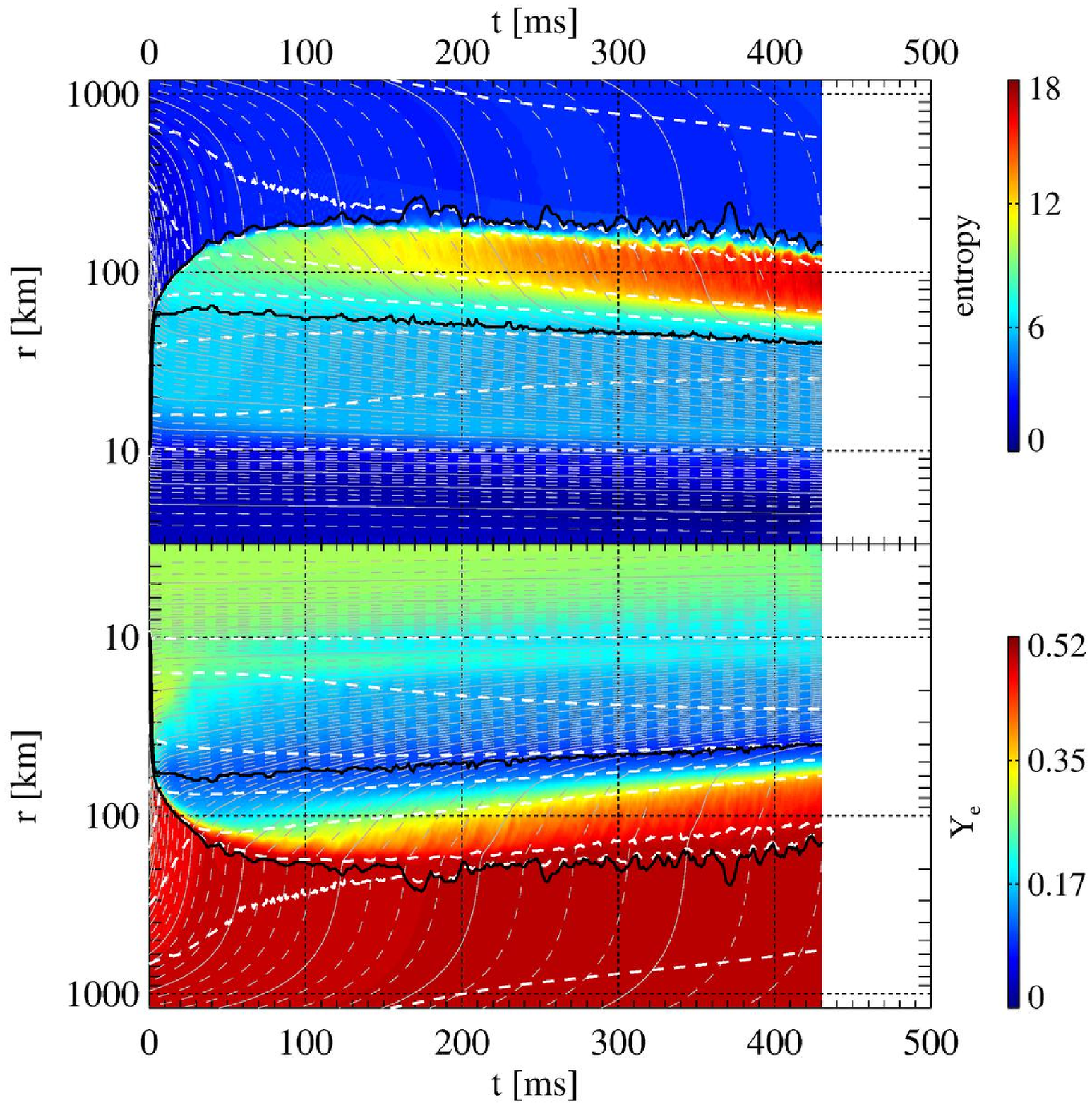}\ \
             \includegraphics[width=0.5\textwidth]{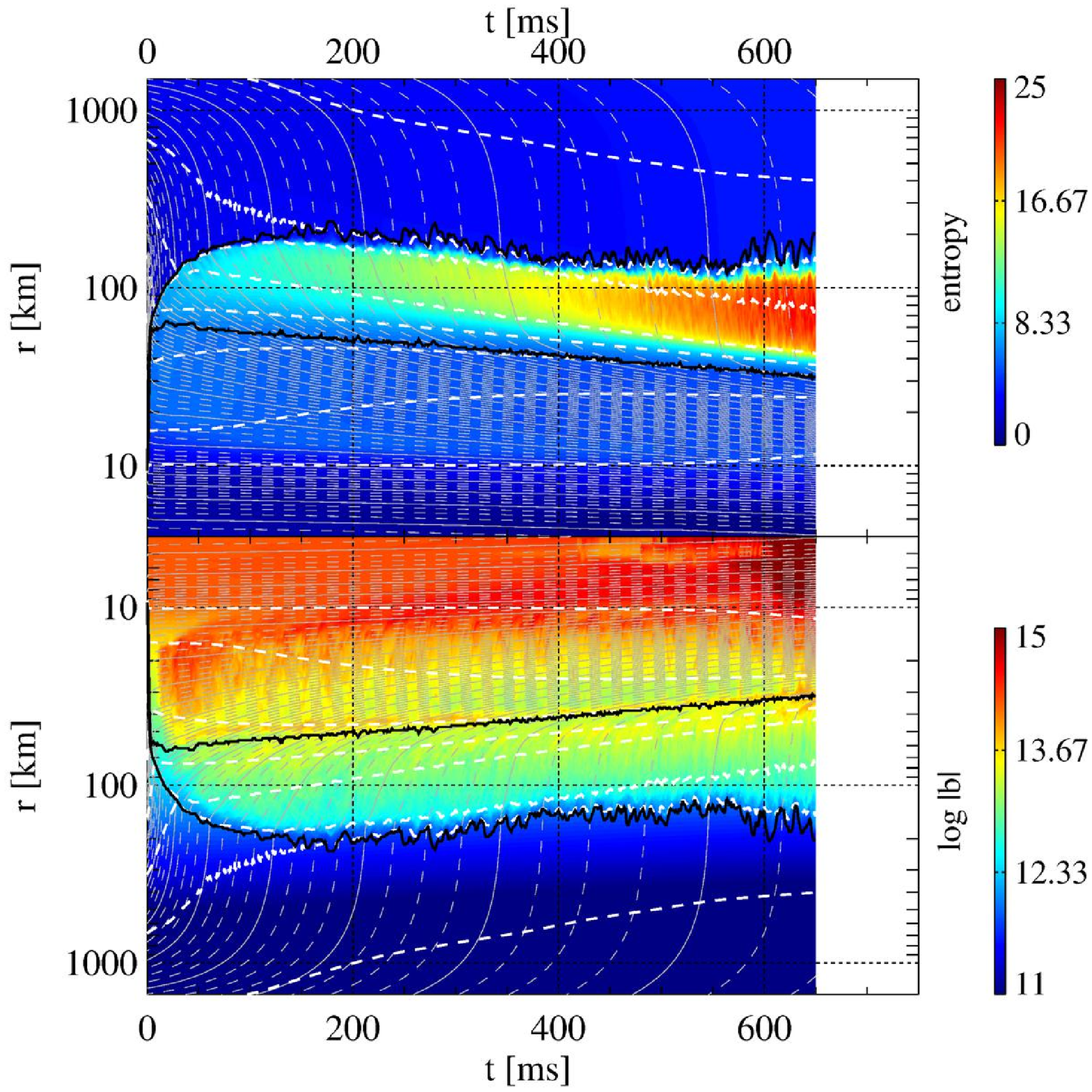}}
 \vspace{5pt}
 \centerline{\includegraphics[width=0.5\textwidth]{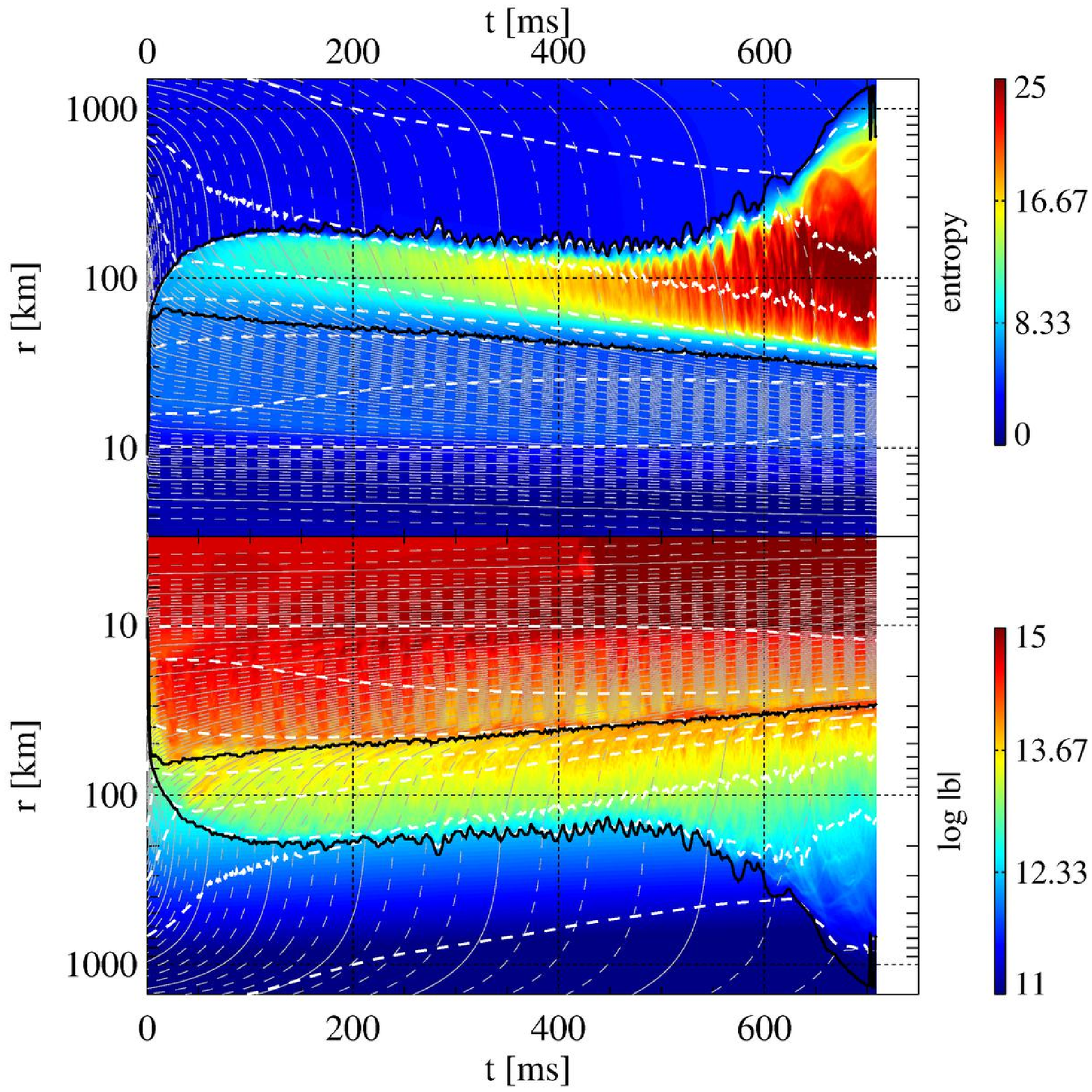}\ \
             \includegraphics[width=0.5\textwidth]{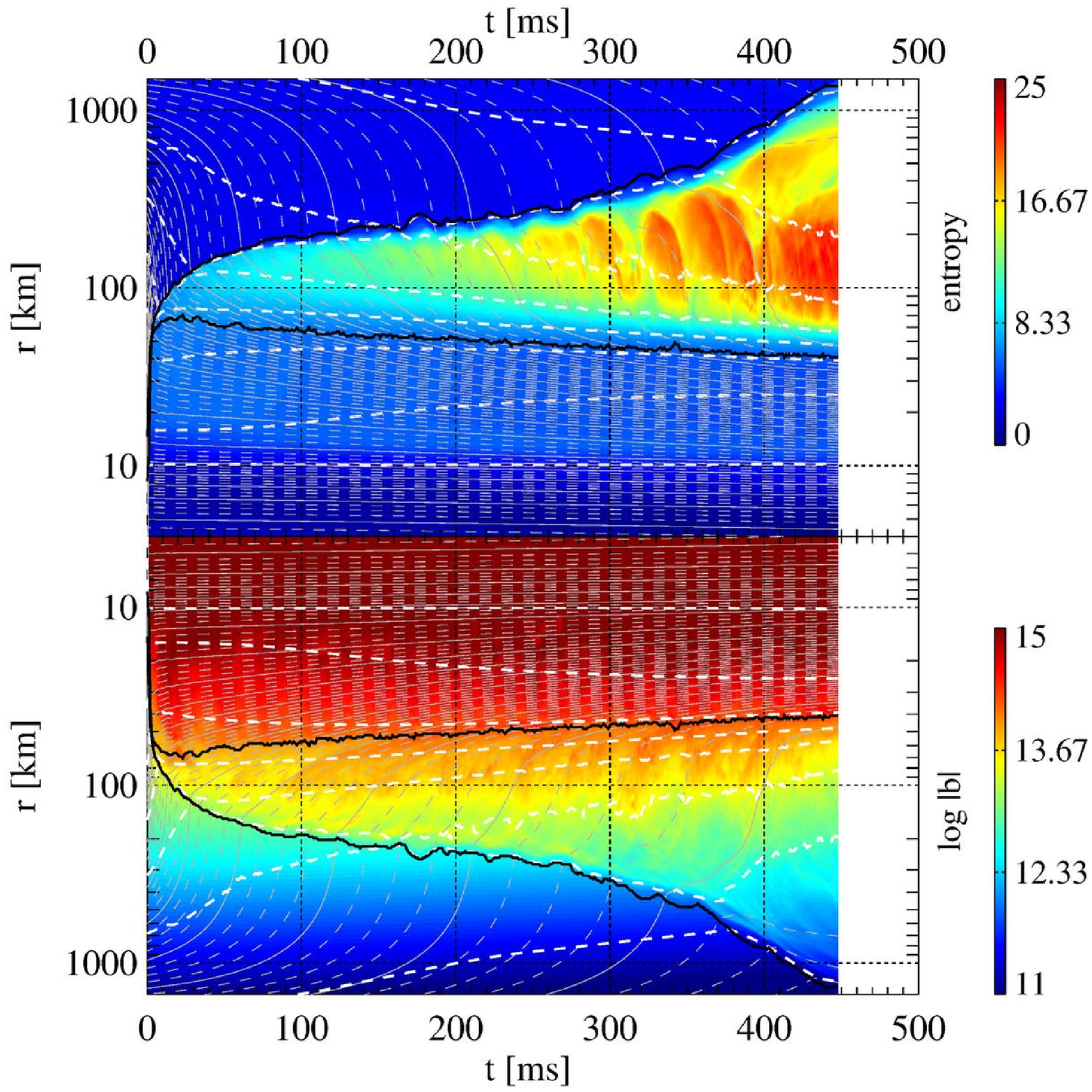}}
  \caption{
    Postbounce evolution of 2D simulations for a 15\,$M_\odot$
    progenitor\cite{rf:Woosley.etal.2002} with different assumptions of the
    initial magnetic field strength for central values of $B_{\mathrm{c}} = 0$
    ({\em top left}), $10^{11}\,$G  ({\em top right}), $10^{11.5}\,$G
    ({\em bottom left}), and $10^{12}\,$G ({\em bottom right}).
    In the case of the nonmagnetized model,
    radial profiles of the angular averages of entropy per baryon (upper
    half panel) and electron fraction, $Y_e$, are color-coded, whereas
    for the magnetized models $Y_e$ is replaced by the
    logarithm of the angularly averaged magnetic field strength.
    Note that the ordinate with the radius has been mirror imaged
    in the lower half-panels. The thin grey lines indicate trajectories
    for different fixed values of the enclosed mass spaced in steps
    of 0.025\,$M_\odot$ (dashed) and 0.1\,$M_\odot$ (solid). The
    two black lines denote the maximum radius of the shock front and
    the outer edge of the convection zone inside of the proto-neutron
    star, respectively. The white dashed lines mark locations of
    chosen constant values of the angle-averaged mass density for
    $\rho = 10^{14}$, $10^{13}$, $10^{12}$\,g\,cm$^{-3}$, etc.
    from the center outwards. The amplification of the initial core
    field during collapse, shock compression, and turbulent folding 
    fosters shock expansion and the 
    initiation of neutrino-powered explosions on a timescale that
    decreases with the strength of the magnetic field (compare
    left lower and right lower panels).
}
  \label{fig:bmodevolution}
\end{figure}

\begin{figure}
 \centerline{\includegraphics[width=\halftext]{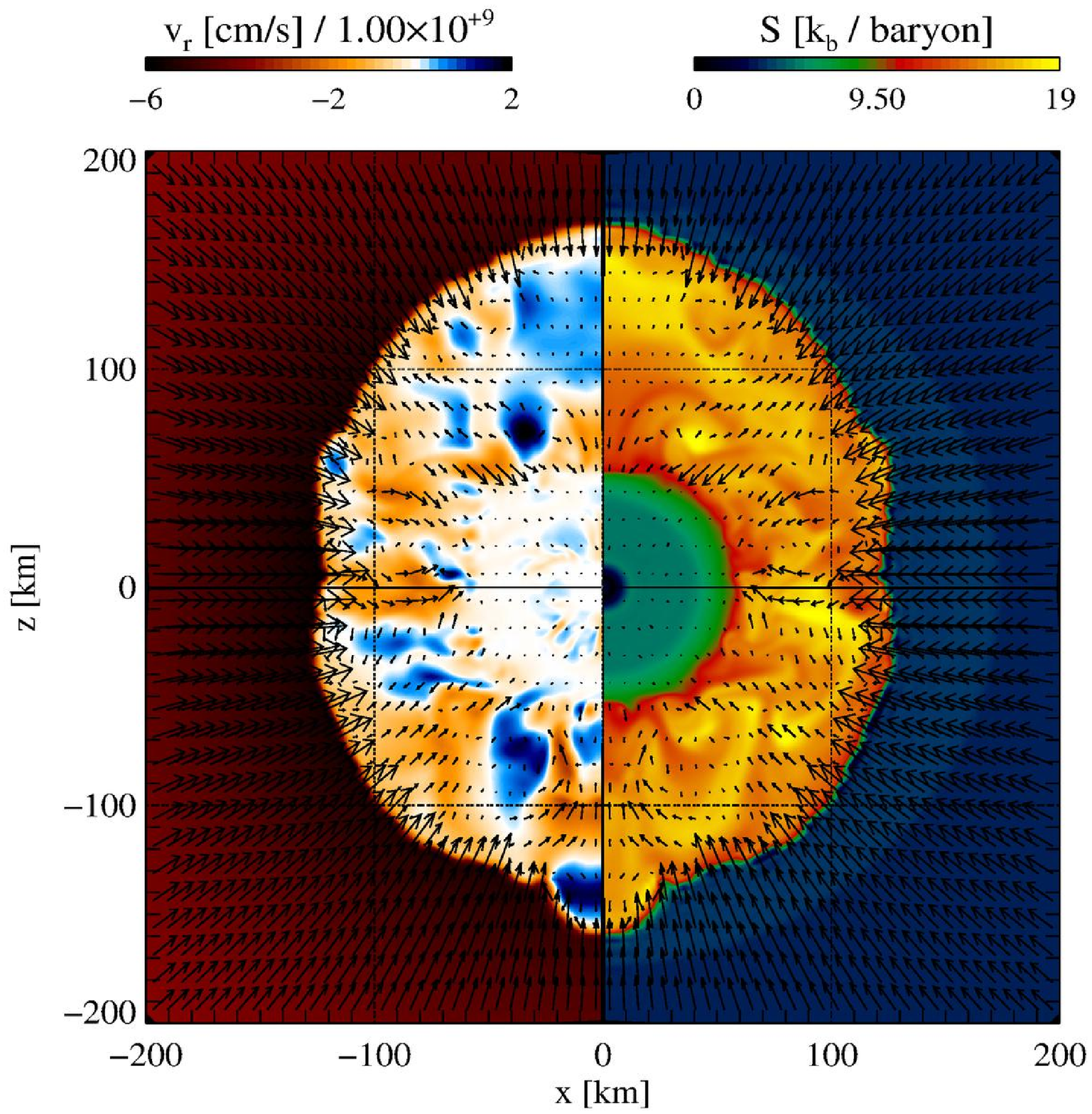}\ \
             \includegraphics[width=\halftext]{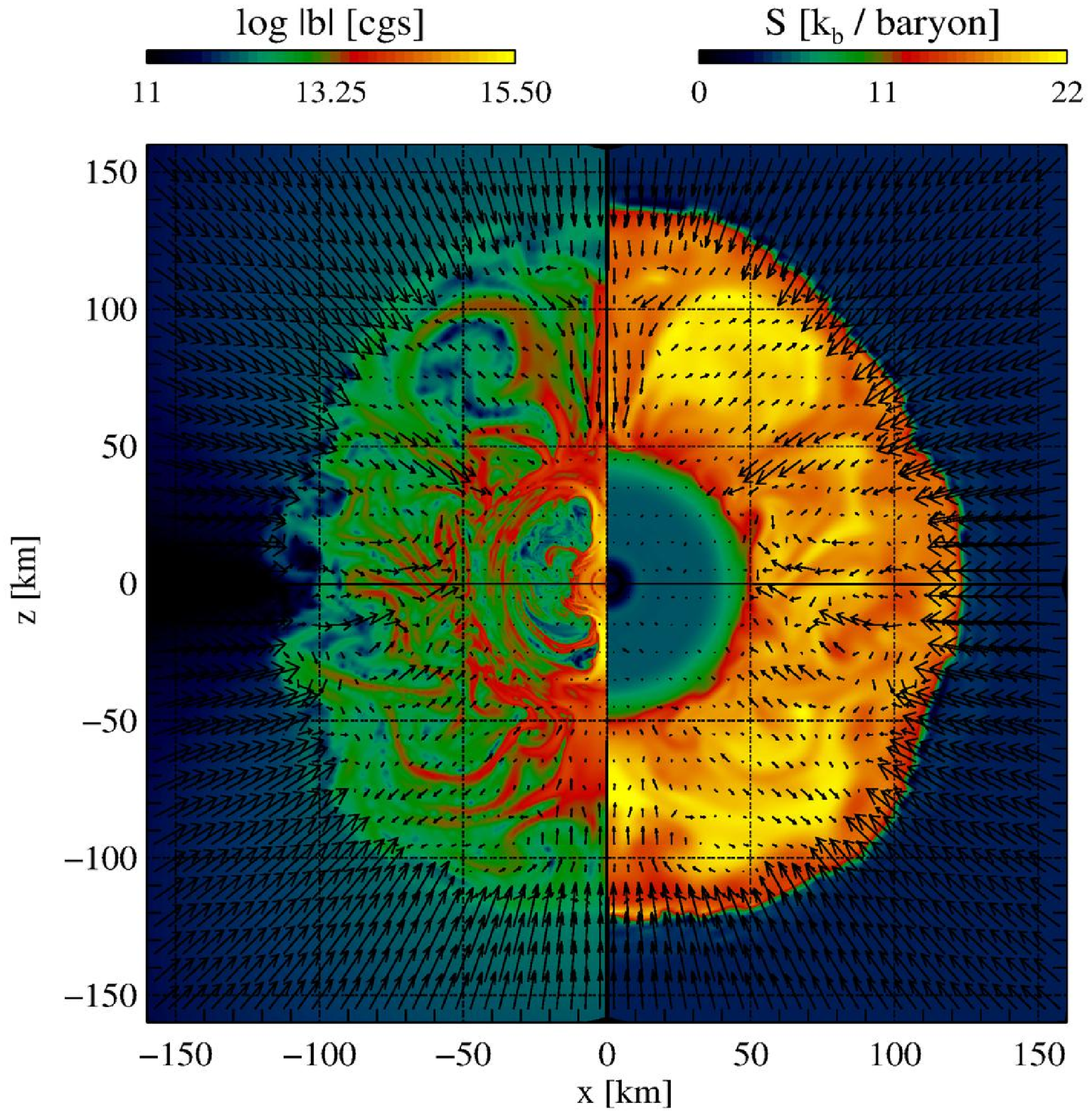}}
 \vspace{5pt}
 \centerline{\includegraphics[width=\halftext]{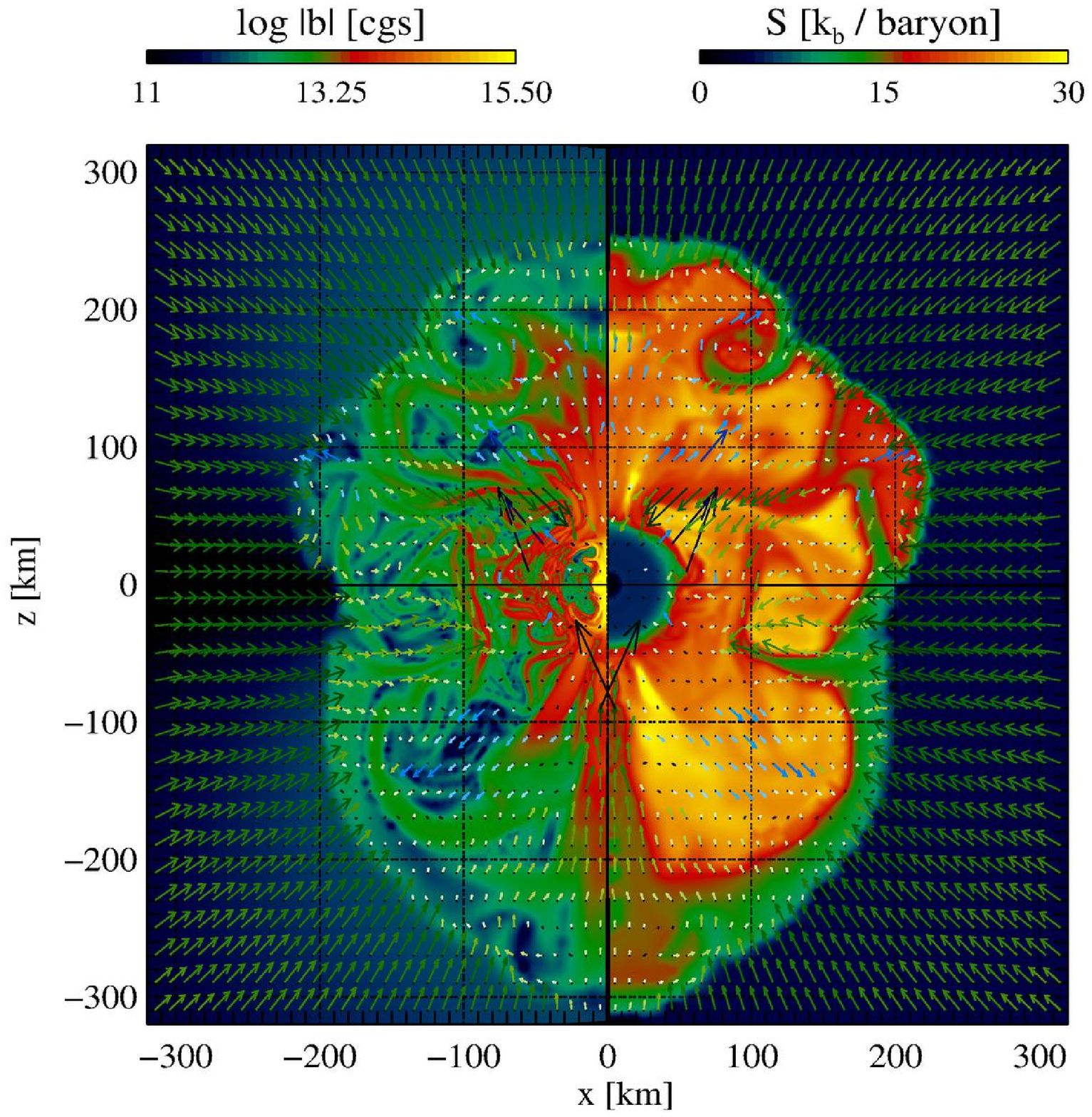}\ \
             \includegraphics[width=\halftext]{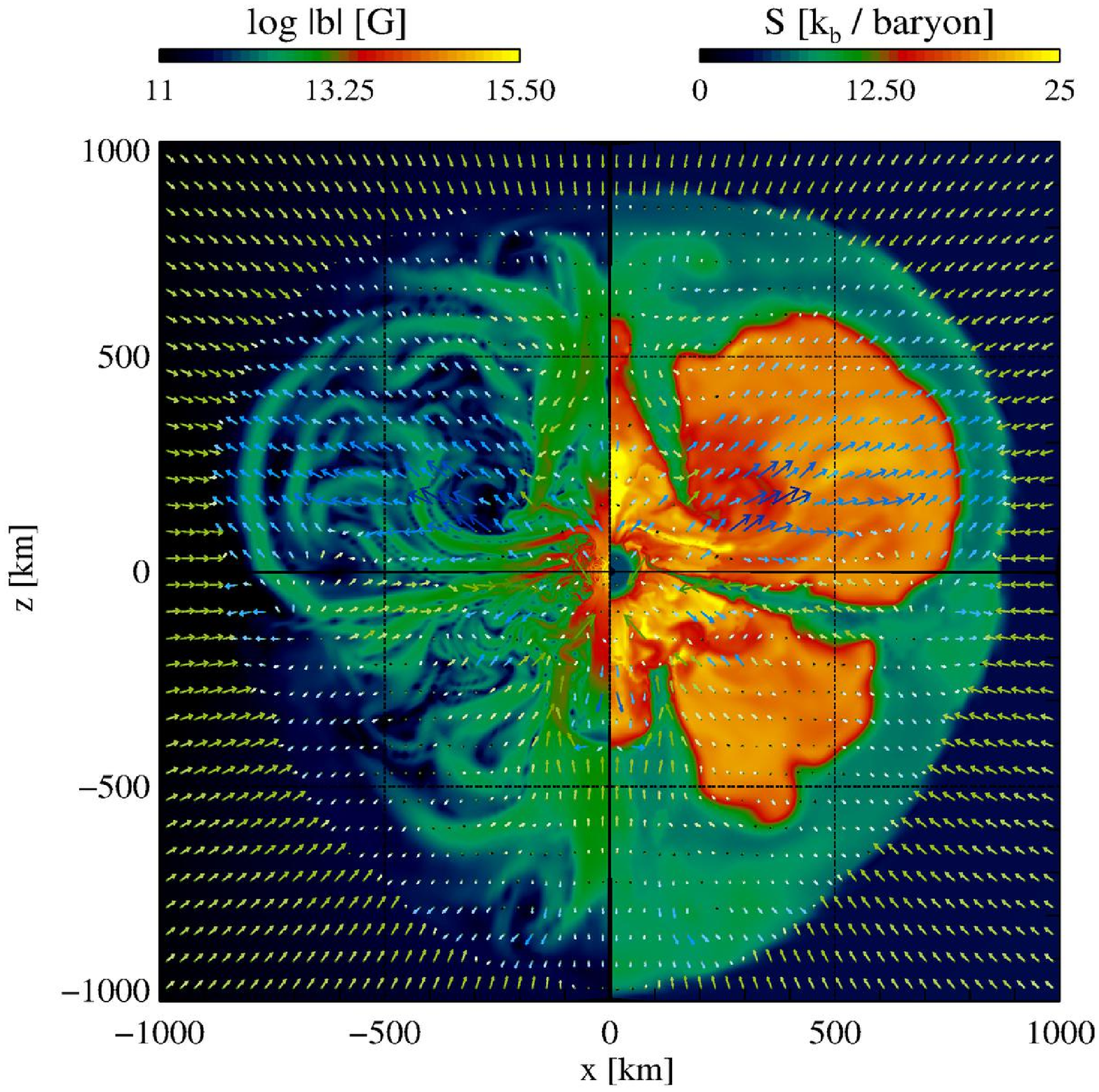}}
  \caption{Representative snapshots for the simulations displayed in
  Fig.~\ref{fig:bmodevolution}. For the nonmagnetized model the radial velocity
  (left half panel) and the entropy per nucleon (right half panel) are shown;
  for the models with magnetic fields the logarithm of the magnetic field strength
  is color-coded instead of the radial velocity. Arrows indicate the flow field.
  The chosen postbounce times are 400\,ms, 512\,ms, 574\,ms, and 412\,ms
  ({\rm from top left to bottom right}), respectively. Neutrino-heated,
  high-entropy gas rises in buoyant bubbles and pushes the shock expansion.
  The magnetic field structure in the postshock layer traces the convective
  pattern of bubbles and downflows and tends to be stronger in the compression
  regions of the downdrafts.
}
  \label{fig:bmodsnapshots}
\end{figure}

\begin{figure}
 \centerline{\includegraphics[width=0.5\textwidth]{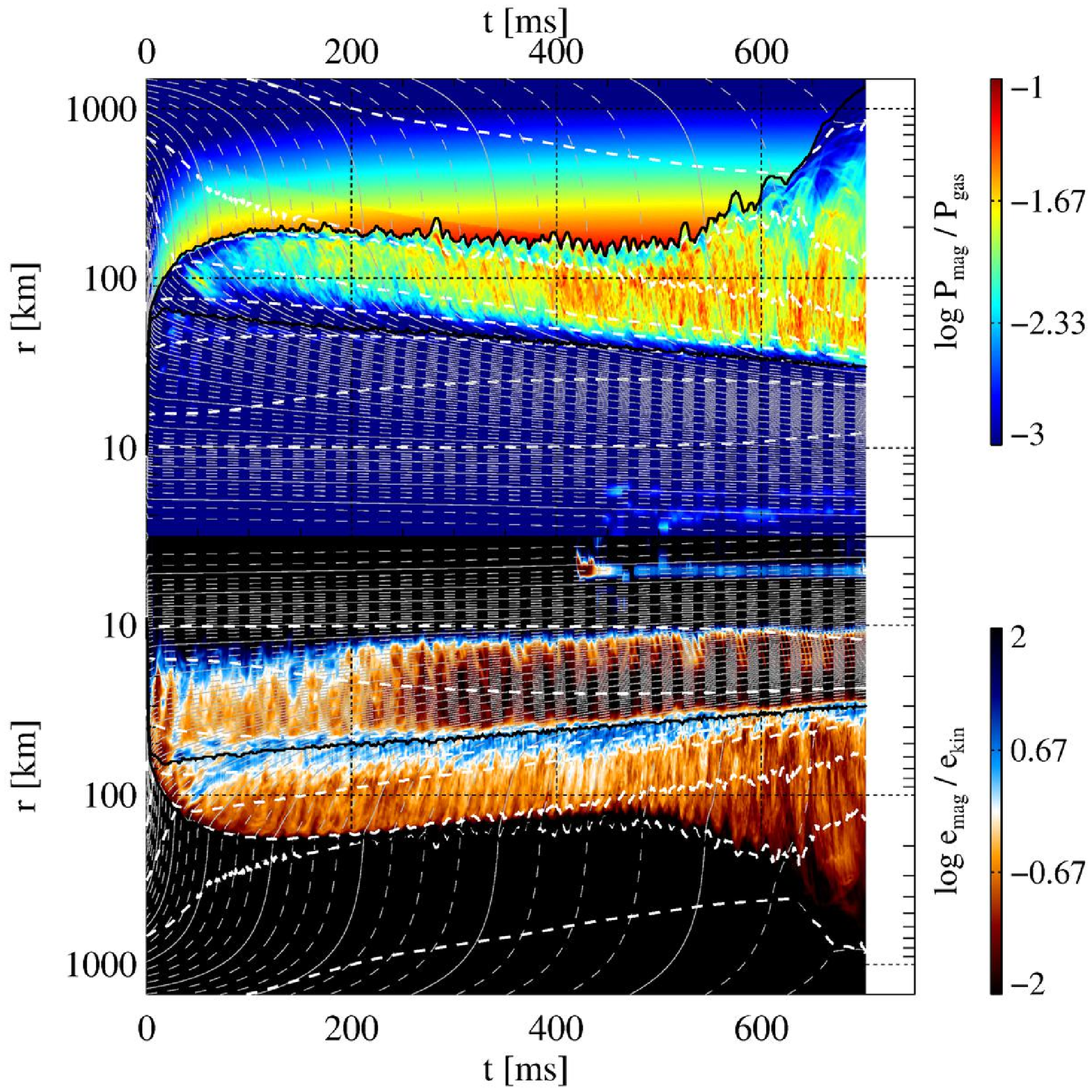}\ \
             \includegraphics[width=0.5\textwidth]{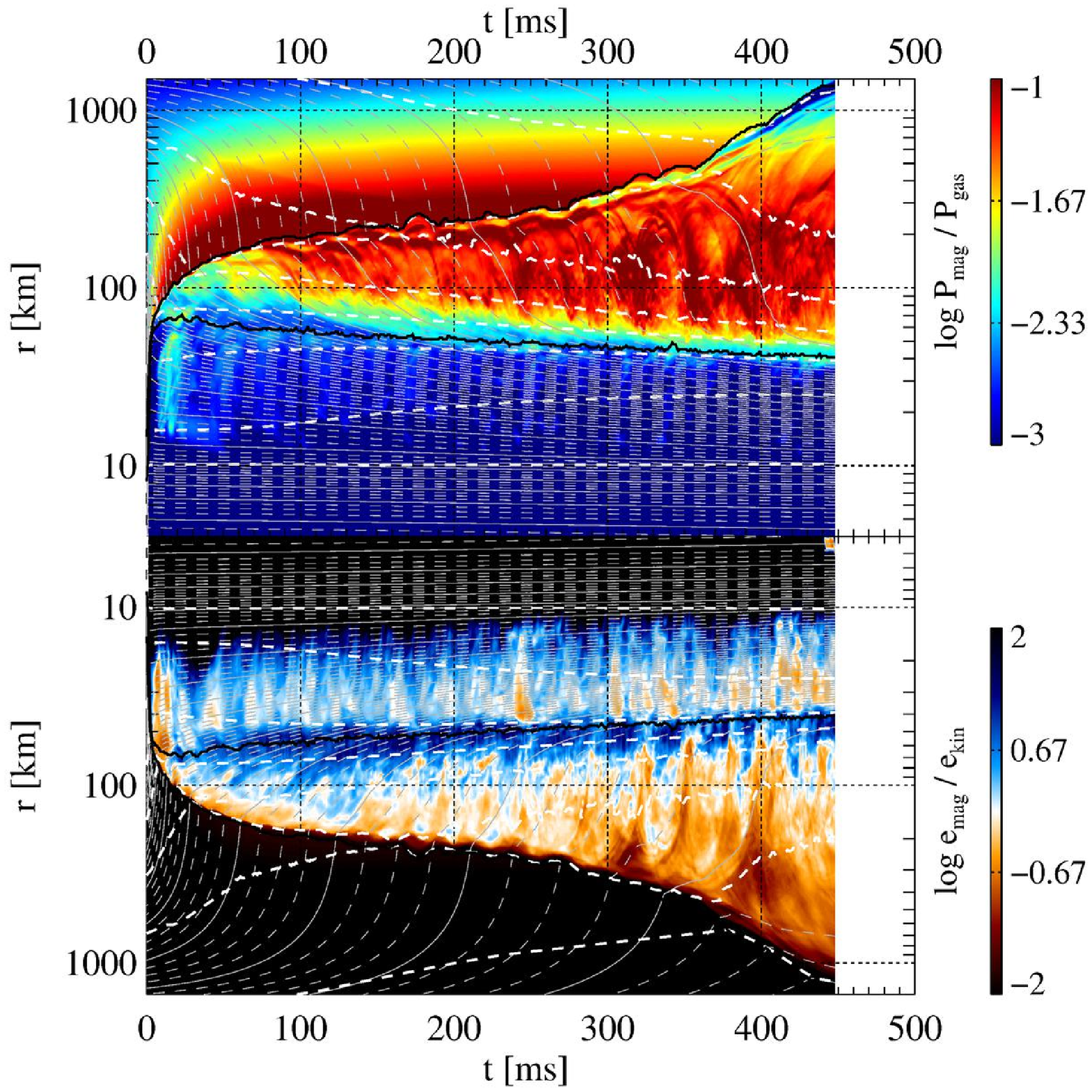}}
  \caption{Postbounce evolution of the 2D simulations with initial central
  magnetic fields of $10^{11.5}$\,G ({\em left}) and $10^{12}$\,G ({\em right}).
  The upper half-panels show the color-coded logarithm of the ratio of magnetic
  pressure to gas pressure, the lower half-panels display the logarithm of
  the ratio of magnetic energy density to kinetic energy density. The shock
  expansion is supported by significant magnetic pressure (between 
  several per cent and $\sim$10\% of the gas pressure) in
  the postshock layer. The field there reaches equipartition strength with
  the kinetic energy of the flow.
}
  \label{fig:bmodpressure}
\end{figure}

Figure~\ref{fig:bmodevolution} shows the postbounce evolution
of four simulations of a 15\,$M_\odot$ progenitor, comparing the
nonmagnetic case with three cases for different initial magnetic
field strengths, corresponding to central field values of 
$10^{11}\,$G, $10^{11.5}\,$G, and $10^{12}\,$G. The assumed
pre-collapse field configuration in the stellar core is displayed in
Fig.~\ref{fig:inifield}. The panels of Fig.~\ref{fig:bmodevolution}
visualize the time dependence of the radial profiles of angularly
averaged entropy and magnetic field values (or $Y_e$ in the nonmagnetic 
case). Spatial distributions for representative times are presented
in Fig.~\ref{fig:bmodsnapshots}.

In the lower two panels of Fig.~\ref{fig:bmodevolution} the
shock evolution is clearly affected by the presence of the magnetic
fields. The shock radius is larger compared to the nonmagnetic
model (upper left panel) and ultimately the shock begins to 
expand, followed by the gas in the postshock region. While for
an initial central field strength of $10^{12}$\,G the shock 
expansion becomes strong about 200\,ms after bounce (bottom right), 
a powerful shock expansion sets in only 500\,ms after bounce for
initial central field strength of $10^{11.5}$\,G (bottom left).
For even weaker initial field ($B_\mathrm{c} = 10^{11}$\,G;
upper right panel) one can see first hints for a transition to
shock expansion after nearly 600\,ms. The onset of shock acceleration
is associated with the development of positive total energy in
the expanding postshock material. This growth of the energy signals 
the beginning of an explosive runaway.

The dynamical relevance of the magnetic field can be expressed
by the ratio of magnetic field pressure to the gas pressure, whose
radial profiles are plotted as function of time for the two
exploding models with highest fields in the upper half-panels 
of Fig.~\ref{fig:bmodpressure}. Postshock fields of about 
$10^{13}$\,G (Fig.~\ref{fig:bmodevolution}) correspond to
a magnetic pressure contribution to the gas pressure of 
several per cent up to about 10\%. This obviously is 
sufficient to play a dynamically important role and to 
support the shock expansion. As a consequence, the inflation
of large bubbles of neutrino-heated, high-entropy gas is 
enabled and the flow in the accretion region between shock 
and gain radius becomes much more ordered, with a dominance of
low-order spherical harmonics modes in the convective pattern
(Fig.~\ref{fig:bmodsnapshots}). As the shock propagates outward,
the push of the rising bubbles of heated matter becomes the
driving force for further expansion, because the relative
importance of the magnetic pressure to the gas pressure
behind the outward rushing shock quickly decreases.
The lower half-panels 
of Fig.~\ref{fig:bmodpressure} demonstrate that in the 
turbulent postshock layer prior to the onset of the explosion
the average magnetic field grows to a value near
equipartition with the amplification-supporting
average density of kinetic energy.

Despite the constraint of our models to 2D and the unavoidable
limitations of the resolution, we think that one can
conclude that the magnetic field amplification in
collapsing stellar cores, even in the absence of any
significant amount of initial core rotation,
can lead to dynamical effects on the evolution of the
supernova shock. For sufficiently strong initial fields
magnetic pressure in the postshock layer can support the
onset of neutrino-powered explosions on relevant timescales
(of order 0.5--1\,s after core bounce). Further simulations,
in particular also 3D models
with more realistic initial conditions for the field
strength and geometry based on stellar evolution 
predictions\cite{rf:Heger.etal.2005}, are needed to clarify
whether this could play a relevant role as ingredient in the 
neutrino-driven explosion mechanism.

\section{Convection or SASI as trigger of shock expansion?}
\label{sec:sasiorconv}

A detailed discussion and analysis of the postbounce dynamics 
leading to the onset of explosions in the 2D models of
Sect.~\ref{sec:2Dexplosions} (including a description of the
associated gravitational-wave signals) is provided in
Refs.~\citen{rf:MarekJanka.2009,rf:Mueller.etal.2012a,rf:Mueller.etal.2012b,rf:Mueller.etal.2012c}. 
Some of the plots of the entropy evolution in 
Figs.~\ref{fig:entropytime} and \ref{fig:entropytime25} 
as well as corresponding movies show violent bipolar sloshing 
motions of the shock and of the matter in the postshock layer.

These bipolar shock oscillations are found to exhibit
growing amplitudes before neutrino-powered explosions 
set in. Such a behavior is visible in the 11.2\,$M_\odot$ and
15\,$M_\odot$ cases\cite{rf:MarekJanka.2009,rf:Mueller.etal.2012a},
but the alternating up and down motions of the shock are especially 
prominent in the 27\,$M_\odot$ model\cite{rf:Mueller.etal.2012b}, 
where the extremely high infall velocities in the postshock flow 
suppress the growth of neutrino-driven convection in the 
first place, unless large initial entropy perturbations drive
non-linear buoyancy from the beginning (for a discussion of 
the necessary conditions, see
Refs.~\citen{rf:Foglizzo.etal.2006,rf:Buras.etal.2006b,rf:Scheck.etal.2008,rf:Mueller.etal.2012b}). 
In this case the SASI-typical oscillatory increase of
the amplitude of low-order shock deformation modes appears 
very clearly. 
In contrast, the dynamics of the lower-mass progenitors 
does not exhibit this characteristic behavior or, at least,
it is not present there in such a pure form 
(Fig.~\ref{fig:entropytime}).
Buras et al.\cite{rf:Buras.etal.2006b} found shock oscillations
with large dipole ($\ell = 1$) amplitudes to be crucial for
the explosion of the 11.2\,$M_\odot$ case because they obtained
successful shock revival in 180-degree (pole-to-pole)
simulations, whereas simulations with a 90-degree
wedge around the equator did not explode.\footnote{Successful
shock revial,
however, could also be obtained for the 11.2\,$M_\odot$ star
with a 90-degree pole-to-equator grid.}.
The violent shock motions and the associated shock expansion
allow matter accreted through the shock to stay in the
neutrino-heating region for a longer time.
The increased abidance timescale in
the gain layer enhances the efficiency of neutrino-energy
transfer and thus supports the development of runaway
conditions.

The shock-oscillation phenomenon exhibits strong similarity 
to the SASI dynamics
that can be observed in 2D as well as 3D simulations of adiabatic 
accretion flows using an ideal-gas EoS and simple representations of 
the effects of neutrino cooling as regulator of (quasi-)stationary 
postshock conditions.\cite{rf:Blondin.etal.2003,rf:BlondinMezzacappa.2007,rf:Fernandez.2010} \ 
An experimental shallow-water analog that considers a hydraulic
jump in a converging two-dimensional water flow has also been
developed\cite{rf:Foglizzo.etal.2012} 
to demonstrate the basic aspects of the 
instability that preferentially leads to low-order
(dipolar, quadrupolar) shock deformation and violent, 
nonradial shock expansion and contraction. While the 
axis of these motions naturally coincides with the artificial
symmetry (polar-grid) axis in 2D-axisymmetric simulations, there 
is no such predetermining constraint in the experimental
setup. This confirms that the grid geometry is not the
crucial aspect that allows the SASI sloshing phenomenon to 
occur. Moreover, the growth behavior of the SASI from the
linear regime to the nonlinear stage in competition with
neutrino-driven convective instability was investigated
for supernova core conditions in 2D by Foglizzo et al.\
and Scheck et al.\ in 
Refs.~\citen{rf:Foglizzo.etal.2006,rf:Scheck.etal.2008}\footnote{Numerous
analytical and numerical studies for the linear growth regime
of the SASI in supernova core like environments been performed. 
For probably the most comprehensive discussion including relevant
references, see Ref.~\citen{rf:GuiletFoglizzo.2012}.}.
In the linear regime both instabilities
can be discriminated by their different growth behavior
(oscillatory vs.\ nonoscillatory) and by the absence or presence
of buoyant flow structures in the postshock shell. Stronger
neutrino heating or slower advection velocities in the
accretion flow were recognized to be favorable for the growth
of neutrino-driven convection, whereas for insufficient
heating and in fast accretion flows the SASI modes were 
seen to grow predominantly. Convective activity can also win 
when its growth is accelerated by strong buoyancy forces
acting on large initial entropy perturbations.
In the fully nonlinear regime, however, a clear separation
of both instabilities could not be achieved and is very
difficult, because SASI shock motions trigger secondary
convection\cite{rf:Scheck.etal.2008}, and inversely,
it is also conceivable that neutrino-driven convection feeds
back into enhanced SASI activity. Fundamental aspects of these
hydrodynamical results seem to be in agreement with global 
linear stability analysis of stalled shocks in accretion 
flows (using a microphysical EoS and taking neutrino reactions
into account) by Yamasaki \& Yamada\cite{rf:YamasakiYamada.2007},
who found the dominant growth of oscillatory or
nonoscillatory radial and nonradial modes, depending on the 
size of the imposed neutrino luminosities. 

For all these reasons Marek \& Janka\cite{rf:MarekJanka.2009}
(see also Ref.~\citen{rf:Buras.etal.2006b})
interpreted the violent shock-sloshing motions with growing
amplitudes, which preceded the explosions they found in 
11.2\,$M_\odot$ and 15\,$M_\odot$ simulations, as a consequence
of SASI activity. They were, however, fully aware that this was 
a speculative 
interpretation, a working hypothesis, which was by no means 
based on unambiguous facts and well-founded theoretical arguments.
A solid theoretical foundation was (and still is)
lacking for the nonlinear stage of the observed
phenomenon. Nevertheless, it is clear that the energy input
for the explosion has to be delivered by neutrino heating (whose
behavior was quantitatively analysed and energetically
evaluated by Marek \& Janka), and that the shock oscillation motions 
are only helpful for improving the heating conditions, but
by themselves are {\em not} the driving agency of the explosion.
In this context it should be kept in mind that the growth of the
SASI depends on an advective-acoustic cycle\cite{rf:GuiletFoglizzo.2012}.
Once shock expansion initiates mass expansion behind the shock
as well, the cycle
tends to lose its support because the inward flow of matter is
decelerated, diminished, or even quenched. It therefore seems 
unlikely that SASI is the dominant mode of nonradial mass motions
up to the stage where the explosion takes off. It may, however,
play a supportive role for establishing favorable neutrino-heating
conditions and for getting close to the final runaway situation.

\begin{figure}
 \centerline{\includegraphics[width=\halftext]{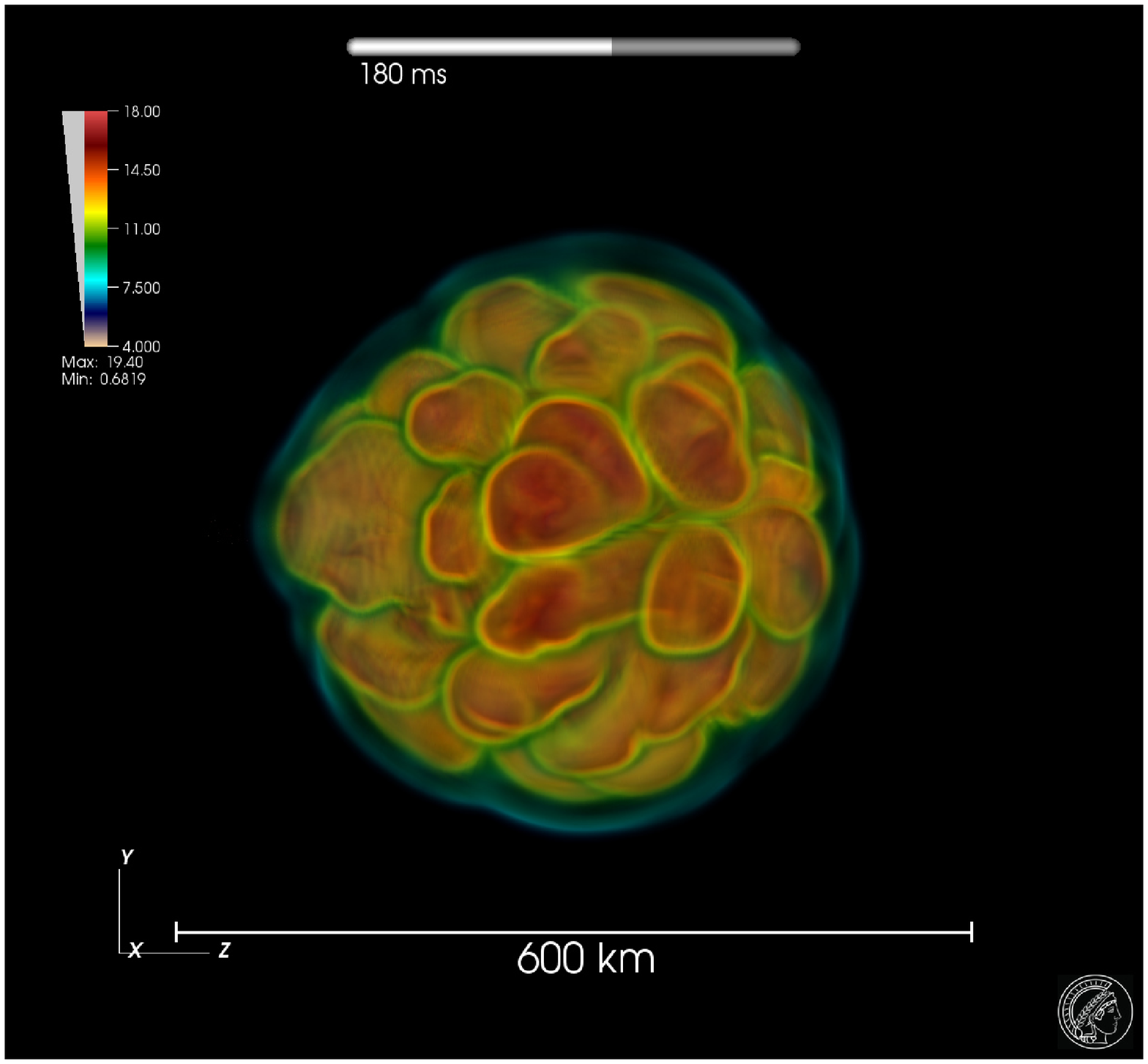}\ \
             \includegraphics[width=0.465\textwidth]{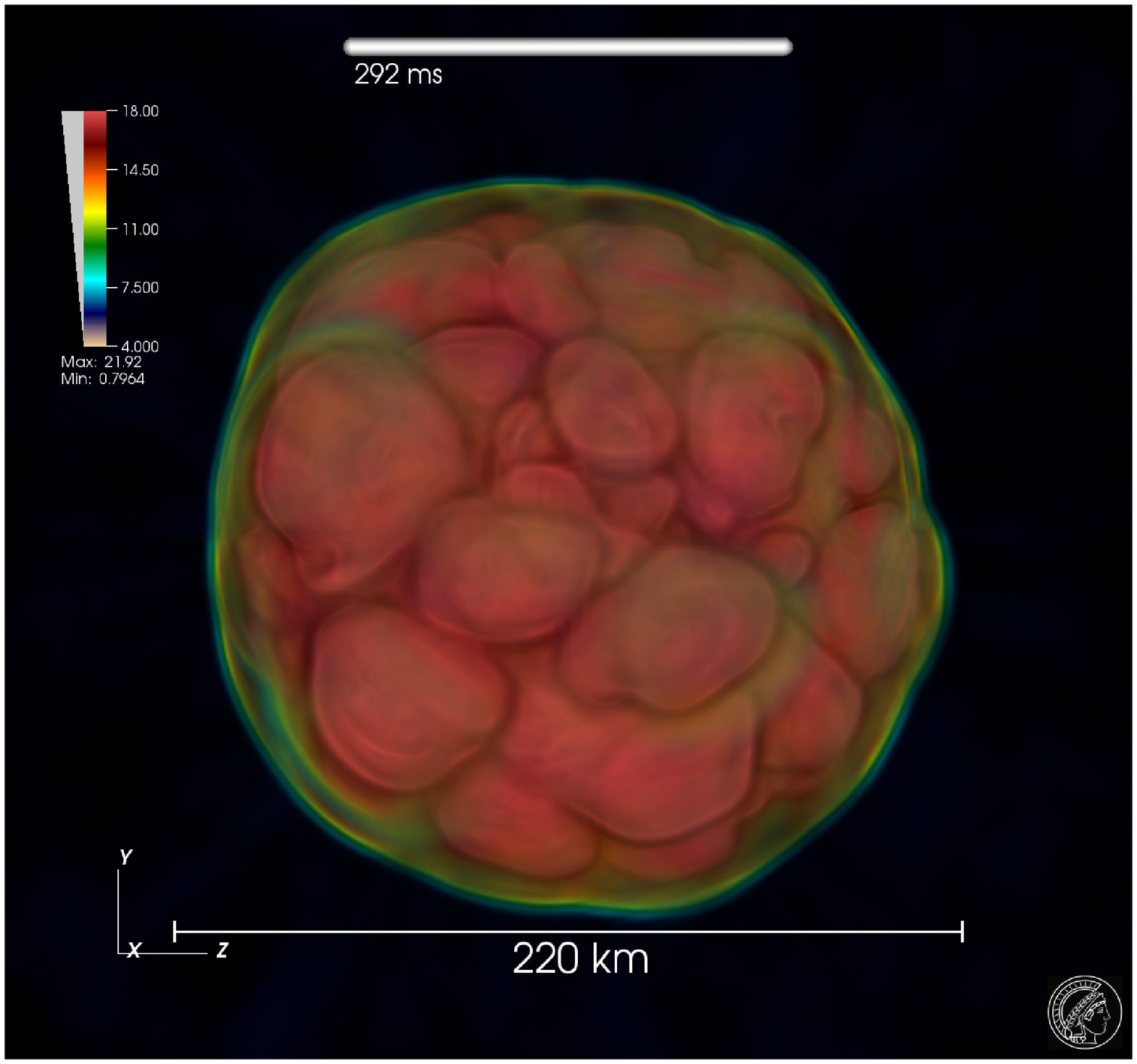}}
  \caption{Volume-rendering images of two instants (180\,ms and 292\,ms post
  bounce) of a first explorative
  3D simulation of an 11.2\,$M_\odot$ progenitor (model s11.2 of
  Ref.~\citen{rf:Woosley.etal.2002}) with full energy-dependent ray-by-ray
  three-flavor neutrino transport and the sophisticated set of neutrino
  opacities as applied also in the 1D and 2D models of the Garching group
  shown in Figs.~\ref{fig:coremasses}--\ref{fig:neutrinotimels} and
  \ref{fig:27msunexplosion}--\ref{fig:expenergies}. Neutrino-heated
  bubbles dominate the asymmetries in the postshock layer. The absence of 
  large-amplitude bipolar SASI shock motions in this simulation challenges
  speculations in the literature\cite{rf:Burrows.2012} that local neutrino
  heating computed with the ray-by-ray approximation might unphysically
  correlate with low-order mode shock
  and matter motions and could be pumping them unphysically.
}
  \label{fig:3Dmodel}
\end{figure}

Burrows\cite{rf:Burrows.2012} classified the SASI as a 
``myth that has crept into modern discourse''. (In fact,
it has sneaked into his own works, too, cf., e.g.,
Refs.~\citen{rf:Ott.etal.2008,rf:Burrows.etal.2007}.)
In his view
the vigorous dipolar shock oscillations seen in 2D 
simulations with full neutrino transport have been 
incorrectly associated with the 
SASI sloshing motions observed in simplified, neutrino-free  
studies. He argued that due to the inverse energy cascade
of 2D turbulence, which transports energy from small to 
large scales, neutrino-heated, buoyant bubble convection
with dominant low-$\ell$ modes
has been confused with the SASI, whose growth rates are
highest on the largest angular scales (i.e., for the
spherical harmonics modes of lowest order $\ell\ge 1$).
Such a possibility cannot be rejected categorically, it is
indeed a viable possibility. Just as neutrino energy input
might play a role as driving force of enhanced SASI 
activity, neutrino-driven buoyancy could also play a 
destructive role by destroying the flow coherence needed
for SASI amplification.

But a proof of a negative
feedback of neutrino-driven convection on the SASI
has so far not been given, and 
a convincing demonstration of an exclusive dominance
of neutrino-triggered buoyancy has also not been provided. 
Burrows et al.\cite{rf:Burrows.etal.2012}, for example,
argued that in their two-dimensional numerical toy model
stronger neutrino heating and thus
enhanced neutrino-driven convection was the cause of 
the growth of the dipolar sloshing amplitude that was
crucial for getting
neutrino-driven explosions. This result, however, may be
tightly connected to the simplified setup chosen for
their simulations, especially with respect to the treatment
of the neutrino source terms, by which energy losses from 
the cooling layer near the neutron star surface were
massively underestimated. Moreover, neutrino emission from
the neutron star interior was ignored and therefore the 
contraction of the neutron star was also underestimated.
Investigations based on such very particular conditions
without varying the progenitor structure, neutron star mass, 
gravity, and contaction behavior, as well as
the efficiency of neutrino cooling, are
endangered to lead to conclusions that hold for this special
case but do not necessarily possess wider validity.

In consequence of the mentioned deficiencies of the
neutrino and neutron star treatment, the accretion
velocities are reduced, which implies more favorable
conditions for the growth of neutrino-driven buoyancy while
at the same time a possible importance of the SASI is
suppressed, because the SASI is amplified more rapidly
for shorter advection timescales in faster accretion
flows (see, e.g., Ref.~\citen{rf:Scheck.etal.2008}).
It is therefore not astonishing that 
Burrows et al.\cite{rf:Burrows.etal.2012} observed only
small SASI amplitudes in the absence of neutrino heating
and the associated buoyant convection.
It is also little astonishing that
Murphy et al.\cite{rf:Murphy.etal.2012} (also 
Ref.~\citen{rf:MurphyMeakin.2011}) diagnosed the 
conditions in the same simplified modeling setup as
fully consistent with neutrino-driven convection
and turbulence in 2D as well as 3D simulations.
The 27\,$M_\odot$ calculation discussed above and
in Sect.~\ref{sec:2Dexplosions}, which includes
a sophisticated and self-consistent treatment of all
relevant physics aspects, demonstrates that there are
situations in collapsing stellar cores ---at least under 
the constraint of axisymmetry and most obvious in the 
25\,$M_\odot$ and 27\,$M_\odot$ cases--- 
where the SASI is the most
quickly growing instability and initially drives 
the shock expansion. It thus establishes, of course,
also better conditions for enhanced neutrino heating.
This may at some point change the situation from 
SASI-dominated to buoyancy-dominated, because it is 
clear that increasingly stronger neutrino heating will 
foster stronger buoyancy and thus will have a feedback
on the flow dynamics (as already discussed in
Ref.~\citen{rf:Scheck.etal.2008}). It is because of
this interdependence of both instabilities
that a dogmatic discussion about a
dominant role of either the one or the other at all
times and for all conditions seems counterproductive and
meaningless.

A more serious possible argument against an important role 
of the SASI in real supernova cores at any time 
follows from the fact that all current
3D simulations that account for neutrino heating and cooling
(Fig.~\ref{fig:3Dmodel}; 
Refs.~\citen{rf:Iwakami.etal.2008,rf:Nordhaus.etal.2010,rf:Wongwathanarat.etal.2010,rf:MuellerE.etal.2012,rf:Wongwathanarat.etal.2012,rf:Hanke.etal.2012,rf:Takiwaki.etal.2012,rf:Dolence.etal.2012,rf:Burrows.etal.2012}) 
do not show the development of large-amplitude dipolar
shock oscillations as seen in 2D models. Even for the
27\,$M_\odot$ progenitor Ott et al.\cite{rf:Ott.etal.2012}
could not find any strong evidence of SASI activity once
neutrino-driven convection has started (whose fast growth, 
according to the authors, is enabled by sizable numerical
perturbations caused by the employed cartesian grid).

Smaller saturation amplitudes of low-order shock oscillation
modes in 3D have been explained by the distribution of turbulent
kinetic energy over a larger number of available modes associated
with the additional spatial degree of freedom in 3D compared
to 2D\cite{rf:Iwakami.etal.2008}. This interpretation may be
consistent with the observation that the direction of the dipolar 
deformation wanders stochastically\cite{rf:Burrows.etal.2012} 
in 3D whereas its axis is artificially fixed to the polar
symmetry axis in 2D models.
Nevertheless, the explanation is neither fully satisfactory
nor generally valid, because it implicitly assumes
that the total kinetic energy stored in turbulent flows
of the gas in the postshock layer, in particular also
in flow structures on the largest possible scales, 
cannot become considerably higher in 3D than in 2D.
A priori a justification of this assumption is not obvious, 
but for some reason higher kinetic energies of the postshock
layer indeed seem to be disfavored as suggested by the results
in Ref.~\citen{rf:Hanke.etal.2012}, which showed roughly the same
nonradial kinetic energies in 2D and 3D models near the explosion.

Although current 3D simulations are still handicapped by
a large number of approximations in the employed microphysics
and neutrino transport, and although they mostly study highly 
simplified, parametrized
setups\cite{rf:Nordhaus.etal.2010,rf:Hanke.etal.2012,rf:Dolence.etal.2012,rf:Burrows.etal.2012} 
or suffer from very limited numerical 
resolution\cite{rf:Takiwaki.etal.2012} or artificial, grid-induced
perturbations\cite{rf:Ott.etal.2012,rf:Dolence.etal.2012,rf:Burrows.etal.2012}
(or even other, so far not realized, numerical problems in codes
newly applied to the stellar core-collapse problem),
the basic similarities of the
outcome of the increasing pool of simulations lends growing support
to expectations that SASI is subdominant in 3D.\cite{rf:Burrows.2012} \
Instead, neutrino 
heating and associated buoyancy appear to be the main trigger of
turbulent mass motions in the postshock layer and of the
shock deformation seen in the numerical models.

Reflecting current results of 2D and 3D simulations, the conclusion
seems to be unavoidable that the turbulent dynamics of the postshock
mass motions in 3D is different from 2D and that this is likely to also
affect the way how the nonradial flows in the neutrino-heating layer
exactly foster the transition of the accretion shock to runaway expansion.
If fully self-consistent, more sophisticated, well resolved, and 
not artificially grid-perturbed 3D models will confirm the current
results of simplified, incomplete, deficient, and parametrized
3D setups, in which various feedback effects are ignored, we will 
be ready to agree with Dolence et al.\cite{rf:Dolence.etal.2012}
that the supernova mechanism in 3D operates differently from 2D.
It may be expected that the basic features and implications of
3D turbulence as seen in current models
will be retained even in more elaborate simulations
(unless the numerical grid of the simulations has a nonneglibile 
influence). It is less clear, however, whether
the quasi-stationary shock expansion, which is less time-variable
and less dynamic than in the 2D models, will survive when all 
relevant feedback effects are taken into account.

In any case, however, we do not think that these aspects
imply a fundamental revision of the overall scenario of how
core-collapse supernovae achieve to explode. 
In the end they only mean refinements
in details, although, unquestionably, such refinements
could be quantitatively relevant and even crucial.
Nobody has rejected and can seriously deny the important
role of neutrino-driven convection also in 2D. Independent of 
whether or not SASI has a supplementing function in 2D, whereas 
potentially being of reduced or minor relevance in 3D,
the basic combination and interplay of 
ingredients in a working supernova mechanism will be same:
Multi-dimensional hydrodynamic instabilities help pushing the 
shock to larger radii, thus stretching the dwell time of 
shock-accreted matter in the gain layer and enhancing the
efficiency of neutrino-energy transfer. In the case of 
successful explosions, bubbles and large-scale plumes 
of neutrino-heated matter will rise
(whether constrained by axisymmetry or not) to 
ultimately bring the shock up to the runaway threshold. 
The mechanism is powered by neutrino-energy deposition, which
also provides the energy input to the blast wave for
unbinding the stellar envelope. The functioning of this
neutrino-driven mechanism 
is crucially supported by hydrodynamic instabilities.

Not each and every single dynamical aspect that needs revision or 
receives refinement in the generalization from 2D to 3D modeling
---as important for a precise and quantitative understanding 
it may be--- will mean a fundamentally new twist of this basic
picture of the mechanism. As an example in this context we mention
the discussion of the runaway growth of buoyant,
neutrino-heating-driven bubbles by
Dolence et al.\cite{rf:Dolence.etal.2012}. The elements of
this picture are not generically linked to the 3D situation
(nor is this so for the discussion of the simplified, analytic toy
model presented in the paper), they also apply to 2D geometry,
and the basic outcome of the analysis 
concerning shock evolution as function 
of time as well as critical luminosity condition remind one of
similar results for the 1D case (apart from a quantitative
reduction of the threshold value of the critical luminosity
by multi-dimensional effects).

\section{Summary and conclusions}

We have discussed results of 2D simulations of the Garching group,
in which for a growing set of progenitor stars successful,
albeit mostly weak, neutrino-driven explosions could be obtained
(Sects.~\ref{sec:2Dexplosions} and \ref{sec:MHDexplosions}).
Most of these simulations were performed with a fully relativistic
treatment and some with an approximate description of 
relativistic effects, but all of them included in a consistent
manner the effects of
hydrodynamics and of sophisticated, energy-dependent neutrino 
transport (either with 
ray-by-ray approximation or two-dimensional, two-moment closure).
Despite interesting and
quantitatively relevant differences in many details, the successful 
runs with relativistic physics are basically compatible with the
simulations of the same progenitors with relativistic approximations.

We also attempted an assessment of these 2D results in the context of 
the current dispute about the dimensional dependence of the 
hydrodynamics of core-collapse supernovae (Sect.~\ref{sec:sasiorconv}).
Since 3D fluid dynamics in the supernova core seems to differ from
2D flows in a variety of aspects, what can one learn from 
self-consistently exploding 2D models? What can 2D modeling 
tell us at all, in particular since the explosions turn out
to be weak and not to be able to reproduce observed supernova
energies?

First, the 2D simulations show that multi-dimensional effects
in the most sophisticated neutrino-hydrodynamics supernova models
indeed lift the core conditions very close to the critical 
threshold for explosions, in most of the cases explored in 
Sect.~\ref{sec:MHDexplosions} even beyond. This is 
obviously different from 1D modeling,
where failures are obtained for all but the lowest-mass supernova 
progenitors. We therefore consider the 2D successes as quickening
progress and as an excellent reason to appreciate the glass of wine
---measuring our understanding of the supernova mechanism---
as half full. Moreover, the successful explosions in 2D are
certainly a very important and necessary
confirmation of the notion that the critical luminosity needed
for runaway conditions is reduced in the multi-dimensional case.
A confirmation by sophisticated and fully self-consistent models
was unquestionably necessary, because all other
analytical and numerical studies that reached this conclusion
were done with highly idealized setups, in which most of
the complex feedback processes in real supernova-core environments,
especially also dissipative effects,
were ignored.\cite{rf:BurrowsGoshy.1993,rf:Fernandez.2012,rf:MurphyBurrows.2008,rf:Nordhaus.etal.2010,rf:Hanke.etal.2012} 

Second, one should keep in mind that the 
neutrino-driven mechanism is powered by neutrino heating,
which therefore is the crucial ingredient. Hydrodynamic instabilities,
as indispensable as they may be, play
``only'' a supportive role. Although the exact way they
operate is, of course, important for our theoretical
understanding and demands detailed 
explorations, dimension-dependent differences like the 
question whether SASI plus neutrino-driven convection or mere
neutrino-driven convection and 3D turbulence are the assisting 
agencies of shock expansion, are possibly more a kind of refinement
than an aspect of central nature in the fundamental problem of the 
explosion mechanism. Despite the constraint to two dimensions,
successfully exploding models therefore lend strong support to
the viability of the neutrino-driven mechanism.

The recent literature on the explosion mechanism in different
dimensions contains a variety of incorrect claims. One of them,
for example, is the argument that local neutrino heating associated
with ray-by-ray neutrino transport unphysically correlates with
low-order mode shock and matter motions and unphysically pumps
them.\cite{rf:Burrows.2012} \ Generally, however, a problem of 
this kind appears highly unlikely because violent,
bipolar motions of the shock and postshock layer were seen
in 2D models with ray-by-ray transport as well as with true
2D transport schemes (cf.\ Fig.~\ref{fig:bmodevolution} and, e.g., 
Refs.~\citen{rf:Burrows.etal.2007,rf:Ott.etal.2008,rf:Brandt.etal.2011})
and even with simple neutrino-heating terms based on a 
spherical neutrino-lightbulb description
(e.g., Refs.~\citen{rf:MurphyBurrows.2008,rf:Nordhaus.etal.2010,rf:Hanke.etal.2012}).
Moreover, vigorous sloshing motions of the shock were neither 
found for all investigated progenitors in 2D simulations
(see Fig.~\ref{fig:entropytime}) nor were they obtained in 3D models 
with ray-by-ray neutrino treatment (Fig.~\ref{fig:3Dmodel}), 
although such shock dynamics was present in 2D.
Ray-by-ray transport and
large-amplitude SASI sloshing of the shock 
therefore do not seem to have any tight causal connection.
Another incorrect claim concerns an alleged proposition by 
Hanke et al.~\cite{rf:Hanke.etal.2012} that 3D simulations
exhibit the tendency to become more similar to 1D results. 
Hanke et al.\ reported their observations and partly 
speculative interpretation of results
for a very specific, highly artificial
and simplified modeling setup adopted from other works
for comparison, in which they found less 
readiness for explosions in 3D simulations with increasing grid
resolution. This result is still not understood in a broader
context, but in the unsuccessful 3D cases the shock evolution
and some, but not all, explosion-relevant quantities
(naturally and undisputably) became more similar to 1D 
simulations, and the 
postshock flow showed less vigorous mass motions than in
the corresponding, successful 2D models. The statement 
by Hanke et al.\ was therefore made in reference to these special
results but was not meant to be a prediction for 3D supernova
dynamics in general.

Current simulations in different dimensions suggest
interesting possible differences of the flow dynamics
in supernova cores between 1D, 2D, and 3D. In (artifically 
exploded) 1D models the transition from shock stagnation and
accretion to runaway shock expansion takes
place either through a number of radial shock pulsations or
by a continuous growth of the shock radius until 
accelerating expansion sets in.\cite{rf:Buras.etal.2006a,rf:Fernandez.2012} \ 
While self-consistent 1D simulations do not produce explosions,
2D models show a reduced neutrino-luminosity threshold
for the onset of explosions. In this case violent, nonradial
shock oscillations (potentially connected to SASI activity)
with increasing amplitude are seen for many but not for all 
progenitors. The conditions in the progenitor seem to play
a crucial role for the type of observed behavior. More violent
shock oscillations appear to occur in more massive stars with
compact cores. Less massive
progenitors with their less compact cores and faster decay of 
the mass accretion rate exhibit more signs of classical 
neutrino-driven buoyancy, in which case the shock expansion
is more spherical and more continuous. In 3D, at least as 
far as current models of different sophistication ---but only
few with self-consistent coupling of hydrodynamics and neutrino
transport\cite{rf:MuellerE.etal.2012,rf:Wongwathanarat.etal.2010,rf:Wongwathanarat.etal.2012,rf:Takiwaki.etal.2012,rf:Ott.etal.2012},
and all of them still with many simplications and deficiencies---
allow one to conclude, the shock expansion appears to be more
quasi-stationary, driven by the inflation of neutrino-heated
bubbles. The turbulent fragmentation of these bubbles leads 
to potentially large shock deformation but seems to push the 
shock outward in all directions more continuously than in 2D.
 
While the first 3D results are inspiring and show new possible
twists and directions in the interplay of different physical 
ingredients of the highly complex problem of the supernova explosion
mechanism, these first 3D simulations leave more open questions
than they can answer. While general agreement exists about the
supportive influence of multi-dimensional postshock flows in 
2D, present results are inconclusive whether there is a further
reduction of the critical luminosity in 3D compared to 2D.
The results of Hanke et al.\cite{rf:Hanke.etal.2012} are in 
conflict with corresponding claims by 
Nordhaus et al.\cite{rf:Nordhaus.etal.2010}, and the more
recent results of the Princeton 
group\cite{rf:Burrows.etal.2012,rf:Dolence.etal.2012} also show
a considerable weakening of the effect advocated previously: 
Although the newest 3D simulations of the Princeton group
still yield somewhat earlier 
($\sim$100--200\,ms) explosions than their 2D models, the
critical luminosity differences found before have 
nearly disappeared. A general trend to faster explosions
in 3D, however, is neither clearly supported by the results of
Ref.~\citen{rf:Hanke.etal.2012} nor by the outcome of more
sophisticated models by L.~Scheck\cite{rf:Scheck.PhD2006}.
So does 3D turbulence foster the onset of explosions even
better than 2D flows? And if so, how large is the difference
and by which effects is it caused?
Turbulence in 3D pumps energy into smaller scales, but the
energy stored on these scales is much lower than that on
large scales. How can this have a considerable, supportive
influence on the shock revival? Only
a smaller fraction of the fluid elements (tracer particles)
in 3D simulations\cite{rf:Takiwaki.etal.2012,rf:Dolence.etal.2012} 
is found to possess longer dwell times in the gain layer, but
a significantly larger fraction has shorter ones than in 2D,
which is fully compatible with the longer neutrino-heating 
timescales and lower net heating rates seen in 3D. If it is
not more efficient neutrino heating, which 3D effects 
could account for faster explosions than in 2D models? 
Is it possible that different dimension-dependent effects
in opposite directions compete with each other and partially
compensate each other, which might explain seemingly
contradictory trends of 2D vs.\ 3D found by different groups?
Conclusive answers have not been given yet.
In view of the turbulent energy cascade, are vigorous 
low-order modes of postshock mass motions generally
excluded in 3D? Does neutrino-driven convection possibly 
destroy the
coherence of SASI sloshing and spiral modes? Does SASI play
a role at all in supernova cores or are all relevant 
mass motions in 2D and 3D pure neutrino-driven convection
as hypothesized in Ref.~\citen{rf:Burrows.2012}. Again,
conclusive answers do not exist. 
If 3D turbulence on small scales is the crucial phenomenon
that discriminates the onset of explosions in 3D from those
in 2D, what is the influence of numerical resolution, what
are the consequences of different grid treatments (spherical
vs.\ cartesian, adaptive meshes vs.\ fixed ones) and 
of different hydro solvers?
And, on the level of physics, what is the influence of 
neutrino viscosity and in particular of magnetic viscosity?
If small-scale turbulence is relevant, the amplification
of the fields may not be negligible and may have an influence
on the dissipation scale and on the transport of energy.

In order to receive reliable answers, self-consistent
models with all relevant ingredients will be needed.
Not every simplified approach that may be sufficient for 
studying observable consequences and measurable signals
connected to supernova explosions
initiated by the neutrino-driven mechanism,
is equally justified when the explosion mechanism
itself and its determining factors are to be explored.
The lack of self-consistency and of important feedback effects
may produce misleading results where subdominant effects show
up much more strongly than they would if all relevant physics
were included. In order to reliably account for the feedback
of neutrino transport, generic transport effects have to
be included and sufficient resolution is indispensable, in
particular also in the surface-near neutron star layers where
the density gradient steepens with time and the 
accretion-regulating cooling region is squeezed into a 
narrow shell with tremendously increasing cooling rates
per nucleon (in Ref.~\citen{rf:Buras.etal.2006a} the 
cooling rates at late postbounce times were at least 10 times 
higher than those shown in Ref.~\citen{rf:Dolence.etal.2012}!).
Also numerical artifacts potentially caused by grid effects 
and grid perturbations or numerical diffusion, e.g.\ connected
to adaptive mesh refinement, will have to be investigated.
Finally, if different hydrodynamic instabilities compete
in their growth, the results might not be independent of the
initial seed perturbations, and the still unsettled 
variations of physical quantities in the convective regions
of the progenitor cores might predetermine the growth of the 
instabilities after core
bounce.\cite{rf:Ott.etal.2012,rf:ArnettMeakin.2011} \ 
Not only is a reliable definition of these initial conditions
needed; also (noise-free) dynamical simulation codes 
that allow for a tracking of the growth behavior of these
initial asymmetries during infall and postbounce evolution,
will be necessary.

Supernova modelers have only now begun to touch the
vast wealth of dynamical phenomena that may play a role in the
explosion mechanism of neutrino-heated supernova cores
in the third dimension. Astonishing discoveries may be waiting
for us when we continue to move forward into this unexplored
territory. Don't let us freeze in humility in the face of this
age-old challenge but let us get down to work! A lot of fun will
be our reward!

\section*{Acknowledgements}
We thank Elena Erastova and Markus Rampp
(Max-Planck-Rechenzentrum Garching) for their help in the visualization
of our 3D data in Fig.~\ref{fig:3Dmodel}.
Data from simulations by the Garching group are accessible either
openly or upon request at {\tt http://www.mpa-garching.mpg.de/\-ccsnarchive/}.
This work was supported by Deutsche Forschungsgemeinschaft through
grants SFB/TR7 ``Grav\-i\-tational-\-Wave Astronomy'' and
EXC~153 ``Cluster of Excellence: Origin and Structure of the
Universe''. M.O.\ is grateful for support
from the European Research Council (grant CAMAP-259276).
We acknowledge that results in this paper have been achieved 
using the PRACE (Tier-0) Research Infrastructure resources
CURIE (France/CEA) and SuperMUC (Germany/LRZ). Moreover,
HPC resources (Tier-1) at the NIC/J\"ulich through a PRACE-2IP/DECI-7
grant are acknowledged.

%

\end{document}